\documentclass[twocolumn,letterpaper,aps,prc,longbibliography,superscriptaddress,nofootinbib,floatfix]{revtex4-1}

\usepackage{graphicx}
\usepackage{xspace}
\usepackage{amsmath}
\usepackage{placeins}
\usepackage{tabularx}
\newcommand{\sqsntwo}{\mbox{$\sqrt{s_{_{NN}}}=200$~GeV}\xspace}
\newcommand{\pp}{\mbox{p$+$p}\xspace}
\newcommand{\ppb}{\mbox{p$+$Pb}\xspace}
\newcommand{\pbpb}{\mbox{Pb$+$Pb}\xspace}
\newcommand{\auau}{\mbox{Au$+$Au}\xspace}
\newcommand{\m}[1]{\mathrm{#1}}
\let\oldc=\c
\renewcommand{\c}[1]{\mathcal{#1}}
\newcommand{\Eq}[1]{Eq.~(\ref{#1})}
\newcommand{\up}{\uparrow}
\newcommand{\dn}{\downarrow}
\newcommand{\mT}{m_{T}}
\newcommand{\pT}{p_{T}}
\newcommand{\kT}{K_{T}}
\newcolumntype{Y}{>{\centering\arraybackslash}X}

\begin{document}

\title{L\'evy-stable two-pion Bose-Einstein correlations in 
$\sqrt{s_{_{NN}}}=200$~GeV Au$+$Au collisions }

\newcommand{\abilene}{Abilene Christian University, Abilene, Texas 79699, USA}
\newcommand{\augie}{Department of Physics, Augustana University, Sioux Falls, South Dakota 57197, USA}
\newcommand{\banaras}{Department of Physics, Banaras Hindu University, Varanasi 221005, India}
\newcommand{\barc}{Bhabha Atomic Research Centre, Bombay 400 085, India}
\newcommand{\baruch}{Baruch College, City University of New York, New York, New York, 10010 USA}
\newcommand{\bnlcoll}{Collider-Accelerator Department, Brookhaven National Laboratory, Upton, New York 11973-5000, USA}
\newcommand{\bnlphys}{Physics Department, Brookhaven National Laboratory, Upton, New York 11973-5000, USA}
\newcommand{\caucr}{University of California-Riverside, Riverside, California 92521, USA}
\newcommand{\charlesczech}{Charles University, Ovocn\'{y} trh 5, Praha 1, 116 36, Prague, Czech Republic}
\newcommand{\chonbuk}{Chonbuk National University, Jeonju, 561-756, Korea}
\newcommand{\cns}{Center for Nuclear Study, Graduate School of Science, University of Tokyo, 7-3-1 Hongo, Bunkyo, Tokyo 113-0033, Japan}
\newcommand{\colorado}{University of Colorado, Boulder, Colorado 80309, USA}
\newcommand{\columbia}{Columbia University, New York, New York 10027 and Nevis Laboratories, Irvington, New York 10533, USA}
\newcommand{\czechtech}{Czech Technical University, Zikova 4, 166 36 Prague 6, Czech Republic}
\newcommand{\dapnia}{Dapnia, CEA Saclay, F-91191, Gif-sur-Yvette, France}
\newcommand{\debrecen}{Debrecen University, H-4010 Debrecen, Egyetem t{\'e}r 1, Hungary}
\newcommand{\elte}{ELTE, E{\"o}tv{\"o}s Lor{\'a}nd University, H-1117 Budapest, P{\'a}zm{\'a}ny P.~s.~1/A, Hungary}
\newcommand{\eszterhazy}{Eszterh\'azy K\'aroly University, K\'aroly R\'obert Campus, H-3200 Gy\"ongy\"os, M\'atrai \'ut 36, Hungary}
\newcommand{\ewha}{Ewha Womans University, Seoul 120-750, Korea}
\newcommand{\fsu}{Florida State University, Tallahassee, Florida 32306, USA}
\newcommand{\gsu}{Georgia State University, Atlanta, Georgia 30303, USA}
\newcommand{\hanyang}{Hanyang University, Seoul 133-792, Korea}
\newcommand{\hiroshima}{Hiroshima University, Kagamiyama, Higashi-Hiroshima 739-8526, Japan}
\newcommand{\howard}{Department of Physics and Astronomy, Howard University, Washington, DC 20059, USA}
\newcommand{\ihepprot}{IHEP Protvino, State Research Center of Russian Federation, Institute for High Energy Physics, Protvino, 142281, Russia}
\newcommand{\illuiuc}{University of Illinois at Urbana-Champaign, Urbana, Illinois 61801, USA}
\newcommand{\inrras}{Institute for Nuclear Research of the Russian Academy of Sciences, prospekt 60-letiya Oktyabrya 7a, Moscow 117312, Russia}
\newcommand{\instpasczech}{Institute of Physics, Academy of Sciences of the Czech Republic, Na Slovance 2, 182 21 Prague 8, Czech Republic}
\newcommand{\isu}{Iowa State University, Ames, Iowa 50011, USA}
\newcommand{\jaea}{Advanced Science Research Center, Japan Atomic Energy Agency, 2-4 Shirakata Shirane, Tokai-mura, Naka-gun, Ibaraki-ken 319-1195, Japan}
\newcommand{\jyvaskyla}{Helsinki Institute of Physics and University of Jyv{\"a}skyl{\"a}, P.O.Box 35, FI-40014 Jyv{\"a}skyl{\"a}, Finland}
\newcommand{\kek}{KEK, High Energy Accelerator Research Organization, Tsukuba, Ibaraki 305-0801, Japan}
\newcommand{\korea}{Korea University, Seoul, 136-701, Korea}
\newcommand{\kurchatov}{National Research Center ``Kurchatov Institute", Moscow, 123098 Russia}
\newcommand{\kyoto}{Kyoto University, Kyoto 606-8502, Japan}
\newcommand{\labllr}{Laboratoire Leprince-Ringuet, Ecole Polytechnique, CNRS-IN2P3, Route de Saclay, F-91128, Palaiseau, France}
\newcommand{\lahorelums}{Physics Department, Lahore University of Management Sciences, Lahore 54792, Pakistan}
\newcommand{\lawllnl}{Lawrence Livermore National Laboratory, Livermore, California 94550, USA}
\newcommand{\losalamos}{Los Alamos National Laboratory, Los Alamos, New Mexico 87545, USA}
\newcommand{\lpc}{LPC, Universit{\'e} Blaise Pascal, CNRS-IN2P3, Clermont-Fd, 63177 Aubiere Cedex, France}
\newcommand{\lund}{Department of Physics, Lund University, Box 118, SE-221 00 Lund, Sweden}
\newcommand{\lyon}{IPNL, CNRS/IN2P3, Univ Lyon, Université Lyon 1, F-69622, Villeurbanne, France}
\newcommand{\maryland}{University of Maryland, College Park, Maryland 20742, USA}
\newcommand{\mass}{Department of Physics, University of Massachusetts, Amherst, Massachusetts 01003-9337, USA}
\newcommand{\michigan}{Department of Physics, University of Michigan, Ann Arbor, Michigan 48109-1040, USA}
\newcommand{\muhlenberg}{Muhlenberg College, Allentown, Pennsylvania 18104-5586, USA}
\newcommand{\myongji}{Myongji University, Yongin, Kyonggido 449-728, Korea}
\newcommand{\nagasaki}{Nagasaki Institute of Applied Science, Nagasaki-shi, Nagasaki 851-0193, Japan}
\newcommand{\nara}{Nara Women's University, Kita-uoya Nishi-machi Nara 630-8506, Japan}
\newcommand{\natmephi}{National Research Nuclear University, MEPhI, Moscow Engineering Physics Institute, Moscow, 115409, Russia}
\newcommand{\newmex}{University of New Mexico, Albuquerque, New Mexico 87131, USA}
\newcommand{\nmsu}{New Mexico State University, Las Cruces, New Mexico 88003, USA}
\newcommand{\ohio}{Department of Physics and Astronomy, Ohio University, Athens, Ohio 45701, USA}
\newcommand{\ornl}{Oak Ridge National Laboratory, Oak Ridge, Tennessee 37831, USA}
\newcommand{\orsay}{IPN-Orsay, Univ.~Paris-Sud, CNRS/IN2P3, Universit\'e Paris-Saclay, BP1, F-91406, Orsay, France}
\newcommand{\pnpi}{PNPI, Petersburg Nuclear Physics Institute, Gatchina, Leningrad region, 188300, Russia}
\newcommand{\riken}{RIKEN Nishina Center for Accelerator-Based Science, Wako, Saitama 351-0198, Japan}
\newcommand{\rikjrbrc}{RIKEN BNL Research Center, Brookhaven National Laboratory, Upton, New York 11973-5000, USA}
\newcommand{\rikkyo}{Physics Department, Rikkyo University, 3-34-1 Nishi-Ikebukuro, Toshima, Tokyo 171-8501, Japan}
\newcommand{\saispbstu}{Saint Petersburg State Polytechnic University, St.~Petersburg, 195251 Russia}
\newcommand{\saopaulo}{Universidade de S{\~a}o Paulo, Instituto de F\'{\i}sica, Caixa Postal 66318, S{\~a}o Paulo CEP05315-970, Brazil}
\newcommand{\seoulnat}{Department of Physics and Astronomy, Seoul National University, Seoul 151-742, Korea}
\newcommand{\stonybrkc}{Chemistry Department, Stony Brook University, SUNY, Stony Brook, New York 11794-3400, USA}
\newcommand{\stonycrkp}{Department of Physics and Astronomy, Stony Brook University, SUNY, Stony Brook, New York 11794-3800, USA}
\newcommand{\sungskku}{Sungkyunkwan University, Suwon, 440-746, Korea}
\newcommand{\tenn}{University of Tennessee, Knoxville, Tennessee 37996, USA}
\newcommand{\titech}{Department of Physics, Tokyo Institute of Technology, Oh-okayama, Meguro, Tokyo 152-8551, Japan}
\newcommand{\tsukuba}{Tomonaga Center for the History of the Universe, University of Tsukuba, Tsukuba, Ibaraki 305, Japan}
\newcommand{\vandy}{Vanderbilt University, Nashville, Tennessee 37235, USA}
\newcommand{\weizmann}{Weizmann Institute, Rehovot 76100, Israel}
\newcommand{\wigner}{Institute for Particle and Nuclear Physics, Wigner Research Centre for Physics, Hungarian Academy of Sciences (Wigner RCP, RMKI) H-1525 Budapest 114, POBox 49, Budapest, Hungary}
\newcommand{\yonsei}{Yonsei University, IPAP, Seoul 120-749, Korea}
\newcommand{\zagreb}{Department of Physics, Faculty of Science, University of Zagreb, Bijeni\v{c}ka c.~32 HR-10002 Zagreb, Croatia}
\affiliation{\abilene}
\affiliation{\augie}
\affiliation{\banaras}
\affiliation{\barc}
\affiliation{\baruch}
\affiliation{\bnlcoll}
\affiliation{\bnlphys}
\affiliation{\caucr}
\affiliation{\charlesczech}
\affiliation{\chonbuk}
\affiliation{\cns}
\affiliation{\colorado}
\affiliation{\columbia}
\affiliation{\czechtech}
\affiliation{\dapnia}
\affiliation{\debrecen}
\affiliation{\elte}
\affiliation{\eszterhazy}
\affiliation{\ewha}
\affiliation{\fsu}
\affiliation{\gsu}
\affiliation{\hanyang}
\affiliation{\hiroshima}
\affiliation{\howard}
\affiliation{\ihepprot}
\affiliation{\illuiuc}
\affiliation{\inrras}
\affiliation{\instpasczech}
\affiliation{\isu}
\affiliation{\jaea}
\affiliation{\jyvaskyla}
\affiliation{\kek}
\affiliation{\korea}
\affiliation{\kurchatov}
\affiliation{\kyoto}
\affiliation{\labllr}
\affiliation{\lahorelums}
\affiliation{\lawllnl}
\affiliation{\losalamos}
\affiliation{\lpc}
\affiliation{\lund}
\affiliation{\lyon}
\affiliation{\maryland}
\affiliation{\mass}
\affiliation{\michigan}
\affiliation{\muhlenberg}
\affiliation{\myongji}
\affiliation{\nagasaki}
\affiliation{\nara}
\affiliation{\natmephi}
\affiliation{\newmex}
\affiliation{\nmsu}
\affiliation{\ohio}
\affiliation{\ornl}
\affiliation{\orsay}
\affiliation{\pnpi}
\affiliation{\riken}
\affiliation{\rikjrbrc}
\affiliation{\rikkyo}
\affiliation{\saispbstu}
\affiliation{\saopaulo}
\affiliation{\seoulnat}
\affiliation{\stonybrkc}
\affiliation{\stonycrkp}
\affiliation{\sungskku}
\affiliation{\tenn}
\affiliation{\titech}
\affiliation{\tsukuba}
\affiliation{\vandy}
\affiliation{\weizmann}
\affiliation{\wigner}
\affiliation{\yonsei}
\affiliation{\zagreb}
\author{A.~Adare} \affiliation{\colorado} 
\author{C.~Aidala} \affiliation{\losalamos} \affiliation{\michigan} 
\author{N.N.~Ajitanand} \altaffiliation{Deceased} \affiliation{\stonybrkc} 
\author{Y.~Akiba} \email[PHENIX Spokesperson: ]{akiba@rcf.rhic.bnl.gov} \affiliation{\riken} \affiliation{\rikjrbrc} 
\author{R.~Akimoto} \affiliation{\cns} 
\author{J.~Alexander} \affiliation{\stonybrkc} 
\author{M.~Alfred} \affiliation{\howard} 
\author{H.~Al-Ta'ani} \affiliation{\nmsu} 
\author{A.~Angerami} \affiliation{\columbia} 
\author{K.~Aoki} \affiliation{\kek} \affiliation{\riken} 
% ...
\collaboration{PHENIX Collaboration}  \noaffiliation

\date{\today}

%-----------------------------------------------------------------------------|

\begin{abstract}

%%\linenumbers

We present a detailed measurement of charged two-pion correlation 
functions in 0\%--30\% centrality $\sqrt{s_{_{NN}}}=200$~GeV Au$+$Au 
collisions by the PHENIX experiment at the Relativistic Heavy Ion 
Collider.  The data are well described by Bose-Einstein correlation 
functions stemming from L\'evy-stable source distributions. Using a 
fine transverse momentum binning, we extract the correlation strength 
parameter $\lambda$, the L\'evy index of stability $\alpha$ and the 
L\'evy length scale parameter $R$ as a function of average transverse 
mass of the pair $m_T$. We find that the positively and the negatively 
charged pion pairs yield consistent results, and their correlation 
functions are represented, within uncertainties, by the same 
L\'evy-stable source functions. The $\lambda(m_T)$ measurements 
indicate a decrease of the strength of the correlations at low $m_T$. 
The L\'evy length scale parameter $R(m_T)$ decreases with increasing 
$m_T$, following a hydrodynamically predicted type of scaling behavior. 
The values of the L\'evy index of stability $\alpha$ are found to be 
significantly lower than the Gaussian case of $\alpha=2$, but also 
significantly larger than the conjectured value that may characterize 
the critical point of a second-order quark-hadron phase transition.

\end{abstract}

\maketitle

\section{Introduction\label{s:intro}}

Femtoscopy is a well-established sub-field of high energy particle and 
nuclear physics, that encompasses all the methods that allow for 
measuring lengths and time intervals on the femtometer (fm) scale. 
While the name was coined in 2001~\cite{Lednicky:2001qv}, several 
earlier methods were developed in other fields of science that can be 
considered as predecessors. As femtoscopy typically deals with 
intensity correlations of particle pairs (or multiplets), the earliest 
intensity correlation measurements, that were performed in radio and 
optical astronomy to measure the angular diameters of main sequence 
stars by R. Hanbury Brown and R. Q. Twiss 
(HBT)~\cite{HanburyBrown:1956bqd} are considered as the experimental 
foundations of this field. The clear understanding of the HBT effect, 
as well as that of the lack of intensity correlations in lasers, by Roy 
J. Glauber is considered to be the opening of a new and prosperous 
field of science called quantum 
optics~\cite{Glauber:1962tt,Glauber:2006zz,Glauber:2006gd}.

Intensity correlations of identical pions were observed in 
proton-antiproton annihilation while searching for the $\rho$ 
meson~\cite{Goldhaber:1959mj}, and these correlations were explained by 
G. Goldhaber, S. Goldhaber, W-Y. Lee and A. Pais on the basis of the 
Bose-Einstein symmetrization of the wave-function of identical pion 
pairs~\cite{Goldhaber:1960sf}. Hence, in particle physics these 
correlations are also called GGLP or simply Bose-Einstein correlations. 
Because the two-particle Bose-Einstein correlation function is related to 
the Fourier transform of the phase-space density of the particle 
emitting source, by measuring the correlation function one can readily 
map out the particle source on a femtometer scale.

The discovery of the strongly coupled quark gluon plasma (sQGP) at the 
Relativistic Heavy Ion 
Collider~\cite{Adcox:2004mh,Adams:2005dq,Arsene:2004fa, Back:2004je} 
(RHIC) relied also on the contribution from Bose-Einstein correlation 
studies, beyond other important observables, many of which were 
confirmed and further elaborated at the Large Hadron Collider (LHC). 
The approximate transverse mass ($\mT$) dependence of the measured 
Gaussian source radii ($R_{\rm Gauss}$) is $R_{\rm Gauss}^{-2}\propto 
a+b \mT$ (where $a$ and $b$ are constants), which is almost universal 
across collision centrality, particle type, colliding energy and 
colliding system size~\cite{Adler:2004rq,Afanasiev:2009ii}. This is a 
direct consequence of a strong longitudinal as well as radial 
hydrodynamical 
expansion~\cite{Makhlin:1987gm,Csorgo:1995bi,Chapman:1994yv,Chapman:1994ax,Csanad:2004mm,Bekele:2007ee,Lisa:2008gf}. 
Directional Hubble flows seem to be a crucial property of the sQGP 
formation in heavy ion collisions, or Little 
Bangs~\cite{Makhlin:1987gm,Csorgo:1995bi,Chapman:1994yv,Chapman:1994ax}. 
The so-called RHIC HBT puzzle, the apparent contradiction between 
several hydrodynamical model predictions and the observed ratio of the 
HBT radii~\cite{Adcox:2004mh,Adams:2005dq}, also turned out to be 
resolvable in a hydrodynamical picture with more realistic physics 
conditions and refined models of three dimensional Hubble 
flows~\cite{Csorgo:1995bi,Csanad:2004mm,Bekele:2007ee,Pratt:2008qv,Heinz:2009xj,Bozek:2011ua}. 
For a more detailed introduction and review of Bose-Einstein 
correlations and their application in high energy heavy ion collisions, 
see the review papers in 
Refs.~\cite{Boal:1990yh,Weiner:1999th,Wiedemann:1999qn,Csorgo:1999sj,Lisa:2005dd, 
Tannenbaum:2006ch,Lisa:2008gf,Kisiel:2011jt,Heinz:2013th,Adamczyk:2014mxp}.

To fully exploit the power of HBT correlations (as observables deemed 
to provide insight into the dynamics of the matter produced in 
heavy-ion collisions), one can and must go beyond the Gaussian 
parameterization and the Gaussian source radii, as observed in $e^+e^-$ 
collisions at the Large Electron-Positron Collider 
(LEP)~\cite{Achard:2011zza} and in \pp, \ppb and \pbpb collisions at the 
LHC~\cite{Khachatryan:2011hi,Sikler:2014aea,Astalos:2015}. One of the 
observables that is rather sensitive to the actual shape of the 
Bose-Einstein correlation function is the so-called ``intercept 
parameter'' (or strength) $\lambda$ of the correlation function, as its 
value depends on the result of an extrapolation of the observed 
correlation function to zero relative momentum. The experimental 
determination of the parameter $\lambda$ for pions can provide 
information about the ratio of primordial pions to those that are decay 
products of long lived resonances~\cite{Bolz:1992hc,Csorgo:1994in}, and 
may also give insight into the possibility of coherent pion 
production~\cite{Bolz:1992hc,Weiner:1999th,Csorgo:1999sj}. The shape of 
the correlation functions, in particular their non-Gaussian behavior, 
may also hint at the vicinity of the critical point of the quark-hadron 
phase transition~\cite{Csorgo:2004sr,Csorgo:2003uv}.

In this paper we present a precise measurement of two-pion HBT 
correlation functions in \sqsntwo \auau collisions by the PHENIX 
experiment at RHIC. We use the data recorded in the 2010 data taking 
period. This data sample allows us to use a fine transverse mass 
binning, and to infer the shape of the correlation function more 
precisely than was possible with earlier data sets. The significance of 
this will become evident when we extract the source parameters. It 
turns out that the measured correlation functions cannot be described 
by a Gaussian approximation in a statistically acceptable way. A 
generalized random walk or anomalous diffusion suggests the appearance 
of L\'evy-stable distributions for the phase-space density of the 
particle emitting source~\cite{Metzler:1999zz,Csorgo:2003uv}. We have 
investigated whether a L\'evy-stable generalization of the Gaussian 
source distributions is consistent with our measurements, and found 
that (with the proper treatment of the final state Coulomb interaction) 
L\'evy-stable source distributions -- applied here for the first time 
in heavy ion HBT analyses -- give a high quality, statistically 
acceptable description of the measured correlation functions.

The structure of this paper is as follows. Section~\ref{s:exp} presents 
the PHENIX experimental setup with emphasis on the tracking and 
particle identification detectors that were used for this analysis. In 
Section~\ref{s:analysis} we present the measurement procedure of the 
two-pion correlation functions. In Section~\ref{s:shape} we discuss the 
shape analysis of the measured HBT correlation functions for 
L\'evy-stable source distributions, and the procedure for determining 
the L\'evy parameters. In Section~\ref{s:results} we present our 
results, namely the extracted L\'evy parameters of the source as a 
function of the transverse mass of the pair. We also discuss here some 
of the possible interpretations of these results. Finally we summarize 
and conclude.

\section{Experimental setup\label{s:exp}}

\begin{figure}[tbh]
\includegraphics[width=1.0\linewidth]{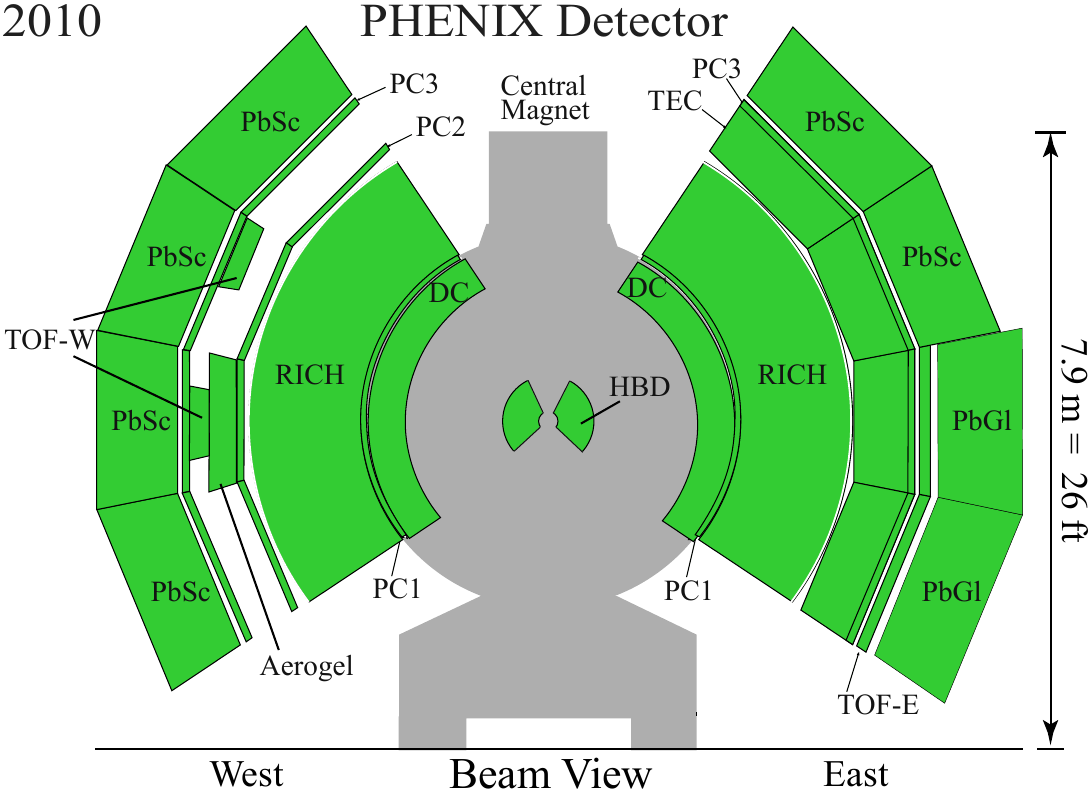}
\caption{View of the PHENIX central arm spectrometer detector setup 
during the 2010 run.}
\label{f:PHENIXsetup2010}
\end{figure}

The PHENIX experiment was designed to study various different particle 
types produced in heavy ion collisions, including photons, electrons, 
muons and charged hadrons, trading spatial acceptance for segmentation, 
good energy and momentum resolution, and high luminosity capability. 
Figure~\ref{f:PHENIXsetup2010} shows a schematic beam view drawing of the 
PHENIX experiment during the 2010 data taking period. The detailed 
description of the basic experimental configuration (without the 
upgrades made after the early 2000s) can be found 
elsewhere~\cite{Adcox:2003zm}; here we give only a brief description of 
the detectors that played a role in this analysis.

%--------------------------------------------- Fig_1

\subsection{Event characterization detectors}\label{ss:bbc}

This analysis uses the beam-beam counters (BBC) for event 
characterization. Its two arms (``North'' and ``South'') are located at 
$\pm144$ cm along the beam axis ($z$ axis) from the center of PHENIX, 
corresponding to the 3.0$<|\eta|<$3.9 pseudorapidity interval. Each arm 
of the BBC comprises 64 quartz \v{C}erenkov counters, covering $2\pi$ in 
azimuth. They provide minimum-bias (MB) triggering; the MB trigger 
condition requires at least two hits in coincidence in both BBC arms, 
thus capturing $92\pm3$\% of the total Au$+$Au inelastic cross 
section~\cite{Adare:2013esx}. The charge sum in both BBC arms is used 
for event centrality determination. The BBCs also measure the average 
hit time in the north and south arm photomultipliers (PMTs), thus 
providing collision vertex position measurements along the $z$ 
direction (from the hit time difference) as well as initial timing 
information for the collision. With an intrinsic timing resolution of 
$\approx$40 ps, the $z$-vertex resolution is $\approx$0.5 cm and $\approx$1.5 cm 
in central and peripheral Au$+$Au collisions, respectively.

\subsection{Central arm tracking}\label{ss:centraltrack}

PHENIX has two central arm spectrometers (``east'' and ``west''), each 
covering $|\eta|<0.35$ in pseudorapidity and $\Delta\varphi = \pi/2$ in 
azimuth, as seen in Fig.~\ref{f:PHENIXsetup2010}. In each central arm, 
charged particle tracks are reconstructed using hit information from 
the drift chamber (DC), the first layer of pad chambers (PC1) and the 
collision $z$-vertex position measured by the BBC~\cite{Adcox:2003zp}.

The DCs are located at a radial distance of 202--246 cm from the beam 
axis. They provide trajectory measurement in the transverse plane, with 
an angular resolution of $\approx$1 mrad. The PC1s are multiwire 
proportional chambers with pad readout, located immediately behind the 
DCs. They provide track position measurement both in the $\varphi$ and in 
the $z$ direction, with a $z$-resolution of $\approx$1.7 mm.

The PHENIX central arm spectrometer magnet generates a magnetic field 
approximately parallel to the beam line. It comprises two pairs of 
independently operable concentric coils, an inner and an outer coil 
pair, located at radial distances of $\approx$60 cm and $\approx$180 cm, 
respectively. The DCs are positioned so that they are in the reduced 
field region. Charged-particle-momentum determination is enabled by the 
measurement of the bending of the track in the magnetic field.  The 
transverse momentum $\pT$ is determined by the bending angle measured 
by the DC, while the polar angle of the momentum is determined by the 
$z$ coordinate measured by PC1 and the $z$-vertex coordinate from the 
BBC. Reconstructed tracks are then projected to the outer detectors 
used for track verification and timing measurement.

Because at not too low $\pT$ the momentum resolution is governed mainly 
by the angular resolution of the DC, high bending fields are desirable. 
Thus usually the two coil pairs are operated with currents flowing in 
the same direction (this is called ``$++$'' or ``$--$'' mode), to 
achieve the designed maximum total field integral of $\int B\cdot dl 
\approx$1.1 T m (this is the relevant quantity for the bending, and in 
turn for the momentum measurement).

In 2010, the Hadron Blind Detector (HBD), a specialized \v{C}erenkov 
counter located around the nominal collision point for the measurement 
of dielectron pairs, was installed~\cite{Anderson:2011jw}. The 
operation of the HBD required a field-free region around the collision 
point, which was achieved by running the inner and outer coils in the 
opposite directions (in ``$+-$'' or ``$-+$'' modes). This reduced the 
field integral to $\approx$40\% of its maximum value. However, the present 
analysis deals with low and intermediate $\pT$ hadrons (up to 
$\pT\approx$0.85 GeV$/c$), so high $\pT$ momentum resolution is not 
crucial. (The momentum resolution for $\pT$ in the dataset used is 
estimated to be $\delta \pT/\pT \approx 1.3\%\oplus 1.2\%\times \pT 
[\m{GeV}/c]$~\cite{Adare:2012vq}. The $p_z$ momentum resolution has, in 
addition, a component stemming from the BBC $z$-vertex resolution.) 
Moreover, the reduced magnetic field had a beneficial side effect for 
the present analysis.  Namely, the low momentum acceptance of this dataset 
is extended to lower values of transverse momentum, enabling a relatively 
clean identified pion sample down to $\pT\approx$0.2 GeV$/c$.  This would 
have been much harder, if not impossible, with the normal $++$ or $--$ 
field setting, because of too large bending angles and residual 
bending outside of the DC nominal radius, which is not taken into 
account in the standard PHENIX track projection algorithm.

\subsection{Particle identification detectors}\label{ss:piddetectors}

In the present analysis, we identify charged pions by their time of 
flight from the collision point to the outer detectors. We use the 
lead-scintillator electromagnetic calorimeter (PbSc) as well as the 
high resolution time-of-flight detectors (TOF east and TOF 
west)~\cite{Aizawa:2003zq}.

The PbSc is a sampling calorimeter located approximately 5.1 m radial 
distance from the beam axis. It covers $|\eta|<$0.35 in both arms, and 
in terms of $\varphi$, it covers all $\pi/2$ acceptance of the west arm, 
and $\pi/4$ (i.e. half) of the east arm, as seen on 
Fig.~\ref{f:PHENIXsetup2010}. It is a finely segmented detector, 
consisting of 15,552 individual channels (``towers''). After careful 
tower-by-tower and energy dependent calibration, a timing resolution of 
$\approx$ 400--600 ps (depending on deposited energy, incident angle, 
individual channel electronics imperfections, etc.) was achieved for 
pions. The part of the east arm acceptance not covered by the PbSc is 
covered by the lead-glass (PbGl) calorimeter, which has a much worse 
timing resolution for hadrons and thus was not used for the present 
analysis.

The TOF east detector is also located at approximately a 5.1 m from the 
beam axis, and covers much of the PbGl acceptance in the east arm. It 
is made of 960 plastic scintillator slats, with 2 PMTs attached to each 
side of them. After calibration, the timing resolution was found to be 
$\approx$150 ps.~\cite{Adare:2015ila}. The TOF west detector takes 
advantage of the multigap resistive plate chamber (MRPC) technology. It 
has two separate panels, each covering $\Delta\varphi\approx \pi/16$ in 
the west arm, at around 4.8 m radial distance from the beam pipe. Each 
panel comprises 64 MRPCs and has 256 individual copper readout 
strips. After calibration, a timing resolution of $\approx$90 ps was 
achieved.

\section{Measurement of two-pion correlation functions\label{s:analysis}}

\subsection{Event and track selection, particle identification}
\label{ss:tracks}

The MB-triggered data sample used in this analysis comprises $\approx 
7.3 \times 10^9$ \sqsntwo \auau events recorded by PHENIX during the 
2010 running period. This sample is reduced to $\approx 2.2 \times 10^9$ 
events when we apply a 0\%--30\% centrality selection. The event 
$z$-vertex position was constrained between $\pm 30$ cm in order to 
have an efficient BBC response as well as to avoid scattering in the 
central magnet steel.

We selected tracks of good quality, i.e.\ those where the DC and PC1 
information was unambiguously matched. To reduce in-flight decays as 
well as random associations between tracks and hits in the PbSc/TOF 
detectors, a track matching cut of $2\sigma$ was applied for the 
difference between the projected track position and the closest hit 
position in these detectors, in both the $\varphi$ and $z$ directions. As 
part of the systematic uncertainty investigation, we studied the 
dependence of the final results on these selection criteria.

For the present analysis, a clean sample of identified pions was 
necessary. Charged pion identification was performed with the help of 
time-of-flight information ($t$) from the PbSc/TOF detectors and the 
BBC, as well as using path length information ($L$) from the track 
model and the momentum value $p$ measured by the DC/PC1. The 
reconstructed squared mass $m^2$ of a track is then

\begin{equation}\label{e:m2pid}
m^2 = \frac{p^2}{c^2}\left[\left(\frac{ct}{L}\right)^2-1\right],
\end{equation}
and pions were selected by applying a $2\sigma$ cut in the $m^2$ 
distribution of the PbSc and the TOF detectors. For the $\pT$ range of 
interest in this analysis, the contamination in the pion sample caused 
by misidentified kaons or protons is negligible. A more important 
contamination in the pion sample comes from the random association of 
tracks and hits in the PbSc or the TOF detectors at low momentum, 
reaching $\approx$2\%--3\% for the TOF detectors, and as high as 8\%--10\% for 
the PbSc at or below $\pT\approx0.2$ GeV$/c$. This background quickly 
diminishes for even slightly higher $\pT$ (at $\pT\approx0.25$ GeV$/c$), 
as inferred from the observed $m^2$ distributions. However, even at low 
$\pT$ this is a gross overestimation of the contamination.  Most 
of the tracks are pions, even those for which the track projection 
algorithm didn't find the proper hit because of the residual bending at 
low momentum.  The systematic uncertainty stemming from mis-identified 
particles is mapped out by varying the mentioned standard $2\sigma$ cut 
on the $m^2$ spectrum of pions, as detailed in 
Section~\ref{s:systematics}.  In this analysis, we apply a 
$\pT>0.16$ GeV$/c$ selection, including all identified pions above this 
threshold into our sample.

\subsection{Construction of the correlation functions\label{ss:corrfunc}}

In general, the two-particle correlation function $C_2(p_1,p_2)$ is 
defined as

\begin{align}
C_{2}^{\rm{spm}}(p_1,p_2)=\frac{N_2(p_1,p_2)}{N_1(p_1)N_1(p_2)},\label{e:c2def}
\end{align}
where $N_1(p_1)$, $N_1(p_2)$ and $N_2(p_1,p_2)$ are the one- and 
two-particle invariant momentum distributions at four-momenta $p_1$ and 
$p_2$, and the superscript ``spm'' denotes that here the correlation 
function is written as a function of the single particle momenta.

There can be many causes of correlated particle production, such as 
collective flow, jets, resonance decays, conservation laws. In heavy ion 
collisions, the main cause of like-sign pion pairs correlation at small 
relative momentum is the quantum-statistical Bose-Einstein or HBT 
correlation stemming from the indistinguishability (and thus the 
symmetrical pair wave-function) of two identical bosons. This source of 
correlations grows with the mean number of pairs at small relative 
momentum, which is approximately proportional to the mean multiplicity 
squared. Other possible sources of correlations (for example pion pair 
production from resonance decays) increase only linearly with the mean 
multiplicity. Hence, for the large multiplicity heavy ion collisions, 
Bose-Einstein correlations dominate the correlation function at small 
relative momenta.

Experimentally the method of the measurement is the so called 
event-mixing. To discuss that in this subsection, let us denote any 
experimental choice for the measure of the two-pion relative momentum 
by $q$, defining our particular choice later in 
subsection~\ref{ss:kinematics}. In the present subsection we discuss 
only those properties of the two-pion Bose-Einstein correlation 
functions that are generally valid, independently of the particular 
experimental choice of $q$ for the measure of the relative momentum of 
the pion pair.

Let us define $A(q,K)$ as the actual $q$ distribution of pion pairs for 
a given average four-momentum $K$, where both members of the pair stem 
from the same event. Note also that our choice for $K$ is detailed 
later in Section~\ref{ss:kinematics}. This $A(q,K)$ distribution will 
contain effects which have to be excluded from the Bose-Einstein 
correlation function (such as resonance decay effects, kinematics, 
acceptance effects etc.). For this purpose, one defines a background 
distribution with pairs of pions from different events. Let us denote 
this background distribution with $B(q,K)$. A usual method is to 
construct the background distribution by keeping an event pool of a 
predefined size, and correlating each pion of the investigated event 
with all same charged pions of the background pool. However, in this 
case, multiple particle pairs will come from the same event pair. In 
this analysis we use the method described in~\cite{Achard:2011zza} that 
eliminates any possible residual correlation of this type as well. For 
each ``actual'' event, we form a ``mixed'' event by choosing pions (of 
the same number as in the actual event for each charge) from other 
randomly selected events within the background pool (that has to be 
larger than the maximal multiplicity of pions of a given charge), under 
the condition that no two tracks may originate from the same event. 
After this procedure, each ``mixed'' event comprises pions 
originating from different events. The background distribution is then 
created from the (same charge) pairs of this mixed event. It must also 
be noted that in order for the background event to exhibit the same 
kinematics and acceptance effects, one has to build the background 
event from the same event class (i.e. from events of similar centrality 
and of similar $z$ coordinate of the collision vertex). We used 3\% 
wide centrality and 2 cm wide $z$-vertex bins to achieve that goal.

If we now take the ratio of the actual and the background 
distributions, we get the prenormalized correlation function as

\begin{align}
C_2(q,K)=\frac{A(q,K)}{B(q,K)}\cdot\frac{\int B(q,K) dq}{\int A(q,K) dq},
\end{align}
where the integral is performed over a range where the correlation 
function is not supposed to exhibit quantum statistical features. Let 
us note that the method described above is applied to pairs belonging 
to a given range of average momenta, and in that case $K$ denotes the 
mean of these average momenta in the given range.  Furthermore,
in the mixing technique described above, the number 
of actual and background pairs is the same -- aside from the effect of 
two-track cuts, which is outlined in the next subsection.

\subsection{Two-track cuts}\label{ss:paircuts}

When forming pairs to construct the aforementioned actual $A(q)$ and 
background $B(q)$ pair distributions, one has to take into account 
detector inefficiencies and peculiarities of the track reconstruction 
algorithm which sometimes doubles or splits one track into two 
(creating so-called ghost tracks). It is also possible that two 
different tracks are not well distinguished when they approach one 
another too closely. To remove these possible track splitting and track 
merging effects, we studied track separation distributions in each 
detector involved, in each of the transverse momentum bins used in this 
analysis. Then we applied the following cuts in the 
$\Delta\varphi-\Delta z$ plane (in units of radians and cm, 
respectively) of pairs of hits in the given detector, associated with 
track pairs:

\begin{align}
\Delta\varphi &> 0.15\left(1-\frac{\Delta z}{11\textnormal{ cm}}\right) \textnormal{ and } \Delta\varphi > 0.025 \textnormal{  (DC),} \label{e:dchpaircut}\\
\Delta\varphi &> 0.14\left(1-\frac{\Delta z}{18\textnormal{ cm}}\right) \textnormal{ and } \Delta\varphi > 0.020 \textnormal{  (PbSc),} \label{e:pbscpaircut}\\
\Delta\varphi &> 0.13\left(1-\frac{\Delta z}{13\textnormal{ cm}}\right) \textnormal{  (TOF east),} \label{e:tofepaircut}\\
\Delta\varphi &> 0.085\textnormal{ or } \Delta z > 15\textnormal{ cm} \textnormal{  (TOF west).} \label{e:tofwpaircut}
\end{align}

We applied these two-track cuts to both the actual and the background sample.

In addition to these cuts, if we found multiple tracks that are 
associated with hits in the same tower of the PbSc, slat of the TOF 
east, or strip of the TOF west detector, we removed all but one of 
them. This ensured that we do not take into account any ghost tracks 
that would have remained in the sample after the above mentioned pair 
cuts.

Our analysis method is somewhat different from those of earlier 
measurements of Bose-Einstein correlations in heavy ion collisions, in 
particular with respect to the kinematic variables and the application 
of L\'evy-stable distributions. Thus we proceed carefully here and 
provide a thorough and detailed description of the concepts and 
procedures that we applied in the determination of the proper kinematic 
variables and the shape analysis of the Bose-Einstein correlation 
functions.

\subsection{Variables of the two-pion correlation function}
\label{ss:kinematics}

The correlation function, as defined in \Eq{e:c2def}, depends on single 
particle and pair momentum distributions. These can be calculated in 
the Wigner function formalism, assuming chaotic particle emission, from 
the single particle and pair wave functions, as detailed in 
Refs.~\cite{Yano:1978gk,Makhlin:1987gm,Pratt:1990zq,Csorgo:1999sj}. For 
the pair momentum distribution, neglecting dynamical two-particle 
correlations, one obtains the Yano-Koonin formula~\cite{Yano:1978gk}

\begin{align}\label{e:yanokoonin}
&N_2(p_1,p_2) =\\
&\int d^4x_1 d^4x_2 S(x_1,p_1) S(x_2,p_2) |\Psi^{(s)}_{p_1,p_2}(x_1,x_2)|^2, \nonumber
\end{align}
by means of the phase-space density of the particle-emitting source 
$S(x,p)$, sometimes referred to as ``source distribution'' or simply as 
``source'', and $\Psi^{(s)}_{p_1,p_2}(x_1,x_2)$, the symmetrized pair 
wave function. Neglecting final state Coulomb and strong interactions, 
as well as possible higher order wave-function symmetrization effects 
on the level of two-particle correlation functions, the pair 
wave-function is a properly symmetrized plane wave, i.e. in this case,

\begin{align}
|\Psi^{(s)}_{p_1,p_2}(x_1,x_2)|^2 = 1+\cos((p_1-p_2)(x_1-x_2)).
\end{align}
This approximation in turn leads to the expression of the pure 
quantum-statistical correlation function ($C^{(0)}_2$) 
as~\cite{Yano:1978gk,Makhlin:1987gm,Pratt:1990zq,Csorgo:1999sj}

\begin{equation}\label{e:Cp1p2:general}
C^{(0),\rm{spm}}_2(p_1,p_2)= 1+ \m{Re}\frac{\widetilde S(q,p_1)\widetilde S^*(q,p_2)}{\widetilde S(0,p_1)\widetilde S^*(0,p_2)} ,
\end{equation}
where complex conjugation is denoted by $^*$, the $(0)$ index signals 
that the Coulomb effect is not taken into account, the superscript 
``spm'' denotes that the correlation function is written as a function 
of the single particle momenta, and from now on

\renewcommand{\v}[1]{\boldsymbol{#1}}

\begin{align}
q\equiv p_1-p_2=(q_0,{\v q}),
\end{align}
stands for the difference of the four-momenta of particles 1 and 2 
($q_0$ denotes energy difference, i.e. the zeroth component of the 
relative four-momentum $q$) and $\widetilde S(q,p)$ denotes the Fourier 
transform of the source

\begin{equation}\label{e:tildeSdef}
\widetilde S(q,p)\equiv\int S(x,p) e^{iqx} d^4x .
\end{equation}

For source distributions and typical kinematic domains encountered in 
heavy ion collisions, the dependence of $\widetilde S(q,p)$ as defined 
in \Eq{e:tildeSdef} is much smoother~\cite{Lisa:2005dd} in the original 
$p$ momentum variable than in the relative momentum $q$, coming from 
the Fourier transform. Hence, it is customary to apply the $p_1\approx 
p_2\approx K$ approximation in \Eq{e:Cp1p2:general}, where

\begin{align}
K\equiv \frac{1}{2}(p_1+p_2)=(K_0,{\v K}),
\end{align}

is the average four-momentum of the pair ($K_0$ denotes the average 
energy of the pair, i.e. the zeroth component of the average 
four-momentum $K$). With this,

\begin{equation}\label{e:C2expr}
C^{(0)}_2(q,K)\approx 1+\frac{|\widetilde S(q,K)|^2}{|\widetilde S(0,K)|^2}.
\end{equation}
The validity of these approximations was reviewed in 
Refs.~\cite{Csorgo:1999sj,Wiedemann:1999qn} and for typically 
exponential single particle spectra the approximation was found to be 
within 5\% of the more detailed and substantiated calculations.

If the above approximations are justified, the two-particle 
Bose-Einstein correlation function is unity plus a positive definite 
function of the relative momentum $q$. In the \sqsntwo 0\%--30\% 
centrality \auau data reported in this analysis, we found that 
\Eq{e:C2expr} is consistent with the data; we did not observe the 
nonpositive definite, oscillatory behavior that was observed in $e^+e^-$ 
collisions at LEP~\cite{Achard:2011zza}, and in $p$$+$$p$ collisions at 
the LHC~\cite{Khachatryan:2011hi,Astalos:2015}. Note that in $e^+e^-$ 
collisions at LEP and in $p$$+$$p$ collisions at the LHC the smoothness 
approximation indicated above is not valid, but the Yano-Koonin formula 
of \Eq{e:yanokoonin} still 
holds~\cite{Achard:2011zza,Khachatryan:2011hi}.

In general, as described above, the correlation function depends on 
four-momenta $p_1$ and $p_2$ or, equivalently, on $q$ and $K$. However, 
the Lorentz product of $q$ and $K$ is zero, i.e. $qK=q_0K_0-{\v q}{\v 
K}=0$. Here $\v q$ and $\v K$ are defined as three-vector components of 
$q$ and $K$ as

\begin{align}
{\v q}\equiv(q_x,q_y,q_z), \qquad {\v K}\equiv(K_x,K_y,K_z)
\end{align}
This in turn implies

\begin{align}
q_0={\v q}\frac{\v K}{K_0}.
\end{align}
Based on this relation, one may transform the $q$-dependent correlation 
function to depend on $\v q$ instead. If the particles contributing to 
the correlation function are similar in energy, then $K$ is 
approximately on-shell; thus the correlation function can be measured 
as a function of $\v K$ and $\v q$.

As the dependence on $\v K$ in heavy ion reactions is typically 
smoother than on $\v q$, one may think of $\v q$ as the ``main'' 
kinematic variable. Then one may assume a parameterization of the $\v q$ 
dependence, and explore the dependence of the parameters on $\v K$. 
Close to midrapidity, instead of $\v K$, the dependence on

\begin{align}
\kT\equiv0.5\sqrt{K_x^2+K_y^2},
\end{align}
or, alternatively, on the transverse mass

\begin{align}\label{e:mTdef}
\mT\equiv\sqrt{m^2+(\kT/c)^2}
\end{align}
may be investigated, with $m$ being the particle (e.g.\ pion) mass. 
Note that the average four-momentum $K$ is not on mass-shell, but $\mT$ 
would be the transverse mass of a particle with momentum $K$. 
Furthermore, $\mT$ also corresponds to the average transverse mass of 
the particle pair, $M_{T} = 0.5 (m_{T,1}+ m_{T,2})$ in the limit of 
vanishing relative momentum $|{\bf q}|\rightarrow 0$. As earlier 
results were frequently given in terms of $\kT$, which is a unique 
function of $\mT$ of Eq.~\eqref{e:mTdef}, we decided to use $\mT$ 
instead of $M_{T}$ to characterize the transverse momentum of a pair of 
identical pions.

Let us also note that \Eq{e:C2expr} can be reinterpreted if we 
introduce the pair distribution as

\begin{align}\label{e:S2Dexpr}
D(r,K) \equiv \int S(\rho+r/2,K) S(\rho-r/2,K) d^4\rho,
\end{align}
where $r$ is the pair separation four-vector and $\rho$ is the four-vector of the
center of mass of the pair. Then the correlation function can be expressed as

\begin{align}
C^{(0)}_2(q,K)=1+\frac{\widetilde D(q,K)}{\widetilde D(0,K)},
\label{e:C2D}
\end{align}
where $\widetilde D$ is defined with the Fourier transformation as

\begin{align}
\widetilde D(q,K) \equiv \int D(r,K) e^{iqr} d^4r.
\end{align}
Thus the two-particle Bose-Einstein correlation function is connected 
to the pion pair distribution $D(r,K)$, so this is the quantity that 
can be reconstructed from two-particle correlation data directly. 
Different source distributions that keep $D(r,K)$ invariant yield 
equivalent results from the point of view of two-particle Bose-Einstein 
correlation measurements.

At any fixed value of the average pair momentum $K$, the correlation 
function $C_2(q,K)$ can be measured as a function of various 
decompositions of the components of the relative momentum $\v q$. The 
Bertsch-Pratt (BP) or side-out-longitudinal 
decomposition~\cite{Pratt:1986ev,Bertsch:1988db} is frequently used. 
Here

\begin{align}
\v q_\m{BP} \equiv (q_\m{\rm out},q_\m{\rm side},q_\m{\rm long}),
\end{align}
with $q_\m{\rm long}$ pointing in the beam direction, $q_\m{\rm out}$ in the 
direction of the average transverse momentum $(K_x,K_y)$, and the 
``side'' direction orthogonal to these two directions. The 
transformation to the BP variables corresponds to a rotation in the 
transverse plane, depending on the direction of the average momentum. 
For the BP decomposition, it is particularly favorable to use the 
longitudinal co-moving system (LCMS) of the pair, where the average 
momentum is perpendicular to the beam axis. Here the BP decomposition 
of the average momentum is simply $\v K_\m{BP}\equiv(\kT,0,0)$, as 
$\kT=K_\m{\rm out}$, and the temporal information of the source is coupled 
to the {\it out} component of the Bose-Einstein correlation 
function~\cite{Wiedemann:1999qn,Csorgo:1999sj}.

However, the Bertsch-Pratt variables require three-dimensional 
Bose-Einstein correlation measurements, so a detailed shape analysis in 
terms of them can suffer from a lack of statistical precision. For 
example, it is very difficult to identify any non-Gaussian structure in 
a three-dimensional analysis of correlation functions. For this reason, 
sometimes the two-particle correlation function is measured as a 
function of a one-dimensional momentum 
variable~\cite{Achard:2011zza,Sikler:2014aea}. The Lorentz invariant 
relative momentum, corresponding to the Lorentz length of $q^{\mu}$, is 
defined as

\begin{equation}\label{e:qinvdef}
q_\m{inv} \equiv \sqrt{-q^\mu q_\mu} = \sqrt{q_x^2+q_y^2+q_z^2-(E_1-E_2)^2} .
\end{equation}
In the LCMS, using the Bertsch-Pratt variables $q_\m{inv}$ is expressed as

\begin{equation}\label{e:qinv:BP}
q_\m{inv}^2=(1-\beta_t^2)q_\m{\rm out}^2+q_\m{\rm side}^2+q_\m{\rm long}^2,
\end{equation}
where $\beta_t=2\kT/(E_1+E_2)$ is the ``average transverse speed'' 
of the pair.

Let us introduce also the rest frame of the pair, here referred to as 
pair center-of-mass system (PCMS), and define the relative 
three-momentum in this system as $\v q_\m{PCMS}$. Then the variable 
$q_\m{inv}$ can be expressed as

\begin{equation}\label{e:qinvdef2}
q_\m{inv} = |\v q_\m{PCMS}|.
\end{equation}
Equation (\ref{e:qinv:BP}) shows that $q_\m{inv}$ can be very small at moderate 
$\kT$, even for not very small $q_\m{\rm out}$ values. It is also well 
known that the Bertsch-Pratt radii ($R_\m{\rm out}$, $R_\m{\rm side}$, 
$R_\m{\rm long}$) are of similar magnitude in \sqsntwo \auau reactions at 
RHIC, so the Bose-Einstein correlation functions are nearly spherically 
symmetric in the LCMS 
frame~\cite{Adler:2004rq,Adams:2004yc,Afanasiev:2007kk,Afanasiev:2009ii}. 
This also implies that the correlation function boosted to the PCMS 
frame is definitely not spherically symmetric (especially for 
intermediate or high $\kT$, i.e. for $\beta_t$ values approaching 1). 
The conclusion is that $q_\m{inv}$ is not a proper one-dimensional 
variable of Bose-Einstein correlations of pions in \sqsntwo \auau 
collisions.

We look for a novel one-dimensional variable whose small value is 
only possible in the case when $q_\m{\rm out}$, $q_\m{\rm side}$, 
$q_\m{\rm long}$ are all small. Hence, we introduce LCMS 
three-momentum difference $\v q_\m{LCMS}$ This quantity is invariant 
for Lorenz boosts in the beam direction. For the sake of simplicity, 
we hereafter define

\begin{align}\label{e:Qdef}
Q\equiv|\v q_\m{LCMS}|.
\end{align}
which can be expressed with the lab-system components of the individual 
particle momenta as

\begin{gather}
Q\!=\!\sqrt{(p_{1x}\!-\!p_{2x})^2\!+\!(p_{1y}\!-\!p_{2y})^2\!+\!q_\m{long,LCMS}^2},\\
\m{where}\quad q_\m{long,LCMS}^2=\frac{4(p_{1z}E_2-p_{2z}E_1)^2}{(E_1+E_2)^2-(p_{1z}+p_{2z})^2}.\label{e:Qcalc}
\end{gather}

Because the correlation functions are approximately spherically symmetric 
in the LCMS, the measured correlation functions are approximately 
independent of the orientation of $\v q_\m{LCMS}$.

We thus conclude that $Q$ can be introduced in a reasonable manner as 
the proper one-dimensional variable of the Bose-Einstein correlations 
in \sqsntwo \auau collisions.

In order to perform a detailed shape analysis in the LCMS, we thus 
measured them as univariate functions of $Q$ (for $\kT$ values in 
various ranges). Thus this one-dimensional analysis in the LCMS in 
terms of $Q$ can be viewed as an approximation to a three-dimensional 
analysis with the approximation that the three HBT radii are equal.

In principle, a more complete picture of the source geometry can be 
obtained by a three-dimensional L\'evy analysis, utilizing Eqs. 
(49)-(52) of Ref.~\cite{Csorgo:2003uv}. Given that the details of 
these studies go beyond the scope of the current manuscript, let us 
make only some general remarks here. If the source is a symmetric 
three-dimensional Gaussian, then in a one-dimensional analysis (in 
our $Q$ variable, measured in the LCMS), one would obtain $\alpha=2$ 
for the L\'evy shape parameter. If the source is an asymmetric 3D 
Gaussian, then non-Gaussian 1D correlation functions would be 
obtained, but also strong deviations from the L\'evy shape could be 
observed. We investigated this using the method of L\'evy expansion 
of the correlation functions~\cite{Novak:2016cyc} for each $\mT$ 
bin, and found no first order deviations from the L\'evy shape. 
However, an $\mT$ averaged correlation function shows deviations 
from the pure L\'evy shape, which may be attributed to the $\mT$ 
dependence of $\alpha$. These observations suggest that the observed 
L\'evy shapes do not originate from an asymmetric three-dimensional 
Gaussian source.

\section{Strength and shape of two-pion correlation functions}\label{s:shape}

We recapitulate some of the important general properties of the 
two-pion Bose-Einstein correlation functions. First we discuss the 
strength of the correlation functions, and the main features of its 
interpretation, following the lines of 
Refs.~\cite{Bolz:1992hc,Csorgo:1994in}. Then we describe the shape 
assumption used in this paper, and the physical interpretation of the 
relevant parameters.

\subsection{Correlation strength and its implications}\label{ss:lambdamtintro}

If the final-state-strong and Coulomb interactions can be neglected, 
then \Eq{e:C2expr} implies that the correlation function takes the value 
$2$ at vanishing relative momentum, $C^{(0)}_2(Q=0,K)=2$. However, 
experimentally the two-track resolution (corresponding to a minimum 
value of $Q_\m{\rm min}$ of at least 6--8 MeV, depending on track momentum) 
prevents the measurement correlation functions at $Q=0$. So the 
correlation function is measured at nonzero relative momenta and then 
extrapolated to $Q = 0$. This extrapolated value in general can be 
different from the exact value at $Q = 0$, and this can be quantified 
by defining

\begin{align}
\lambda \equiv \lim_{Q \rightarrow 0} C_2(Q,K) -1.
\end{align}
where $\lambda$ may depend on average momentum $K$.

In our analysis we measure the $C_2$ correlation functions as a ratio 
of actual and background distributions $A$ and $B$, and we have 
carefully checked in our dataset that $\lim_{Q \rightarrow 0} 
A(Q,\kT)=0$ and $\lim_{Q \rightarrow 0} B(Q,\kT)=0$ in every transverse 
momentum range, indicating that the split tracks have been removed from 
our data sample. The two-track resolution, embodied into the values of 
two-track cuts as seen in Section~\ref{ss:paircuts}, corresponds to a 
maximum spatial resolution of 
$R_\m{\rm max}\approx\hbar/Q_\m{\rm min}\approx 25-30$~fm. 
In our analysis, source details on spatial scales larger or 
equal to $R_{\rm max}$ cannot be experimentally resolved.

This (perhaps with different $R_\m{\rm max}$ values) is a general feature 
of any similar experiment, and it leads to the core-halo 
picture of Bose-Einstein correlations in high energy heavy ion 
reactions~\cite{Bolz:1992hc,Csorgo:1994in}. The core-halo picture treats the 
particle emitting source as a composite one, corresponding to particle 
emission from a hydrodynamically behaving fireball-type core, 
surrounded by a halo of long-lived resonances. Such a picture is 
particularly relevant for pion production.  Several long-lived 
resonances with decay widths of $\Gamma \ll Q_\m{\rm min}$ (like the 
$\eta$, $\eta'$, $K^0_S$ mesons, and, depending on the experimental 
two-track resolution, maybe the $\omega$ meson) decay to pions that 
contribute to the halo region. The general structure of the core-halo 
model may hold not only for pion production but for the production of 
other mesons as well.

In short, $\lim_{Q\rightarrow 0} C_2(Q,K) = 1 + \lambda(K)$ is in 
general different from the exact value of $C_2(Q=0,K)$ which 
(independently of $K$) is 2 for a thermal, fully chaotic particle 
source. In most data sets, $\lambda < 1$ holds, see again the overview 
papers in 
Refs.~\cite{Boal:1990yh,Weiner:1999th,Wiedemann:1999qn,Csorgo:1999sj,Lisa:2005dd, 
Tannenbaum:2006ch,Lisa:2008gf,Kisiel:2011jt,Heinz:2013th,Adamczyk:2014mxp}.

In the core-halo picture, for thermal particle emission, the intercept 
$\lambda$, the extrapolation of the measured \emph{resolvable} part of 
the correlation function to zero relative momentum, is the square of 
the fraction of pions coming from the core, defined as

\begin{align}
f_c \equiv \frac{N_\m{core}}{N_\m{core}+N_\m{halo}},
\end{align}
because both pions have to come from the core if they are to contribute 
to the resolvable correlation function. This requires a physical 
assumption, that the phase-space density of the pion emitting source is 
made up of two components, i.e.

\begin{align}
S=S_{\rm core}+S_{\rm halo},\label{e:Scoreplushalo}
\end{align}
each component having a Fourier transform defined as

\begin{align}
\widetilde S_\m{core}(q,K)\equiv\int S_\m{core}(x,K) e^{iqx} d^4x,\\
\widetilde S_\m{halo}(q,K)\equiv\int S_\m{halo}(x,K) e^{iqx} d^4x,
\end{align}
where we again used the four-vector variables $q=p_1-p_2$ and 
$K=(p_1+p_2)/2$. Then each component has a space-time integral 
corresponding to the contribution of the given component to the 
momentum distribution. We then may define

\begin{align}
N_{\rm core}(K) &\equiv \int S_{\rm core}(x,K) d^4x = \widetilde S_{\rm core}(0,K),\label{e:NcoreK}\\
N_{\rm halo}(K) &\equiv \int S_{\rm halo}(x,K) d^4x = \widetilde S_{\rm halo}(0,K).\label{e:NhaloK}
\end{align}
Here the first equation in \Eq{e:NcoreK} and \Eq{e:NhaloK} represents 
our physical assumption about the phase-space density of the core and 
the halo, while the second equation in \Eq{e:NcoreK} and \Eq{e:NhaloK} 
indicates a mathematical identity about the Fourier transform. Taking 
these and \Eq{e:Scoreplushalo} into account, we obtain

\begin{align}
\widetilde S(0,K)&=N_{\rm core}(K)+N_{\rm halo}(K).
\end{align}
For the experimentally resolvable $q$ values, this system of physical 
assumptions yields the approximation

\begin{align}
\widetilde S(q,K) \approx \widetilde S_{\rm core}(q,K),
\end{align}
thus the correlation function ($C^{(0)}_2(q,K)$) shown in \Eq{e:C2expr} 
can be expressed as

\begin{align}\label{e:C2:corehalo}
&C^{(0)}_2(q,K)\approx\\
&1+\left(\frac{N_\m{core}(K)}{N_\m{core}(K)+N_\m{halo}(K)}\right)^2\frac{|\widetilde S_{\rm core}(q,K)|^2}{|\widetilde S_{\rm core}(0,K)|^2}.\nonumber
\end{align}
Hence, in the core-halo picture, at any given momentum 

\begin{align}\label{e:lambda:corehalo}
\lambda = f_c^2
\end{align}
holds; see Ref.~\cite{Csorgo:1994in} for details. Thus parameter 
$\lambda$ can be interpreted as the squared fraction of pions from the 
core with respect to the total number of pions with a given average 
momentum $K$. The $q$ dependent part in Eq.~\eqref{e:C2:corehalo}, i.e. 
the shape of the Bose-Einstein correlation function is connected to the 
core, $S_{\rm core}$. This source component is the one that may 
correspond to the perfect fluid, the hydrodynamically evolving central 
part of the fireball created in high energy heavy ion collisions.

If we assume that the source ($S$) is a sum of the core and the halo 
components as shown in \Eq{e:Scoreplushalo}, then it follows that the 
pair distribution ($D$) shown in \Eq{e:S2Dexpr}, is a sum of the three 
components,

\begin{align}
D=D_\m{(c,c)}+D_\m{(c,h)}+D_\m{(h,h)},
\end{align}
where subscript `c' denotes the core and `h' denotes the halo. It can 
be easily shown that the core-core component denoted by $\m{(c,c)}$ is 
resolvable, but the core-halo or $\m{(c,h)}$ type of pion pairs or the 
halo-halo or $\m{(h,h)}$ components are unresolvable (i.e. the width of 
their Fourier transform is below the minimal resolvable momentum 
difference). With this compared to \Eq{e:C2D}, the correlation function 
of \Eq{e:C2:corehalo} can be re-expressed as

\begin{align}
C^{(0)}_2(q,K)=1+\lambda\frac{\widetilde D_\m{(c,c)}(q,K)}{\widetilde D_\m{(c,c)}(0,K)} .
\end{align}
In summary, $\lim_{q\rightarrow 0} C_2(q,K) \neq 2$ is an experimental 
finding, and so it is customary to introduce $\lambda$ as an 
experimental parameter, defined as $\lim_{q\rightarrow 0} C_2(q,K)-1$, 
and measured by extrapolating the correlation function to zero relative 
momentum. The core-halo model is then an interpretation of the value 
$\lambda$. It also relates the relative momentum dependent, resolvable 
part of the Bose-Einstein correlation function to $S_{\rm core}$, the 
core component of particle emission in high energy heavy ion 
collisions. From this interpretation it is particularly clear that 
while long-lived resonance effects dominate the variances of the 
source, they lead to a peak in the unresolvable part of the 
Bose-Einstein correlation function, with measurable effects only on 
$\lambda$. Particle emission from the hydrodynamically expanding 
fireball however, i.e. the core component of the source, is observable 
from the $q$-dependent shape analysis of the Bose-Einstein correlation 
functions.

Thus one of the motivations for measuring the $\lambda$ parameter is 
that it carries indirect information on the decays of long-lived 
resonances to the observable pion spectra. Of particular interest is 
the contribution of the $\eta'$ meson to the low momentum pion yield. 
It is expected~\cite{Kapusta:1995ww} that in the case of chiral 
$U_A(1)$ symmetry restoration in heavy-ion collisions, the in-medium 
mass of the $\eta'$ meson (the ninth pseudoscalar meson, a would-be 
Goldstone boson) is decreased, thus its production cross section is 
heavily enhanced at low momentum. This (because the decay chain of the 
$\eta'$ meson produces many charged pions) implies that at low 
transverse momentum, the $\lambda$ parameter 
decreases~\cite{Vance:1998wd}. A recent study~\cite{Csorgo:2009pa} of 
existing $\lambda(\mT)$ measurements (presented in greater detail in 
Ref~\cite{Vertesi:2009wf}) reported an indirect observation of a mass 
drop of the $\eta'$ meson in \sqsntwo \auau collisions at RHIC.

However, many of the earlier $\lambda(\mT)$ measurements were made with 
the assumption that the shape of the correlation function is a Gaussian 
one. Given the fact that the detailed analysis presented below 
indicates that the Gaussian approximation is a statistically unfavored 
assumption, we attempt here a precise shape analysis of the correlation 
functions. This is required for a precise measurement of the intercept 
parameter $\lambda$, as its value depends on the shape of the 
correlation function through the extrapolation of the measured 
correlation function to vanishing relative momentum.

Let us note here that the modification of the observable intercept 
parameter $\lambda$ from unity can result from various reasons besides 
the core-halo model, for example coherence in the pion 
production~\cite{Weiner:1999th,Csorgo:1999sj}. If a fraction of pions 
are created in a coherent manner, then two- and three-particle 
Bose-Einstein correlation functions at zero relative momentum are 
simply related to the fraction of coherently produced pions and to the 
fraction of pions coming from the core~\cite{Csorgo:1999sj}. Thus a 
simultaneous measurement of $\lambda$ in two- and three-pion 
correlation functions offers the possibility of separating the 
component of a possibly coherent pion production, in addition to the 
resonance decay contribution. Such a simultaneous analysis of second, 
third and higher order correlations was recently reported at the 
LHC~\cite{Adam:2015pbc}. Also, more exotic quantum statistical effects 
like squeezed coherent states may modify the values of the intercept 
parameter (however, in the present analysis we have no compelling 
reason to consider this possibility). Hence, one of the goals of the 
paper is to measure $\lambda(\mT)$ precisely, without any physical 
assumption about the mechanism of the pion production.

In the following, we utilize a generalization of the usual Gaussian 
shape of the Bose-Einstein correlations, namely we analyze our data 
using L\'evy-stable source distributions. We have carefully tested that 
this source model is in agreement with our data in all the transverse 
momentum regions studied.   All the L\'evy fits were statistically 
acceptable, as discussed in Section~\ref{s:results}. We note that using 
the method of L\'evy expansion of the correlation 
functions~\cite{Novak:2016cyc}, we investigated deviations from the 
L\'evy shape. We have found that the coefficient of the first 
correction term is within uncertainties consistent with zero. Hence, we 
restrict the presentation of our results to the analysis of the 
correlation functions in terms of L\'evy-stable source distributions.

\subsection{L\'evy-type correlation functions and critical behavior}\label{ss:Levy}

Past measurements of two-pion Bose-Einstein correlation functions in 
Au$+$Au collisions that went beyond the Gaussian approximation show that 
the precise shape of Bose-Einstein correlations is indeed not 
Gaussian~\cite{Afanasiev:2007kk,Adler:2006as}.   The shape exhibits a 
power-law-like long-range component.  In expanding systems, a generalized 
form of the central limit theorem and investigation of generalized 
random walk (also called anomalous diffusion) suggests the appearance 
of L\'evy distributions as source 
functions~\cite{Metzler:1999zz,Csorgo:2003uv}. The one-dimensional, 
symmetric L\'evy distribution is the generalization of the Gaussian 
distribution defined by the Fourier transform

\begin{align}\label{e:Levydef}
\c L(\alpha,R,\v r)=\frac{1}{(2\pi)^3}\int\m d^3\v q\, e^{i\v q\v r} e^{-\frac{1}{2}|\v qR|^{\alpha}} .
\end{align}
Here $R$ is called the L\'evy length scale parameter, and $\alpha$ is 
called the L\'evy index of stability. In the $\alpha = 2$ case we 
recover a Gaussian form; in the $\alpha = 1$ case, we have a Cauchy 
distribution. For $\alpha<2$, the L\'evy distributions have a 
power-law-like tail, $\c L(\alpha,R,\v r) \propto (r/R)^{-(3+\alpha)}$ 
for $r/R \to\infty$ (with $r\equiv|\v r|$). Equivalently, for the 
angle-averaged L\'evy distribution one gets

\begin{align}\label{e:r2Ldist}
r^2 \c L(\alpha,R,\v r) \propto r^{-1-\alpha}.
\end{align}
Thus L\'evy distributions for $\alpha < 2$ have an infinite second 
moment or root-mean-square (RMS) radius. However, even in this case, 
the scale parameter $R$ provides a measure of the characteristic size 
of the system. In particular, the integral of the L\'evy distribution 
is finite and proportional to $R^3$. Note also that if the core part of 
the source ($S_\m{core}$) has a L\'evy shape, then the core-core pair 
distribution ($D_\m{(c,c)}$) also has a L\'evy shape, due to the fact 
that the autocorrelation of two identical L\'evy distributions is also 
a L\'evy distribution with the same index of stability $\alpha$,

\begin{align}
S_\m{core}(\v r)=\c L(\alpha,R,\v r) \Rightarrow D_\m{(c,c)}(\v r)=\c L(\alpha,2^{\frac{1}{\alpha}}R,\v r) .
\end{align}

\begin{figure}
\includegraphics[width=0.8\linewidth]{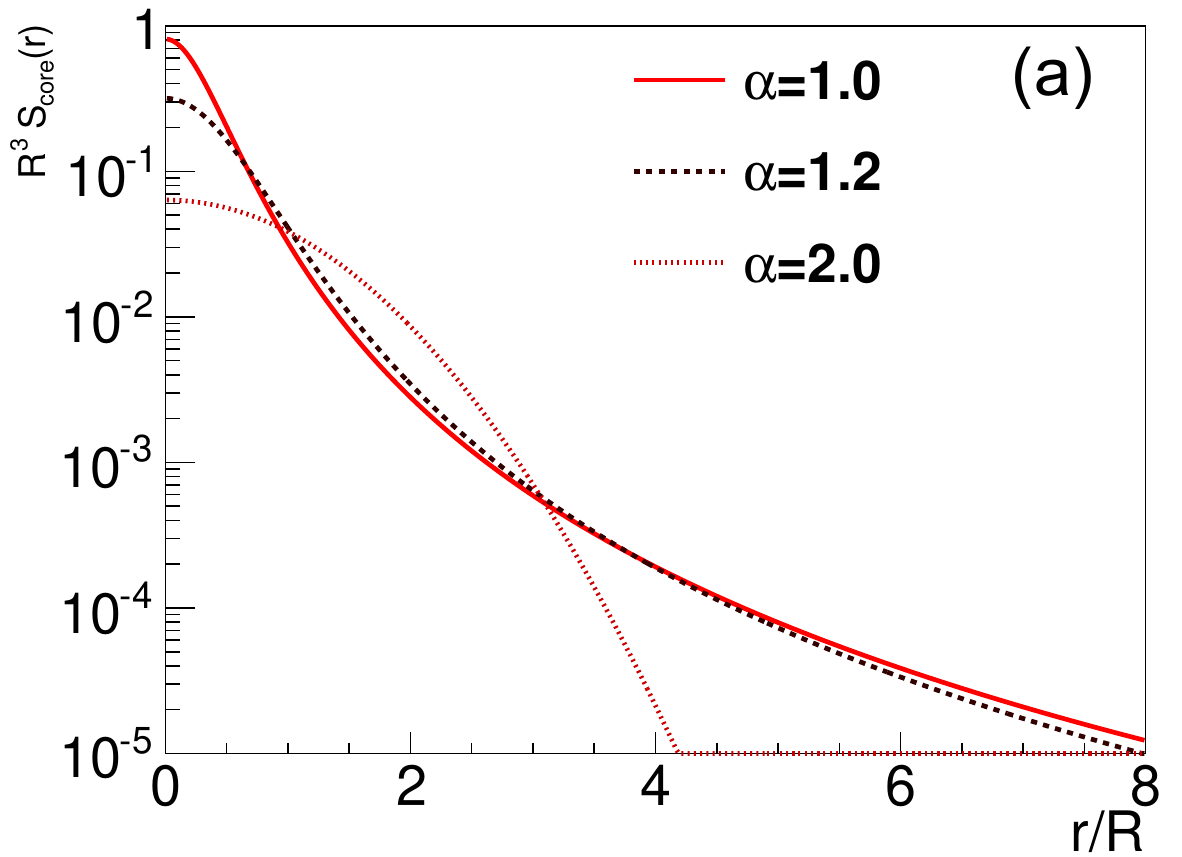}
\includegraphics[width=0.8\linewidth]{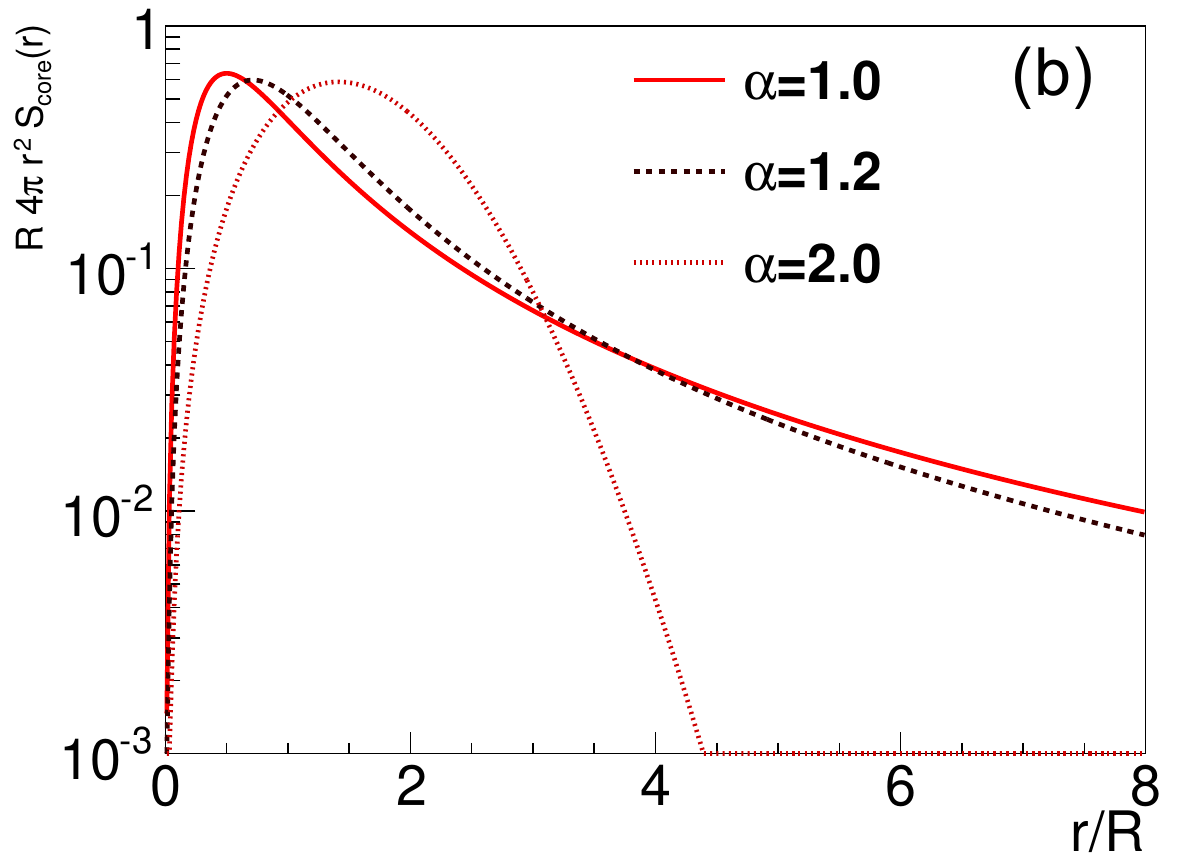}
\caption{L\'evy-stable source distributions with 
(a) $S_\m{core}(\v r)=\c L(\alpha,R,\v r)$ and $r=|\v r|$ for 
$\alpha=$ 1, 1.2, and 2. (b) Radial source distributions 
$4\pi r^2 S_\m{core}$ for $\alpha=$ 1, 1.2, and 2. 
In these plots, the dependence of the source distribution on 
L\'evy scale $R$ is scaled out by using $r\rightarrow r/R$ and 
$S_\m{core}\rightarrow R^3 S_\m{core}$.  With this transformation, 
source distributions coincide for any $R$.
}
\label{f:sourceplot}
\end{figure}

Thus the L\'evy-type source distributions offer a more general 
description of the shape of the correlation function than a Gaussian 
would do. They provide a better handle on the $\lambda$ intercept 
parameter as well. The Gaussian limit corresponds to the special 
$\alpha = 2$ case, so one can experimentally check how far given data 
are from the Gaussian limit. We illustrate the shape of L\'evy-type 
source distributions ($S_\m{core}=\c L(\alpha,R,\v r)$) with various 
$\alpha$ values in Fig.~\ref{f:sourceplot}.

%--------------------------------------------- Fig_2

There is yet another motivation for L\'evy distributions.  Namely, the 
exponent $\alpha$ of the L\'evy distribution (that determines the 
power-law-like behavior of the distribution at large distances) is 
related to the critical exponent $\eta$ of a system at a second order 
phase transition~\cite{Csorgo:2009gb}.  This exponent characterizes the 
power-law structure of the spatial correlation at the critical point.
If an order parameter $\phi$ is introduced, its correlation 
function (in three dimensions, as a function of distance $r$) will be

\begin{align}
\langle\phi(r)\phi(0)\rangle\propto r^{-1-\eta}.
\end{align}
As noted above in \Eq{e:r2Ldist}, the L\'evy source distribution has 
the same limiting behavior, thus in this case, $\eta=\alpha$. According 
to lattice quantum chromodynamics (QCD)~\cite{Aoki:2006we,Bhattacharya:2014ara,Soltz:2015ula} 
the quark-hadron transition is analytic (cross-over) at vanishing 
baryochemical potential $\mu_B= 0$, and is expected to be a first order 
phase transition at high values of $\mu_B$. There may be a critical 
endpoint (CEP) at certain intermediate values of $\mu_B$, where one has 
a second order phase transition, with a specific value of the $\eta$ 
exponent. This value is $0.03631(3)$ in the 3D Ising 
model~\cite{El-Showk:2014dwa}, and $0.50\pm0.05$ in the random field 3D 
Ising model~\cite{Rieger:1995aa}. Given that the second order QCD phase 
transition is expected to be in the same universality class as the 3D 
Ising model~\cite{Halasz:1998qr,Stephanov:1998dy}, the QCD critical 
point may be signaled by L\'evy sources with a specific $\alpha$ 
exponent. To locate and characterize the CEP is one of the most 
pressing present day challenges of experimental heavy-ion physics. It 
is thus desirable to measure $\alpha$ for various colliding systems and 
collision energies, to map various parts of the $(\mu_B,T)$ plane, in a 
quest to find the location of the CEP of the quark-hadron transition. 
We present below the first determination of the L\'evy index of 
stability in \sqsntwo \auau collisions.

\subsection{Coulomb effect}\label{ss:coulomb}

Using the plane-wave approximation, and assuming a spherically 
symmetric, three dimensional L\'evy-type source and using the core-halo 
model, the shape of the two-particle correlation function turns out to 
have the simple form of

\begin{align}\label{e:Levy:FT}
C^{(0)}_2(Q,K)=1+\lambda e^{-Q^\alpha R^\alpha}
\end{align}
with $Q$ being the independent variable as introduced in \Eq{e:Qdef}, 
and with three fit parameters, which may depend on average momentum 
$K$.  The scale parameter $R$, the strength (intercept) $\lambda$ and 
the L\'evy index $\alpha$ (note that the fitting procedure is detailed 
in Section~\ref{ss:fitting}). However, one cannot fit the above 
functional form to the measured correlation functions before properly 
taking the final state Coulomb repulsion of the identically charged 
pions into account.

In the treatment of this effect, we follow the general lines of the 
Sinyukov-Bowler method~\cite{Sinyukov:1998fc,Bowler:1991vx}. Coupling 
this with the core-halo picture, one has to average the modulus squared 
of the final state pair wave-function over the ``core-core'' spatial 
pair distribution $D_\m{(c,c)}(\v r,K)$, obtaining

\begin{align}\label{e:C2Scc}
C_2(\v q,K)=1\!-\!\lambda\!+\!\lambda\int\m d^3\v r\,D_\m{(c,c)}(\v r,K)|\psi^{(2)}_{\v q}(\v r)|^2,
\end{align}
where the Coulomb wave function is defined as

\begin{gather}
\psi^{(2)}_{\v q}(\v r)
= \frac{\c N}{\sqrt 2}\Big\{e^{i\v q\v r}F\left(-i\eta_{_C},1,i(kr-\v q\v r)\right)+\left[\v r\to-\v r\right]\Big\},\nonumber\\
\m{with}\quad \c N = \frac{\Gamma\left(1+i\eta_{_C}\right)}{e^{\pi\eta_{_C}/2}},\quad\quad \eta_{_C}=\frac{m_\pi c^2\alpha_\m{_{f.s.}}}{2\hbar q c}. \label{e:C2expr:confhyp}
\end{gather}
Here $F(\cdot,\cdot,\cdot)$ is the confluent hypergeometric function, 
$\eta_{_C}$ is the Coulomb-parameter, $\alpha_\m{_{f.s.}}$ is the fine 
structure constant, $\Gamma(\cdot)$ is the Gamma function, $\v r$ is a 
spatial integration variable representing the spatial pair separation, 
and $\v q$ is the three dimensional momentum difference in the pair 
rest frame, $\v q_\m{PCMS}$. The $\left[\v r\to-\v r\right]$ term 
represents a term similar to the first one, just with a mirrored $\v 
r$. The above Coulomb wave function formula is a standard result in 
quantum scattering theory. Note that in \Eq{e:C2Scc}, the right side 
does not depend on the direction of $\v q$ if the source is spherically 
symmetric. Hence, we modified the formula of \Eq{e:C2Scc} slightly to 
make it compatible with our analysis.  We substitute $\v q = \v 
q_\m{LCMS}$, and thus obtain $C_2$ as a function of $Q=|\v q|$.  We 
analyzed the error coming from this approximation by averaging $C_2(\v 
q_\m{PCMS},K)$ values for various $\v q_\m{PCMS}$ momenta at a given 
$|\v q_\m{LCMS}|$, and treated it as a source of uncertainty, as 
quantified next in Section~\ref{s:systematics}.

\section{Systematic uncertainties\label{s:systematics}}

%======================================================== Table_I
\begin{table}[tbh]
\caption{\label{t:systdefs}
List of settings that are varied in order to determine the systematic
uncertainties of our results. The individual cut settings are described in
Sections \ref{ss:tracks} and \ref{ss:paircuts}.
}
\begin{ruledtabular} \begin{tabular}{cll}
$n$ & setting name              & settings ($j=0,1,\dots$) \\
\hline
0   & PID arm                   & east, west, both\\
1   & PID cut                   & 3 cut settings\\
2   & PID det. matching cut     & 3 cut settings\\
3   & PC3 matching cut          & 3 cut settings\\
4   & PID det. pair cut         & 3 cut settings\\
5   & DC pair cut               & 3 cut settings\\
6   & Fit range ($Q_{\rm max}$) & 7 ranges\\
7   & Fit range ($Q_{\rm min}$) & 3 ranges\\
8   & Coulomb effect            & 2 versions\\
\end{tabular} \end{ruledtabular}
\end{table}

The extracted Bose-Einstein correlation functions depend on a number of 
experimental parameters and cut values, as discussed e.g.\ in 
subsections~\ref{ss:paircuts} and \ref{ss:tracks}.  The dependence is 
on the cut for $\pi^\pm$ identification in the $m^2$ spectrum (PID cut), 
the track matching cut in the PID detector and in PC3, the pair cuts in 
the PID detectors and in the DC, the choice of fit range and some other 
settings (like the choice of $Q$ and $\mT$ binning, or the settings of 
the Coulomb-calculation) with negligible contributions. When performing 
fits to the correlation functions (note that the fitting procedure is 
detailed in Section~\ref{ss:fitting}), the fit parameters also depend 
on these settings. Then a given fit parameter $P$ (which represents 
here $R$, $\lambda$ or $\alpha$) takes the value $P^0(i)$ (where $i$ 
represents the number of the $\mT$ bin) if all cuts and settings are at 
their default values. However, the resulting fit parameter is 
$P_n^j(i)$, when a different setting (indexed by $j>0$) was chosen for 
the given setting (indexed by $n$). See a summary of the possible $n$ 
and $j$ values in Table~\ref{t:systdefs}. Then the systematic 
uncertainty of parameter $P$ at the given $\mT$ bin is calculated as 
the average deviation from the default value, for lower and upper 
uncertainties separately. This can be illustrated by the following 
formulas:

\begin{align}
\delta P^\uparrow(i)  &=\sqrt{\sum_{n={\rm cuts}}\frac{1}{N_n^{j\uparrow  }}\sum_{j\in J_n^\uparrow  }(P_n^j(i)-P^0(i))^2}\\
\delta P^\downarrow(i)&=\sqrt{\sum_{n={\rm cuts}}\frac{1}{N_n^{j\downarrow}}\sum_{j\in J_n^\downarrow}(P_n^j(i)-P^0(i))^2}
\end{align}
where $J_n^\uparrow$ is the set of $j$ values where $P_n^j(i)>P^0(i)$, 
and $N_n^{j\uparrow}$ is the number of elements in this set. This 
number may vary from 0 (if both changes increase the fitted value of 
the given parameter) to the number of possible settings (if all changes 
decrease the fitted value of the given parameter). Similarly, 
$J_n^\downarrow$ is the set of $j$ values where $P_n^j(i)<P^0(i)$, and 
$N_n^{j\downarrow}$ is the cardinality of this set. In the above 
formulas, summing over $j$ is only done if $N_n^{j\downarrow}>0$ or 
$N_n^{j\uparrow}>0$. The values for $\delta P^\uparrow(i)$ and $\delta 
P^\downarrow(i)$ were then averaged over the neighboring 5 $\mT$ bins 
(two bins at higher, and two bins at lower $\mT$, in addition to the 
central, averaged value). This procedure allowed us to smooth out the 
apparently nonphysical large fluctuations in the upper or lower limits 
on the systematic uncertainties. Let us also note here that we found 
the different systematic uncertainty sources to be uncorrelated with 
each other, so the quadratic sum in the equation above is justified.

In addition to settings in the correlation function measurement, we 
have performed fit range studies by varying the initial and the final 
$Q$ bin locations ($Q_{\rm min}$ and $Q_{\rm max}$). The results were 
remarkably stable for adding or removing the first few (1--5) or the 
last few (10--20) data points at the beginning or the end of the fit. 
In fact we used this stability criteria to define the beginning and the 
end points of the fitted range. We have also investigated the stability 
of the fit results with respect to duplicating or halving the number of 
$\mT$ bins, and also with respect to doubling the bin size in $Q$, or 
splitting the bins into two equal parts. These sources of uncertainty 
had negligible effects on the fit parameters. We also analyzed the 
uncertainty of the fit results originating from the Coulomb calculation 
(as detailed in subsection~\ref{ss:coulomb}).

Now that all the details of the formalism are described in detail, in the
following we outline the experimental procedure of the measurement and the
results on the L\'evy parameters of two-pion ($\pi^+\pi^+$ and $\pi^-\pi^-$)
Bose-Einstein correlation functions in \sqsntwo \auau collisions.

\section{Results\label{s:results}}

We measured Bose-Einstein correlation functions of $\pi^+\pi^+$ and 
$\pi^-\pi^-$ pairs in 31 bins in the pair average transverse mass 
$\mT$, from 228 MeV$/c^2$ to 871 MeV$/c^2$. Our measurement was based 
on 2.2 billion 0\%--30\% centrality \auau collisions at \sqsntwo 
colliding energy, selected from 7.3 billion MB events. 
Further centrality bins and their analysis is outside the scope of 
present manuscript.

\subsection{Fitting procedure}\label{ss:fitting}

The formulas in Eqs.~\eqref{e:C2Scc}--\eqref{e:C2expr:confhyp} cannot 
be evaluated analytically, and the numerical calculation is also 
cumbersome, so to accelerate the fitting process, we created a lookup 
table for this function, and used it for fitting. We denote our fit 
function based on Eqs.~\eqref{e:C2Scc}--\eqref{e:C2expr:confhyp} as 
$C_2(\lambda,R,\alpha;Q)$, and from now on we drop the notation of the 
$K$ dependence, and explicitly write out the parameter values, i.e.

\begin{align}
C_2(\lambda,R,\alpha;Q)\equiv C_2(Q,K).
\end{align}
However, it turned out that fits using this function resulted in a 
numerically fluctuating $\chi^2$-landscape, so we applied an 
``iterative afterburner'' where the fit function contained only 
analytic dependencies on the fit parameters. Our second round fit 
function was

\begin{align}
C^{(0)}_2(\lambda,R,\alpha;Q)\frac{C_2(\lambda_0,R_0,\alpha_0;Q)}{C_2^{(0)}(\lambda_0,R_0,\alpha_0;Q)}\!\times\!N\!\times\!(1\!+\!\epsilon Q),\label{e:c2coulit}\\
\textnormal{with } C^{(0)}_2(\lambda,R,\alpha;Q) \equiv 1+\lambda e^{-R^\alpha Q^\alpha} , \label{e:C2stable}
\end{align}
where $\lambda_0$, $R_0$, and $\alpha_0$ are the fit parameters from 
the first round of fit. Let us call the resulting fit parameters of 
this next fit $R_1$, $\lambda_1$ and $\alpha_1$. If these differ 
substantially (more than 1\% in squared sum) from $R_0$, $\lambda_0$ 
and $\alpha_0$, then we set $R_0=R_1$, $\lambda_0=\lambda_1$ and 
$\alpha_0=\alpha_1$, and do one more round of fitting. We continued 
this iterative procedure with a fit function of

\begin{align}
C^{(0)}_2(\lambda,R,\alpha;Q)\frac{C_2(\lambda_n,R_n,\alpha_n;Q)}{C_2^{(0)}(\lambda_n,R_n,\alpha_n;Q)}\!\times\!N\!\times\!(1\!+\!\epsilon Q),
\end{align}
until the previous parameter vector $(\lambda_n,R_n,\alpha_n)$ and the 
newly obtained parameter vector $(\lambda_{n+1},R_{n+1},\alpha_{n+1})$ 
differed less than 1\% in the squared sum. Note at this point that in 
the actual fits, a normalization parameter $N$ and a parameter 
$\epsilon$ that represents a possible but small background long-range 
correlation effect were also included. In practice $N\approx1$ and 
$\epsilon\approx1$, and these parameters converge earlier in the fit 
than do the physical parameters $\lambda$, $R$, and $\alpha$. For this 
reason only the physical parameters were used in the test of the 
convergence criteria. In this way the physical source parameters were 
extracted from the data in a reliable manner, with a self-consistent 
treatment for the Coulomb effect. Note that our procedure is in fact 
rather similar to the iterative Coulomb correction method applied by 
the NA44 Collaboration in Ref.~\cite{Boggild:1994vk}. However, in our 
implementation, we use this iterative procedure also for the correction 
for the halo effects, by evaluating the Coulomb wave-functions only for 
the experimentally resolvable (core,core) type of pion pairs.

Pair multiplicities allowed us to use a $\chi^2$ minimization method 
(in contrast to the need for log-likelihood fitting methods if the 
value of $C(Q)$ in the given bin is obtained by the ratio of two small 
numbers $A(Q)$ and $B(Q)$; see details in Ref.~\cite{Zajc:1982vf}). We 
applied \textsc{MINUIT2} minimization libraries~\cite{James:1975dr} 
when performing $\chi^2$ fits to the measured correlation functions. We 
accept the fit results if the following criteria are satisfied: (a) the 
status of the fit is ``converged'' (i.e. a valid minimum was reached), 
(b) the error matrix is ``accurate'' (i.e. fully calculable and 
positive definite), (c) the $\chi^2/$NDF values are acceptable, 
corresponding to a confidence level (CL) above 0.1\%. Our fits 
satisfied these conditions, implying that the fit parameters represent 
the measurements in a statistically acceptable manner. We note here 
that fits with an $\alpha=2$ constraint, i.e. fits with a Gaussian 
assumption were not acceptable. The CL of these Gaussian fits were many 
orders of magnitude below 0.1\%, as the $\chi^2$ values ranged from 
100--600 (for the lowest $\mT$ bins, where $\m{NDF}\approx 100$, and 
also for the highest bins, where NDF values are around 350) to 600--1000 
(for $\mT=$300--500 MeV$/c^2$, where NDF is about 150--220). In 
contrast, L\'evy fits resulted in $\chi^2$ values in the 
1--1.3$\times$NDF range. Note that the statistical acceptability of our 
L\'evy fits to \sqsntwo \auau RHIC data also confirms the validity of 
the assumption about the correlation function being unity plus a 
positive definite function.

\begin{figure*}
\includegraphics[width=0.75\linewidth]{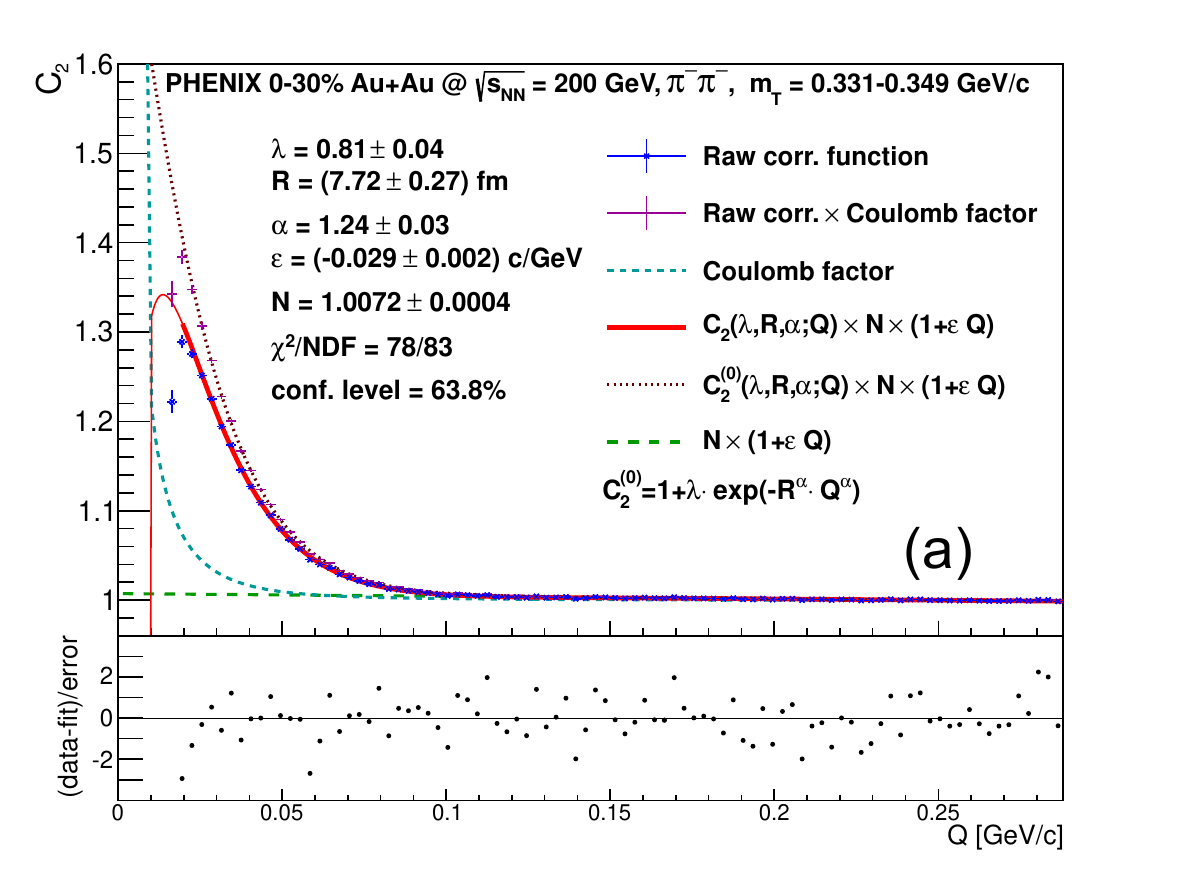}
\includegraphics[width=0.75\linewidth]{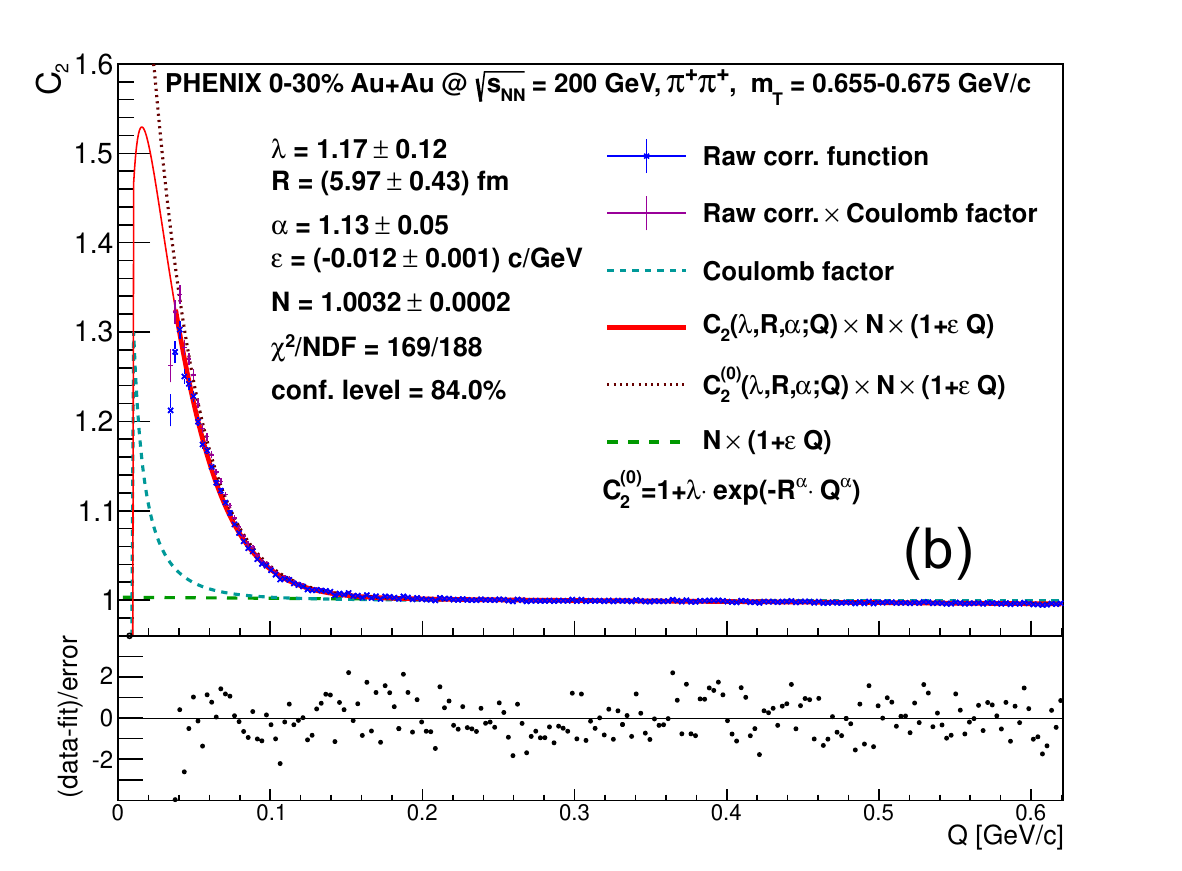}
\caption{Example fits of Bose-Einstein correlation functions of 
(a) $\pi^-\pi^-$ pair with $\mT$ between 0.331 and 0.349 GeV$/c^2$ 
and of (b) $\pi^+\pi^+$ pair with $\mT$ between 0.655 and 0.675 
GeV$/c^2$, as a function $Q\equiv |{\v q}_\m{LCMS}|$, 
defined in~\Eq{e:Qdef}.  Both fits show the measured correlation 
function and the complete fit function (described in~\ref{ss:fitting}), 
while a Bose-Einstein fit function $C_2^{(0)}(Q)$ is also shown, with 
the Coulomb-corrected data, i.e.\ the raw data multiplied by 
$C_2^{(0)}(Q)/C_2(Q)$. In this analysis we measured 62 such correlation 
functions (for $++$ and $--$ pairs, in 31 $\mT$ bins), and fitted all 
of them with the method described in~\ref{ss:fitting}. The first 
visible point on both panels corresponds to $Q$ values below the 
accessible range (based on an evaluation of the two-track cuts), these 
were not taken into account in the fitting.
}
\label{f:fits}
\end{figure*}

We fitted the measured correlation functions with the above outlined 
procedure. Figure~\ref{f:fits} shows some examples of the measured 
Coulomb-distorted two-pion Bose-Einstein correlation function, the 
Coulomb correction factor and the resulting Coulomb-corrected two-pion 
Bose-Einstein correlation functions, together with the fits with 
Eqs.~\eqref{e:c2coulit}-\eqref{e:C2stable} that define the parameters 
of the L\'evy-stable Bose-Einstein correlation functions.

%--------------------------------------------- Fig_3

In Section \ref{ss:mtdepresults}, we present our results for the fits 
and for the trends of the fit parameters, versus average pair 
$\mT=\sqrt{m^2+(\kT/c)^2}$ calculated from the $\kT$ of the pair.

\subsection{Results for the transverse momentum dependence of the fit 
parameters}\label{ss:mtdepresults}

\begin{figure}[htb]
\includegraphics[width=1.0\linewidth]{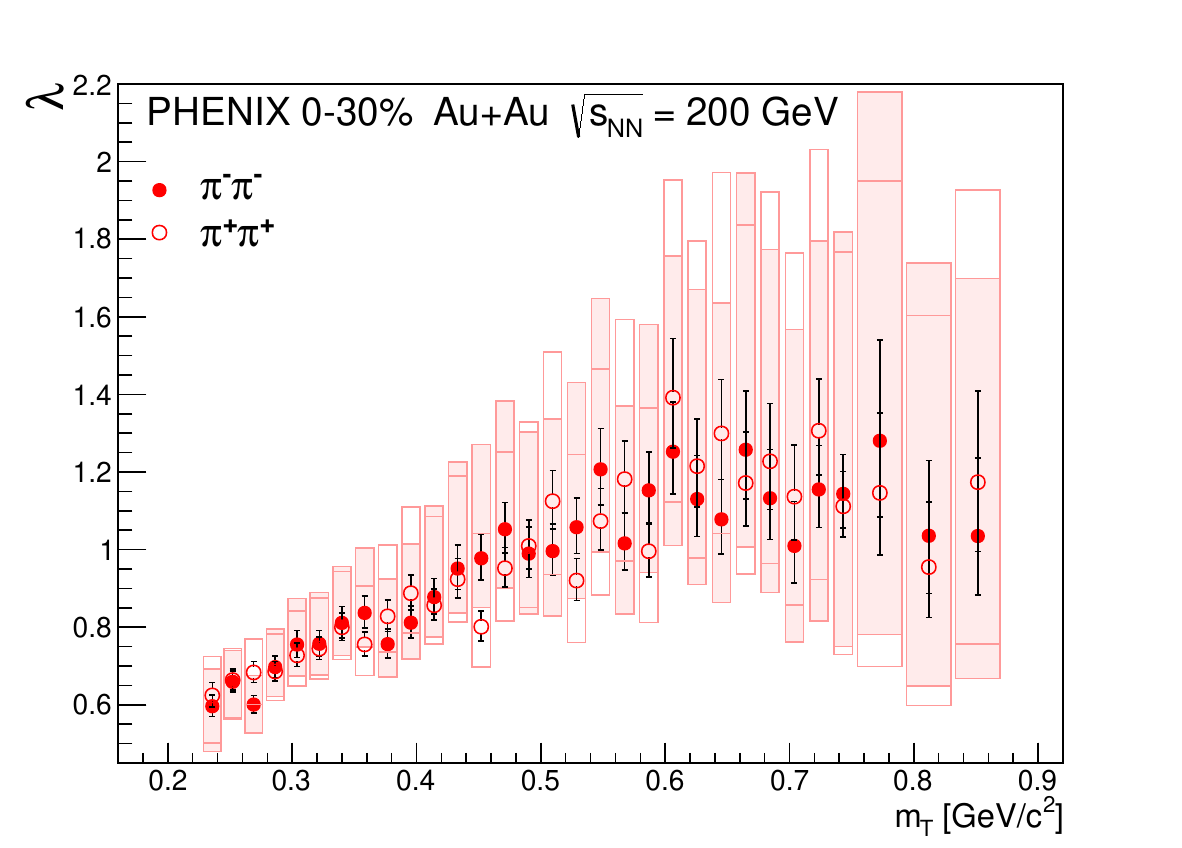}
\caption{Correlation strength parameter $\lambda$ versus average $\mT$ 
of the pair, for 0\%--30\% centrality collisions. Statistical and 
systematic uncertainties are shown as bars and boxes.}
\label{f:lambdamt}
\end{figure}

\begin{figure}[htb]
\includegraphics[width=1.0\linewidth]{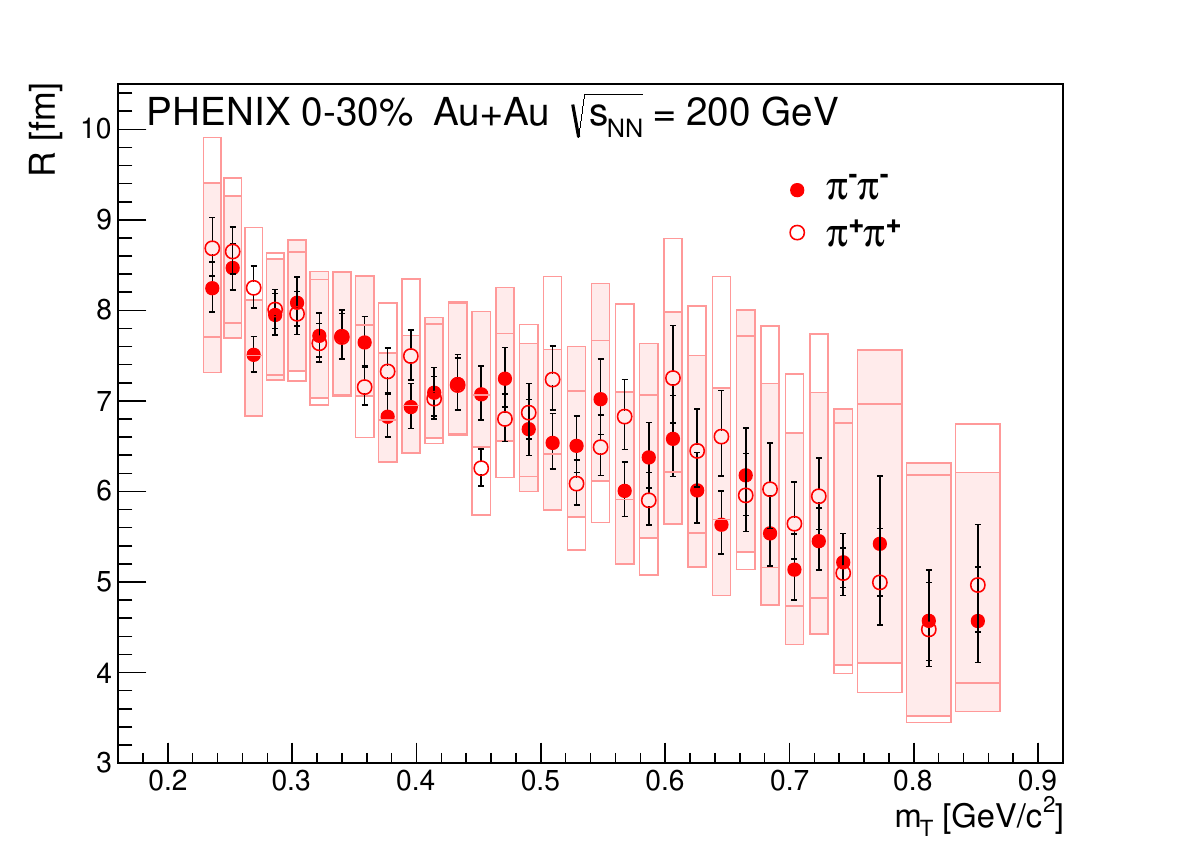}
\caption{L\'evy scale parameter $R$ versus average $\mT$ of the pair. 
The graphical representation of statistical and systematic 
uncertainties is the same as in Fig.~\ref{f:lambdamt}.}
\label{f:Rmt}
\end{figure}

\begin{figure}[htb]
\includegraphics[width=1.0\linewidth]{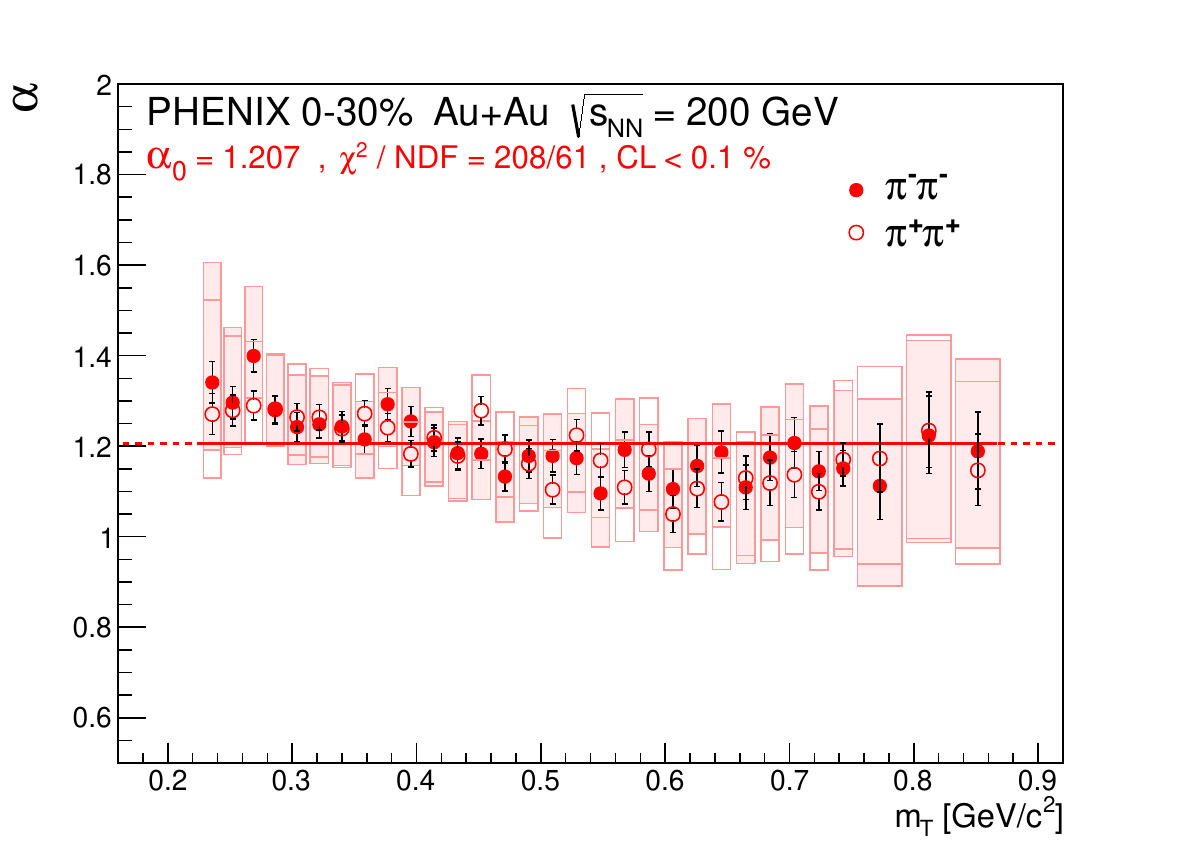}
\caption{L\'evy index parameter $\alpha$ versus average $\mT$ of the 
pair. Statistical and systematic uncertainties are indicated similarly 
to Fig.~\ref{f:lambdamt}. The horizontal line, $\alpha=1.207$, 
represents the 0\%--30\% centrality average value of $\alpha$.}
\label{f:alphamt}
\end{figure}

Parameters $\lambda$, $\alpha$ and $R$ are the physical parameters of 
the fit, while $N \approx 1 $ and $\epsilon \approx 0$ are the 
normalization and background-slope parameters. The $\mT$ dependence of 
the physical parameters $(\lambda, R, \alpha)$ is shown in 
Figs.~\ref{f:lambdamt}, \ref{f:Rmt} and \ref{f:alphamt}. The parameter 
values for $++$ and $--$ pairs in 0\%--30\% centrality collisions are 
given in Table~\ref{t:pars}, while the decomposition of their 
systematic uncertainties is detailed below in Table~\ref{t:systtable}.

The intercept parameter $\lambda$ seems to saturate at high $\mT$. Even 
within the sizable systematic uncertainties of the measurement, a 
decrease of $\lambda(\mT)$ is clearly visible at low values of the 
average transverse mass $\mT$, where the uncertainties of the analysis 
are reduced significantly.

The L\'evy scale parameter $R(\mT)$ indicates a characteristic 
decreasing trend, that is similar to the decrease predicted by 
hydrodynamical calculations of a three-dimensionally expanding source 
for the $\alpha = 2$ Gaussian 
case~\cite{Makhlin:1987gm,Csorgo:1995bi,Chapman:1994yv,Chapman:1994ax}. 
Note that for $\alpha<2$ we are not aware of any theoretical 
predictions for the $m_T$ dependence of the L\'evy scale parameter $R$.

The values of $\alpha(\mT)$ are significantly below the Gaussian limit 
of 2. In certain measurements of two-particle Bose-Einstein 
correlations, if the $\alpha = 2$ Gaussian approximation fails, the 
$\alpha = 1$ exponential approximation is attempted. In our analysis, 
we observe that our $\alpha(\mT)$ data are systematically above 1. 
Although the case of $\alpha=1$ is closer to the measured $\alpha$ 
values than the case of $\alpha=2$, it also is disfavored by the data. 
When we repeat the fits with $\alpha=1$ fixed, the fits become 
statistically unacceptable in most of the $\mT$ bins.

\begin{figure*}[htb]
\includegraphics[width=0.32\linewidth]{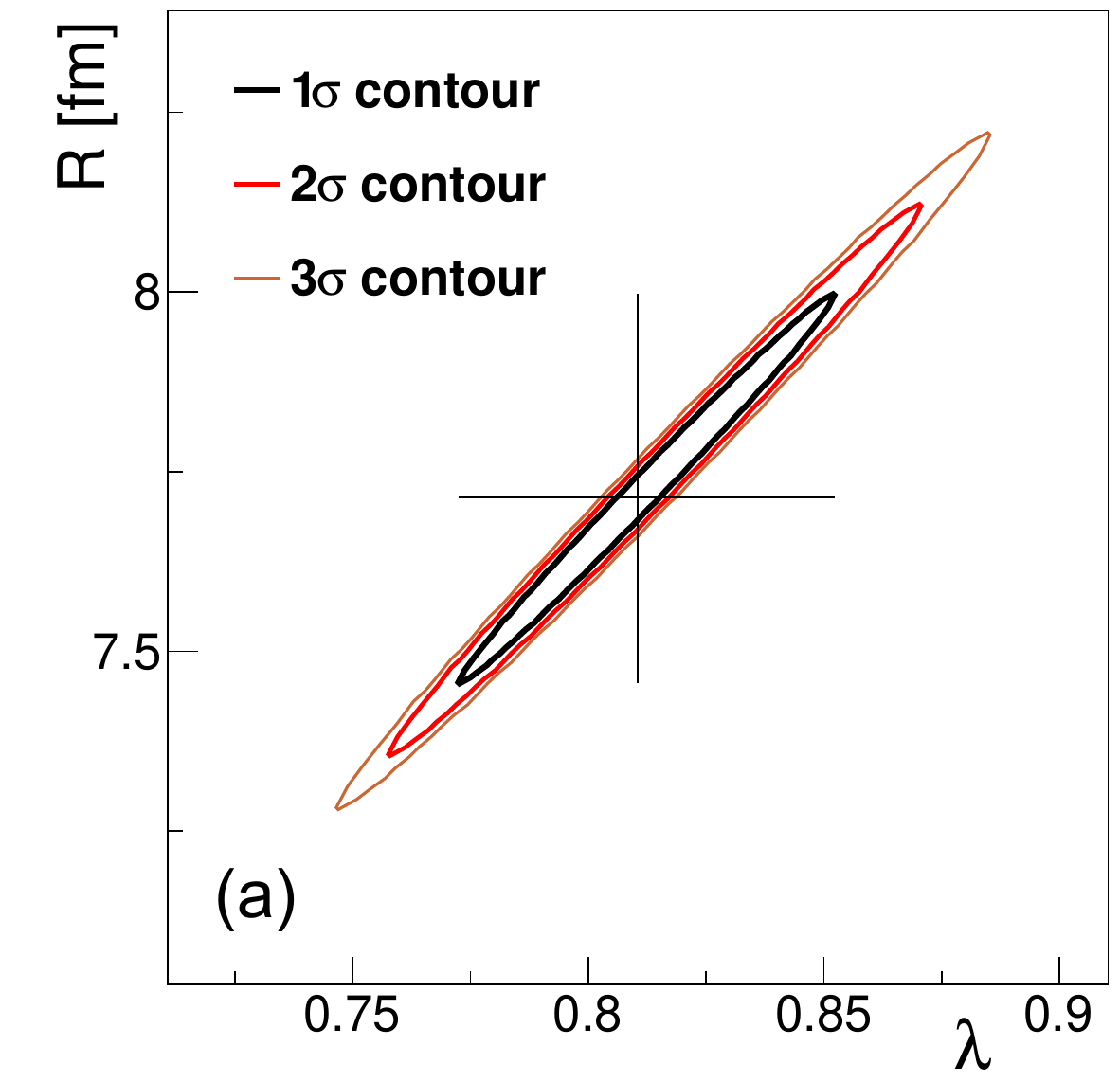}
\includegraphics[width=0.32\linewidth]{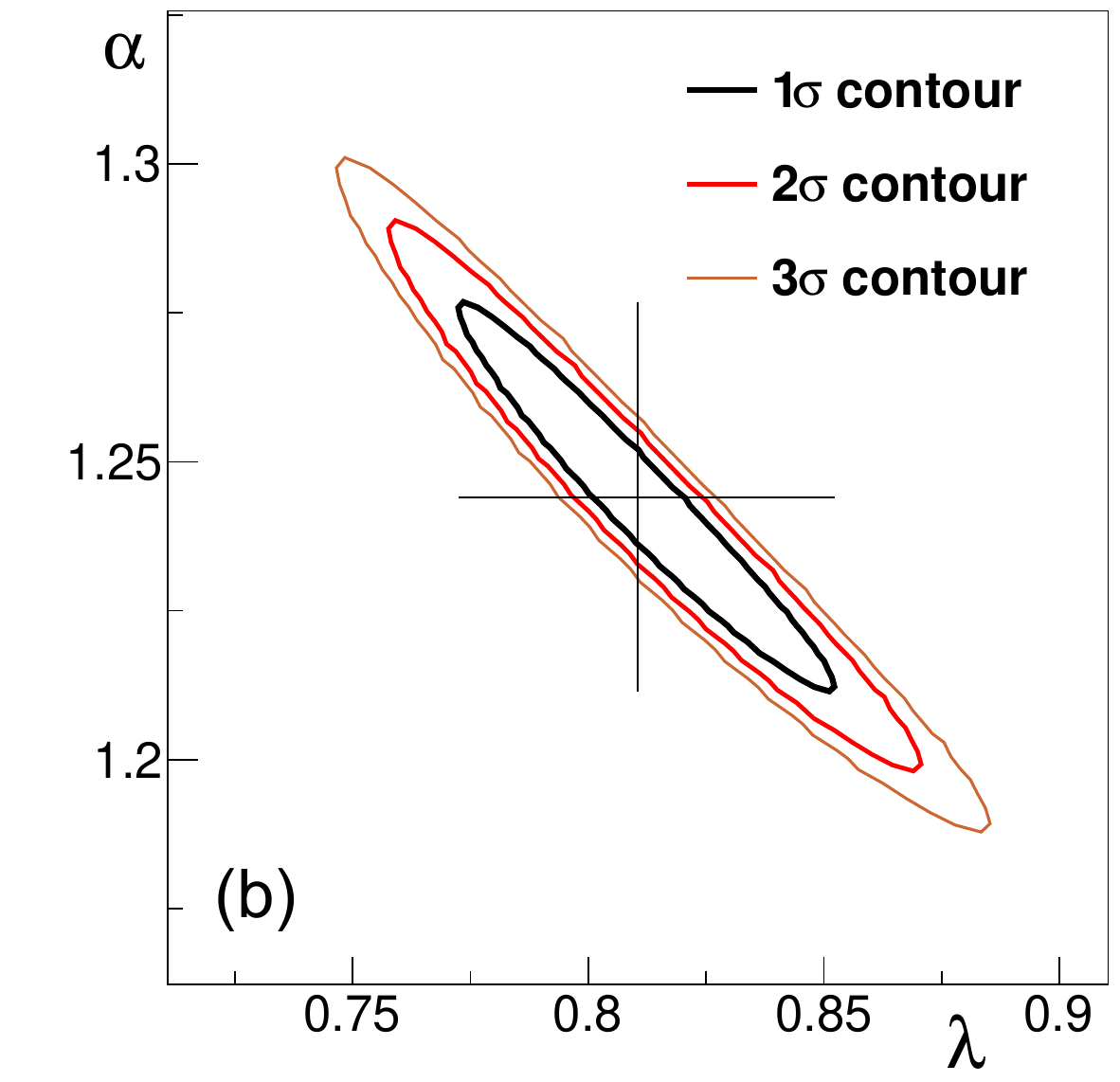}
\includegraphics[width=0.32\linewidth]{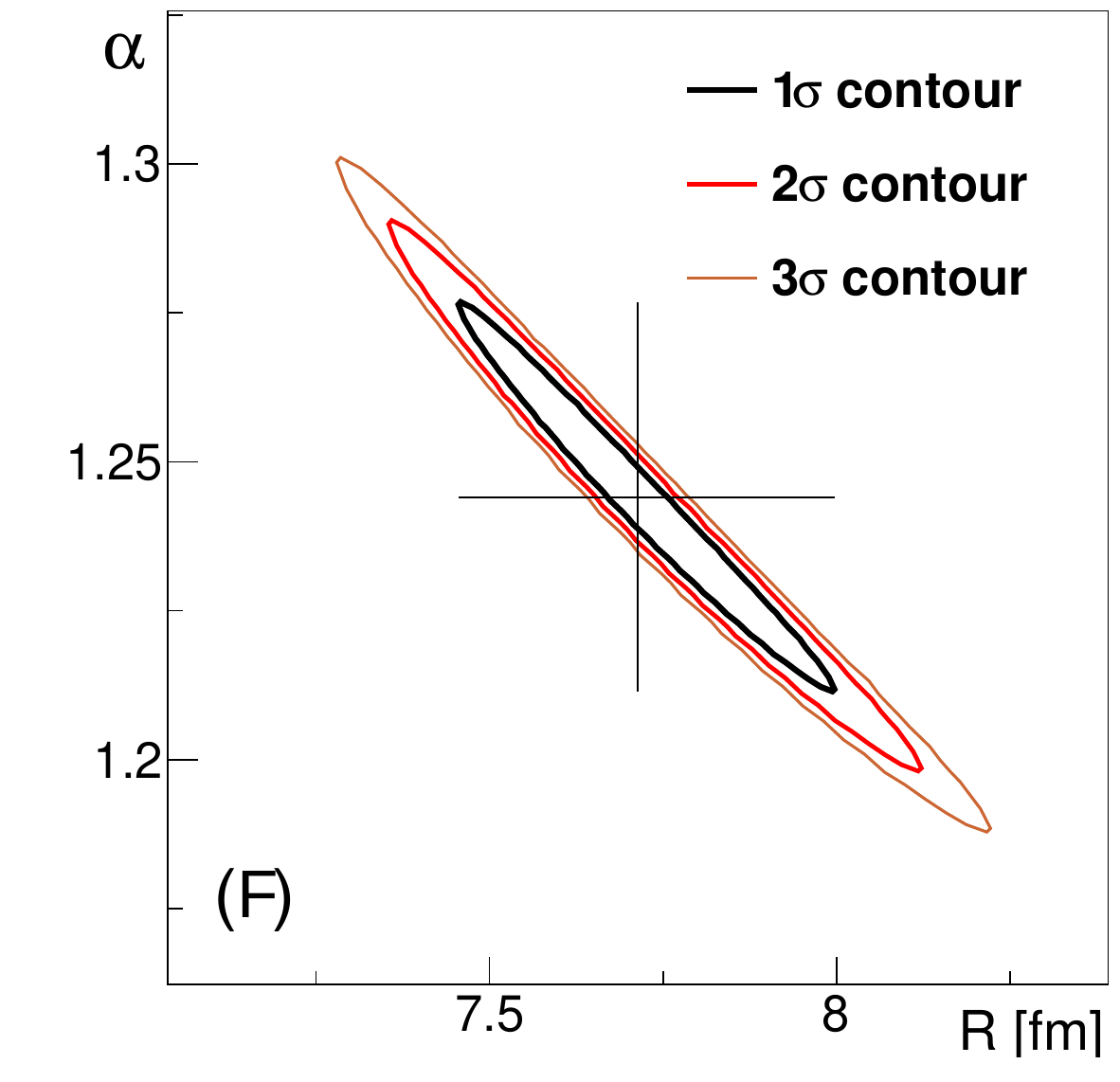}
\caption{Contour lines of the $\chi^2$ map in the (a) $\lambda,R$ and 
(b) $\lambda,\alpha$ and (c) $R,\alpha$ planes for fits to $\pi^-\pi^-$ 
correlation functions of pairs with $\mT$ between 0.331 and 0.349 
GeV$/c^2$. The horizontal and vertical lines represent the MINOS fit 
uncertainties.}
\label{f:contours}
\end{figure*}

Let us also note that the error contours are all narrow tilted ellipses 
on the two-dimensional $\chi^2$ maps in the $(\lambda,R)$, 
$(\lambda,\alpha)$ and $(R,\alpha)$ planes, as shown in 
Fig.~\ref{f:contours}. This illustrates that the parameters of the 
L\'evy-stable fits of \Eq{e:c2coulit} are highly correlated. Typical 
values of the correlation coefficients for the $(\lambda,R)$, 
$(\lambda,\alpha)$ and $(R,\alpha)$ coefficients are around $99$\%, 
$-97$\% and $-99$\%, respectively.

As discussed in Section \ref{s:systematics}, the extracted parameters 
of Bose-Einstein correlation functions depend on a number of 
experimental parameters and settings. In 
Figs.~\ref{f:lambdamt}--\ref{f:alphamt} and Table~\ref{t:systtable}, we 
indicate the corresponding total systematic uncertainty, bin by bin. A 
charge averaged, and (in two $\mT$ regions) $\mT$ averaged 
decomposition of the systematic uncertainties is given in Table 
\ref{t:systtable} (both for the parameters introduced above, and those 
defined in the next subsections). Let us note here that the systematic 
uncertainties contain both $\mT$-correlated and uncorrelated 
components. Uncertainties coming from the variations of pair-cuts are 
mostly uncorrelated, while the ones from the PID arm and fit 
extrapolation are $\mT$-correlated. As for the other sources of 
uncertainties, they have an $\mT$-correlated effect on $\lambda$, but 
an uncorrelated effect on $R$ and $\alpha$. There are clear differences 
in the systematic uncertainties between the two $\mT$ regions both in 
relative size and in distribution among the sources of uncertainty. 
This translates into differences in the $\mT$-correlated nature of the 
systematic uncertainties as well. Let us also note here that the 
systematic uncertainties are further $\mT$-correlated because of the 
averaging process described in Section \ref{s:systematics}.

%======================================================== Table_II
\begin{table*}
\begin{minipage}{0.99\linewidth}
\caption{\label{t:pars} Physical fit parameters $\lambda$, $R$ and 
$\alpha$, as a function of bin $\mT$, for $\pi^+\pi^+$ and $\pi^-\pi^-$ 
pairs measured in 0\%--30\% centrality collisions. Statistical 
uncertainties (corresponding to $1\sigma$ contours, determined by 
Minuit's Minos algorithm) are indicated, followed by systematic 
uncertainties.}
\begin{ruledtabular} \begin{tabular}{ccccccc}
$m_T$  & $\lambda(\pi^-)$       & $R(\pi^-)$     & $\alpha(\pi^-)$        
& $\lambda(\pi^+)$       & $R(\pi^+)$     & $\alpha(\pi^+)$ \\
(GeV/$c^2$) &       &   (fm)     &      &     &  (fm)     &  \\\hline
0.236 & $0.60^{+0.03+0.10}_{-0.03-0.12}$ & $8.2^{+0.3+1.2}_{-0.3-0.9}$ & $1.34^{+0.05+0.27}_{-0.05-0.15}$ & $0.62^{+0.03+0.10}_{-0.03-0.12}$ & $8.7^{+0.3+1.2}_{-0.3-1}$ & $1.27^{+0.05+0.25}_{-0.04-0.14}$ \\
0.252 & $0.66^{+0.03+0.08}_{-0.03-0.10}$ & $8.5^{+0.3+0.8}_{-0.2-0.8}$ & $1.30^{+0.04+0.17}_{-0.04-0.10}$ & $0.66^{+0.03+0.08}_{-0.03-0.10}$ & $8.7^{+0.3+0.8}_{-0.2-0.8}$ & $1.28^{+0.03+0.16}_{-0.03-0.10}$ \\
0.269 & $0.60^{+0.02+0.08}_{-0.02-0.07}$ & $7.5^{+0.2+0.6}_{-0.2-0.7}$ & $1.40^{+0.04+0.15}_{-0.04-0.09}$ & $0.68^{+0.03+0.09}_{-0.03-0.08}$ & $8.2^{+0.2+0.7}_{-0.2-0.7}$ & $1.29^{+0.03+0.14}_{-0.03-0.09}$ \\
0.286 & $0.70^{+0.03+0.10}_{-0.03-0.08}$ & $7.9^{+0.2+0.6}_{-0.2-0.7}$ & $1.28^{+0.03+0.12}_{-0.03-0.08}$ & $0.69^{+0.03+0.10}_{-0.02-0.08}$ & $8.0^{+0.2+0.6}_{-0.2-0.7}$ & $1.28^{+0.03+0.12}_{-0.03-0.08}$ \\
0.304 & $0.76^{+0.04+0.12}_{-0.03-0.08}$ & $8.1^{+0.3+0.7}_{-0.3-0.8}$ & $1.24^{+0.03+0.12}_{-0.03-0.08}$ & $0.73^{+0.03+0.12}_{-0.03-0.08}$ & $8.0^{+0.2+0.7}_{-0.2-0.7}$ & $1.26^{+0.03+0.12}_{-0.03-0.08}$ \\
0.322 & $0.76^{+0.03+0.13}_{-0.03-0.08}$ & $7.7^{+0.3+0.7}_{-0.2-0.7}$ & $1.25^{+0.03+0.11}_{-0.03-0.09}$ & $0.74^{+0.03+0.13}_{-0.03-0.08}$ & $7.6^{+0.2+0.7}_{-0.2-0.7}$ & $1.26^{+0.03+0.11}_{-0.03-0.09}$ \\
0.340 & $0.81^{+0.04+0.15}_{-0.04-0.08}$ & $7.7^{+0.3+0.7}_{-0.3-0.7}$ & $1.24^{+0.03+0.10}_{-0.03-0.09}$ & $0.80^{+0.04+0.14}_{-0.03-0.08}$ & $7.7^{+0.3+0.7}_{-0.2-0.6}$ & $1.24^{+0.03+0.10}_{-0.03-0.09}$ \\
0.358 & $0.84^{+0.04+0.17}_{-0.04-0.09}$ & $7.6^{+0.3+0.7}_{-0.3-0.6}$ & $1.21^{+0.03+0.08}_{-0.03-0.08}$ & $0.76^{+0.03+0.15}_{-0.03-0.08}$ & $7.2^{+0.2+0.7}_{-0.2-0.6}$ & $1.27^{+0.03+0.09}_{-0.03-0.09}$ \\
0.377 & $0.76^{+0.04+0.17}_{-0.04-0.08}$ & $6.8^{+0.2+0.7}_{-0.2-0.5}$ & $1.29^{+0.03+0.08}_{-0.03-0.09}$ & $0.83^{+0.04+0.18}_{-0.04-0.09}$ & $7.3^{+0.3+0.8}_{-0.2-0.5}$ & $1.24^{+0.03+0.08}_{-0.03-0.09}$ \\
0.395 & $0.81^{+0.04+0.20}_{-0.04-0.09}$ & $6.9^{+0.3+0.8}_{-0.2-0.5}$ & $1.25^{+0.03+0.07}_{-0.03-0.10}$ & $0.89^{+0.05+0.22}_{-0.04-0.10}$ & $7.5^{+0.3+0.9}_{-0.3-0.5}$ & $1.18^{+0.03+0.07}_{-0.03-0.09}$ \\
0.414 & $0.88^{+0.05+0.23}_{-0.04-0.10}$ & $7.1^{+0.3+0.8}_{-0.3-0.5}$ & $1.21^{+0.03+0.07}_{-0.03-0.10}$ & $0.86^{+0.04+0.23}_{-0.04-0.10}$ & $7.0^{+0.2+0.8}_{-0.2-0.5}$ & $1.22^{+0.03+0.07}_{-0.03-0.10}$ \\
0.433 & $0.95^{+0.06+0.27}_{-0.05-0.11}$ & $7.2^{+0.3+0.9}_{-0.3-0.6}$ & $1.18^{+0.03+0.07}_{-0.03-0.10}$ & $0.92^{+0.05+0.26}_{-0.05-0.11}$ & $7.2^{+0.3+0.9}_{-0.3-0.6}$ & $1.18^{+0.03+0.07}_{-0.03-0.10}$ \\
0.452 & $0.98^{+0.06+0.29}_{-0.06-0.13}$ & $7.1^{+0.3+0.9}_{-0.3-0.6}$ & $1.18^{+0.03+0.07}_{-0.03-0.10}$ & $0.80^{+0.04+0.24}_{-0.04-0.10}$ & $6.3^{+0.2+0.8}_{-0.2-0.5}$ & $1.28^{+0.03+0.08}_{-0.03-0.11}$ \\
0.471 & $1.05^{+0.07+0.33}_{-0.06-0.15}$ & $7.2^{+0.3+1.0}_{-0.3-0.7}$ & $1.13^{+0.03+0.08}_{-0.03-0.10}$ & $0.95^{+0.05+0.30}_{-0.05-0.14}$ & $6.8^{+0.3+0.9}_{-0.2-0.6}$ & $1.19^{+0.03+0.08}_{-0.03-0.11}$ \\
0.490 & $0.99^{+0.07+0.31}_{-0.06-0.16}$ & $6.7^{+0.3+0.9}_{-0.3-0.7}$ & $1.18^{+0.04+0.09}_{-0.04-0.11}$ & $1.01^{+0.07+0.32}_{-0.06-0.16}$ & $6.9^{+0.3+1.0}_{-0.3-0.7}$ & $1.16^{+0.03+0.08}_{-0.03-0.10}$ \\
0.509 & $1.00^{+0.07+0.34}_{-0.06-0.17}$ & $6.5^{+0.3+1.0}_{-0.3-0.7}$ & $1.18^{+0.04+0.09}_{-0.04-0.11}$ & $1.12^{+0.08+0.38}_{-0.07-0.19}$ & $7.2^{+0.4+1.1}_{-0.3-0.8}$ & $1.10^{+0.03+0.09}_{-0.03-0.11}$ \\
0.529 & $1.06^{+0.08+0.37}_{-0.07-0.18}$ & $6.5^{+0.3+1.1}_{-0.3-0.8}$ & $1.17^{+0.04+0.10}_{-0.04-0.12}$ & $0.92^{+0.06+0.32}_{-0.05-0.16}$ & $6.1^{+0.3+1.0}_{-0.2-0.7}$ & $1.22^{+0.03+0.10}_{-0.03-0.12}$ \\
0.548 & $1.21^{+0.10+0.44}_{-0.09-0.21}$ & $7.0^{+0.4+1.3}_{-0.4-0.9}$ & $1.10^{+0.04+0.10}_{-0.04-0.12}$ & $1.07^{+0.08+0.39}_{-0.07-0.19}$ & $6.5^{+0.4+1.2}_{-0.3-0.8}$ & $1.17^{+0.04+0.11}_{-0.04-0.13}$ \\
0.567 & $1.02^{+0.08+0.35}_{-0.07-0.18}$ & $6.0^{+0.3+1.1}_{-0.3-0.8}$ & $1.19^{+0.04+0.11}_{-0.04-0.13}$ & $1.18^{+0.10+0.41}_{-0.09-0.21}$ & $6.8^{+0.4+1.2}_{-0.4-0.9}$ & $1.11^{+0.04+0.10}_{-0.04-0.12}$ \\
0.587 & $1.15^{+0.10+0.43}_{-0.09-0.21}$ & $6.4^{+0.4+1.3}_{-0.3-0.9}$ & $1.14^{+0.04+0.11}_{-0.04-0.13}$ & $1.00^{+0.07+0.37}_{-0.07-0.18}$ & $5.9^{+0.3+1.2}_{-0.3-0.8}$ & $1.19^{+0.04+0.11}_{-0.04-0.13}$ \\
0.606 & $1.25^{+0.13+0.50}_{-0.11-0.24}$ & $6.6^{+0.5+1.4}_{-0.4-0.9}$ & $1.11^{+0.04+0.10}_{-0.04-0.13}$ & $1.39^{+0.15+0.56}_{-0.13-0.27}$ & $7.3^{+0.6+1.6}_{-0.5-1.0}$ & $1.05^{+0.04+0.10}_{-0.04-0.12}$ \\
0.626 & $1.13^{+0.11+0.54}_{-0.10-0.22}$ & $6.0^{+0.4+1.5}_{-0.4-0.8}$ & $1.16^{+0.05+0.10}_{-0.05-0.15}$ & $1.22^{+0.12+0.58}_{-0.10-0.24}$ & $6.4^{+0.5+1.6}_{-0.4-0.9}$ & $1.11^{+0.04+0.10}_{-0.04-0.14}$ \\
0.645 & $1.08^{+0.10+0.56}_{-0.09-0.21}$ & $5.6^{+0.4+1.5}_{-0.3-0.8}$ & $1.19^{+0.05+0.11}_{-0.05-0.16}$ & $1.30^{+0.14+0.67}_{-0.12-0.26}$ & $6.6^{+0.5+1.8}_{-0.4-0.9}$ & $1.08^{+0.04+0.10}_{-0.04-0.15}$ \\
0.665 & $1.26^{+0.15+0.71}_{-0.13-0.25}$ & $6.2^{+0.5+1.8}_{-0.4-0.8}$ & $1.11^{+0.05+0.10}_{-0.05-0.17}$ & $1.17^{+0.13+0.66}_{-0.11-0.23}$ & $6.0^{+0.5+1.8}_{-0.4-0.8}$ & $1.13^{+0.05+0.10}_{-0.05-0.17}$ \\
0.684 & $1.13^{+0.13+0.64}_{-0.11-0.24}$ & $5.5^{+0.4+1.7}_{-0.4-0.8}$ & $1.17^{+0.05+0.11}_{-0.05-0.18}$ & $1.23^{+0.15+0.70}_{-0.12-0.26}$ & $6.0^{+0.5+1.8}_{-0.4-0.9}$ & $1.12^{+0.05+0.11}_{-0.05-0.17}$ \\
0.704 & $1.01^{+0.11+0.56}_{-0.10-0.25}$ & $5.1^{+0.4+1.5}_{-0.3-0.8}$ & $1.21^{+0.06+0.13}_{-0.06-0.19}$ & $1.14^{+0.13+0.63}_{-0.11-0.28}$ & $5.6^{+0.5+1.6}_{-0.4-0.9}$ & $1.14^{+0.05+0.12}_{-0.05-0.18}$ \\
0.724 & $1.16^{+0.11+0.64}_{-0.10-0.34}$ & $5.5^{+0.4+1.6}_{-0.3-1.0}$ & $1.14^{+0.04+0.14}_{-0.04-0.18}$ & $1.31^{+0.13+0.73}_{-0.11-0.38}$ & $5.9^{+0.4+1.8}_{-0.4-1.1}$ & $1.10^{+0.04+0.14}_{-0.04-0.17}$ \\
0.743 & $1.14^{+0.10+0.67}_{-0.09-0.39}$ & $5.2^{+0.3+1.7}_{-0.3-1.1}$ & $1.15^{+0.04+0.17}_{-0.04-0.19}$ & $1.11^{+0.09+0.65}_{-0.08-0.38}$ & $5.1^{+0.3+1.7}_{-0.2-1.1}$ & $1.17^{+0.04+0.18}_{-0.04-0.20}$ \\
0.773 & $1.28^{+0.26+0.90}_{-0.20-0.50}$ & $5.4^{+0.7+2.1}_{-0.6-1.3}$ & $1.11^{+0.08+0.19}_{-0.07-0.22}$ & $1.15^{+0.21+0.81}_{-0.16-0.45}$ & $5.0^{+0.6+2.0}_{-0.5-1.2}$ & $1.17^{+0.08+0.20}_{-0.07-0.23}$ \\
0.812 & $1.04^{+0.19+0.71}_{-0.15-0.39}$ & $4.6^{+0.6+1.7}_{-0.4-1.1}$ & $1.22^{+0.09+0.21}_{-0.08-0.24}$ & $0.96^{+0.17+0.65}_{-0.13-0.36}$ & $4.5^{+0.5+1.7}_{-0.4-1.0}$ & $1.23^{+0.08+0.21}_{-0.08-0.24}$ \\
0.852 & $1.04^{+0.20+0.67}_{-0.15-0.37}$ & $4.6^{+0.6+1.6}_{-0.5-1.0}$ & $1.19^{+0.09+0.20}_{-0.08-0.21}$ & $1.17^{+0.23+0.75}_{-0.18-0.42}$ & $5.0^{+0.7+1.8}_{-0.5-1.1}$ & $1.15^{+0.08+0.20}_{-0.08-0.21}$ \\
\end{tabular} \end{ruledtabular} 
%\end{table*}
\end{minipage}
\begin{minipage}{0.99\linewidth}
%======================================================== Table_III
%\begin{table*}
\caption{\label{t:systtable}
$\mT$ and charge averaged asymmetric systematic uncertainties of the 
physical parameters, separately for the low $\mT$ bins (180--500 
MeV$/c^2$) and the high $\mT$ bins (500--850 MeV$/c^2$). The arrows 
$\up$ and $\dn$ represent the up and down systematic uncertainties.}
\begin{ruledtabular} \begin{tabular}{lccccccccccccccccccccccc}
                          & \multicolumn{11}{c}{$\mT<500$ MeV$/c^{2}$ average uncertainties [\%]}
                        & & \multicolumn{11}{c}{$\mT>500$ MeV$/c^{2}$ average uncertainties [\%]}
\\
                          & \multicolumn{2}{c}{$\lambda$}
                          & \multicolumn{2}{c}{$R$}
                          & \multicolumn{2}{c}{$\alpha$}
                          & \multicolumn{2}{c}{$1/\hat{R}$}
                        & & \multicolumn{2}{c}{$\lambda/\lambda_{\rm max}$}
                        & & \multicolumn{2}{c}{$\lambda$}
                          & \multicolumn{2}{c}{$R$}
                          & \multicolumn{2}{c}{$\alpha$}
                          & \multicolumn{2}{c}{$1/\hat{R}$}
                        & & \multicolumn{2}{c}{$\lambda/\lambda_{\rm max}$}
\\
                          &$\up$&$\dn$&$\up$&$\dn$&$\up$&$\dn$&$\up$&$\dn$&&$\up$&$\dn$
                        & &$\up$&$\dn$&$\up$&$\dn$&$\up$&$\dn$&$\up$&$\dn$&&$\up$&$\dn$
\\ \hline
PID arm                    & 8.9 & 9.6 & 8.5 & 5.8 & 9.2 & 4.9 & 5.4 & 6.0 && 12. & 20. 
                        &  & 28. & 12. & 17. & 6.9 & 4.9 & 7.4 & 5.6 & 4.2 && 16. & 12. \\
PID cut                    & 4.4 & 3.8 & 1.8 & 2.2 & 2.0 & 1.3 & 4.0 & 3.8 && 3.8 & 5.9 
                        &  & 11. & 7.7 & 6.0 & 4.2 & 2.9 & 3.4 & 3.6 & 3.5 && 6.0 & 5.7 \\
PID det. matching cut      & 4.0 & 13. & 2.2 & 1.8 & 1.4 & 1.5 & 2.9 & 2.0 && 1.8 & 1.8 
                        &  & 3.8 & 22. & 2.4 & 4.2 & 2.7 & 1.6 & 1.2 & 0.5 && 2.4 & 1.9 \\
PID det. paircut           & 4.4 & 3.0 & 2.2 & 1.8 & 1.5 & 1.5 & 3.1 & 2.3 && 8.0 & 4.3 
                        &  & 7.7 & 7.5 & 4.3 & 5.1 & 3.5 & 2.5 & 2.9 & 2.1 && 4.1 & 4.5 \\
PC3 matching cut           & 14. & 0.6 & 4.7 & 2.2 & 1.9 & 3.0 & 8.9 & 0.0 && 0.2 & 19. 
                        &  & 38. & 0.1 & 17. & 1.5 & 0.9 & 8.7 & 13. & 0.0 && 9.1 & 7.6 \\
DC paircut                 & 3.0 & 3.4 & 1.9 & 2.5 & 1.9 & 1.5 & 0.7 & 0.7 && 13. & 1.7 
                        &  & 2.1 & 16. & 7.7 & 9.9 & 7.7 & 0.8 & 0.5 & 4.0 && 10. & 10. \\
Fit range ($Q_{\rm min}$)  & 4.4 & 4.8 & 3.1 & 3.3 & 2.3 & 2.0 & 0.5 & 0.5 && 12. & 5.7 
                        &  & 7.8 & 14. & 6.2 & 9.3 & 6.2 & 3.2 & 1.4 & 2.4 && 5.1 & 5.4 \\
Fit range ($Q_{\rm max}$)  & 3.2 & 3.2 & 2.2 & 2.2 & 2.0 & 2.0 & 0.2 & 0.2 && 4.3 & 4.3 
                        &  & 4.5 & 4.5 & 3.2 & 3.2 & 2.1 & 2.1 & 0.5 & 0.5 && 6.6 & 6.6 \\
Coulomb effect             & 9.4 & 0.0 & 4.2 & 0.0 & 0.0 & 3.4 & 3.8 & 0.0 && 0.0 & 10. 
                        &  & 21. & 0.0 & 13. & 0.0 & 0.0 & 8.1 & 2.0 & 0.0 && 1.6 & 2.0 \\
Total                      & 21. & 18. & 12. & 8.5 & 11. & 7.8 & 13. & 7.8 && 24. & 31. 
                        &  & 54. & 35. & 30. & 18. & 12. & 15. & 15. & 7.5 && 24. & 21. \\
\end{tabular} \end{ruledtabular}
\end{minipage}
\end{table*}

%--------------------------------------------- Fig_4

%--------------------------------------------- Fig_5

%--------------------------------------------- Fig_6

%--------------------------------------------- Fig_7

\subsection{Discussion and interpretation of the results}

In this subsection we discuss more subtle physical interpretations of 
the measured trends of the parameters of the two-pion Bose-Einstein 
correlation functions.

Starting with the L\'evy exponent, we observe that in each of the 
investigated cases, $\alpha$ values were slightly above 1. It is known 
that the value of the critical exponent of the random field 3D Ising 
model is 0.5~\cite{Rieger:1995aa}, much larger than the value of the 
critical exponent in the 3D Ising model~\cite{El-Showk:2014dwa} 
(without random external fields). It is also known that the 3D Ising 
model is expected to be in the same universality class as the second 
order QCD phase transition~\cite{Halasz:1998qr,Stephanov:1998dy}. 
Therefore, we observe that the measured values of the L\'evy exponent in 
0\%--30\% centrality Au$+$Au collisions at \sqsntwo do not correspond to the 
conjectured value ($\leq 0.5$) of the exponent of the two-particle 
correlation function at the QCD critical point~\cite{Csorgo:2009wc}. 
The appearance of the critical point is not expected near \sqsntwo, 
thus we emphasize the need for similar measurements at lower collision 
energies.

Hydrodynamic calculations typically predict Gaussian shapes (i.e. 
$\alpha=2$) for the Bose-Einstein correlation 
functions~\cite{Akkelin:1996sg,Akkelin:1995gh,Csorgo:1994fg,Csizmadia:1998ef,Csorgo:1995bi,Csanad:2009wc}. 
We may also note that in certain cases the freeze-out criteria may 
alter this behavior, interference terms between two different extrema 
in the source may lead to small deviations from Gaussian Bose-Einstein 
correlations~\cite{Csorgo:1999sj,Csorgo:2000vs}. The measured 
correlation functions discussed in the present paper show large 
deviations from the Gaussian assumption. Our observations show that the 
source of charged pions in the investigated momentum range is a L\'evy 
distribution with an average index of stability of $\alpha \approx 
1.2$, see Fig.~\ref{f:alphamt}.

Various scenarios may lead to such a source with a long power-law like 
tail, e.g.\ rescattering in an expanding medium with time-dependent 
mean free path, which is also called anomalous diffusion or L\'evy 
flight. In such a scenario, the smaller the cross section, the longer 
the mean free path (at a given time), thus the longer the tail of the 
source distribution. This might be tested by comparing the L\'evy 
source distributions for pions, kaons and 
protons~\cite{Akiba:1992cj,Csanad:2007fr}.

\begin{figure}
\includegraphics[width=1.0\linewidth]{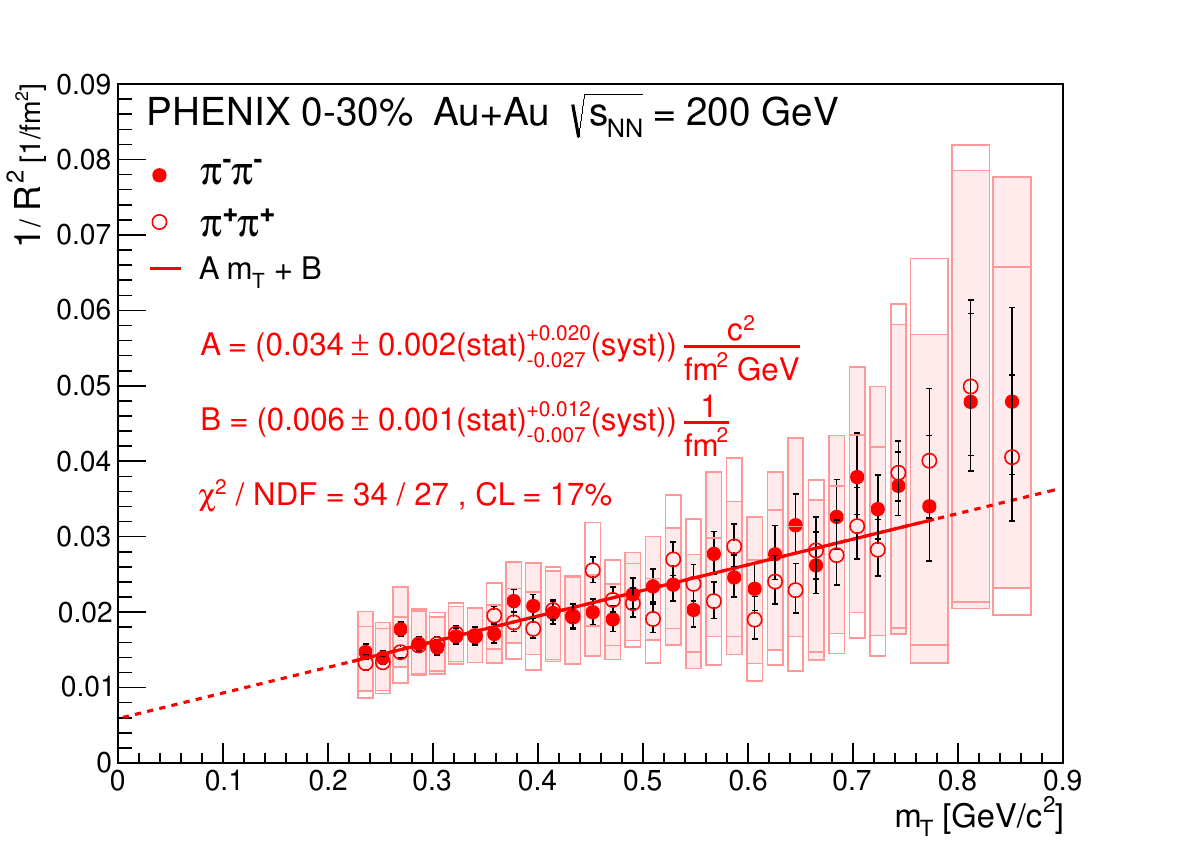}
\caption{Inverse square of the L\'evy scale parameter $1/R^2$ versus 
average $\mT$ of the pair. Statistical and systematic uncertainties 
shown as bars and boxes, respectively.}
\label{f:Rcontextmt}
\end{figure}

As the L\'evy scale parameter $R$ defines the length scales of the 
particle-emitting source for particle emission with heavy tails, the 
$\mT$ dependence of these parameters is worth investigating in greater 
detail. It turns out (shown in Fig.~\ref{f:Rcontextmt}) that a 
hydrodynamical type of $1/R^2\propto \mT$ scaling holds approximately, 
especially in the low $\mT$ region. This corresponds to the scaling 
predictions for the HBT radii from hydrodynamical 
calculations~\cite{Makhlin:1987gm,Csorgo:1995bi,Chapman:1994yv,Chapman:1994ax,Akkelin:1996sg,Akkelin:1995gh,Csorgo:1994fg,Csizmadia:1998ef,Csanad:2009wc}. 
Although these predictions assumed $\alpha=2$, the scaling seems to 
hold remarkably even in this case of $\alpha<2$. We also show a linear 
$A\mT+B$ fit to $1/R^2$ versus $\mT$, taking into account only the 
statistical uncertainties when determining the best values and the 
statistical errors of the fit parameters. The resulting parameters 
turned out to be

\begin{align}
A&= 0.034\pm0.002\textnormal{ (stat)}^{+0.020}_{-0.027}\textnormal{ (syst) } \frac{c^2}{\m{fm}^2\m{GeV}},\\
B&= 0.006\pm0.001\textnormal{ (stat)}^{+0.012}_{-0.007}\textnormal{ (syst) } \frac{1}{\m{fm}^2},
\end{align}
as noted in Fig.~\ref{f:Rcontextmt}. Systematic uncertainties of the 
fit parameters were determined by performing a linear fit to $1/R^2$ 
versus $\mT$ obtained from measurements and fits with varied settings 
(listed e.g. in Table~\ref{t:systtable}). The $A$ and $B$ parameters 
above can be converted to a simple

\begin{align}\label{e:Rlinear_physical}
R(\mT) = \frac{R_\xi}{\sqrt{\mT/m_\pi+\xi}}
\end{align}
dependence, where one then gets $R_\xi=(14.55\pm 0.43)$ fm and 
$\xi=1.27\pm0.22$.

%--------------------------------------------- Fig_8

Because the estimators of L\'evy parameters $\alpha$, $R$ and $\lambda$ 
are strongly correlated, reasonably good (although not necessarily 
statistically acceptable) fits can be obtained with multiple sets of 
co-varied parameters. This motivated us to search for less correlated 
combinations of these parameters. Unexpectedly, and without any 
theoretical motivation for this new scaling law except perhaps the 
suggestions of Ref.~\cite{Zajc:1992sz}, we indeed found such a 
parameter, defined as

\begin{align}
\hat{R} = \frac{R}{\lambda(1+\alpha)}.\label{e:rhatdef}
\end{align}

\begin{figure*}
\includegraphics[width=0.32\linewidth]{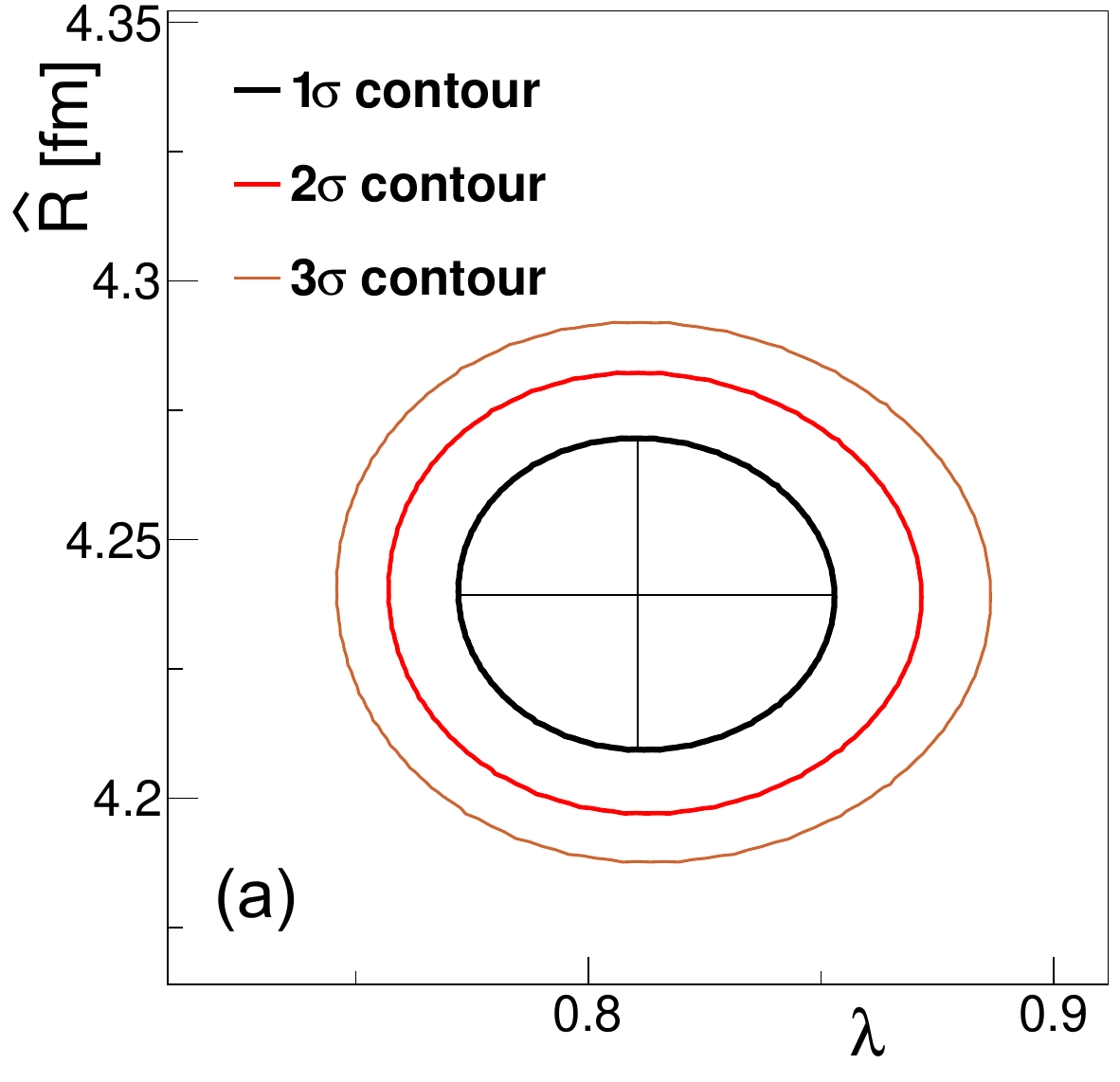}
\includegraphics[width=0.32\linewidth]{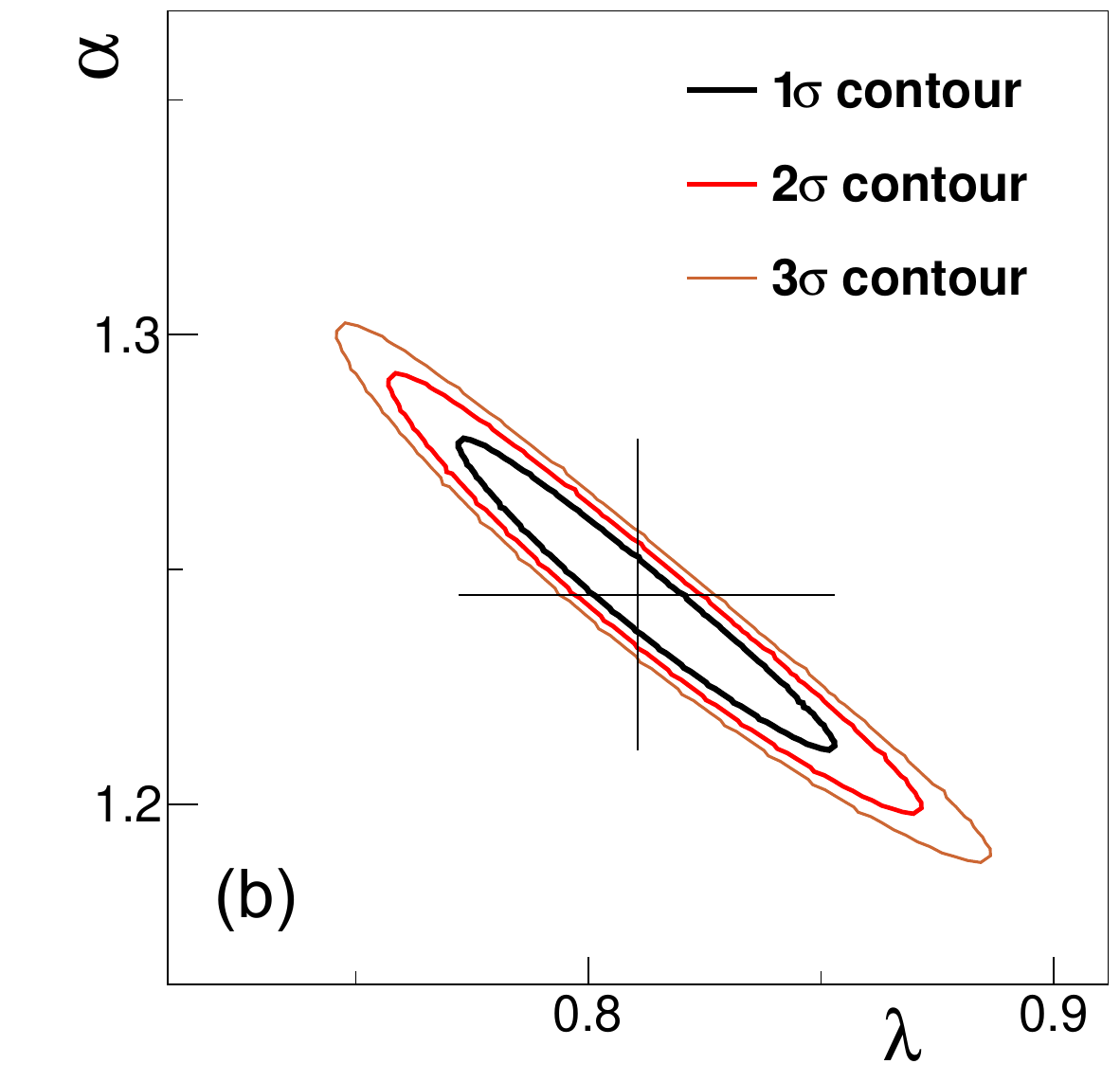}
\includegraphics[width=0.32\linewidth]{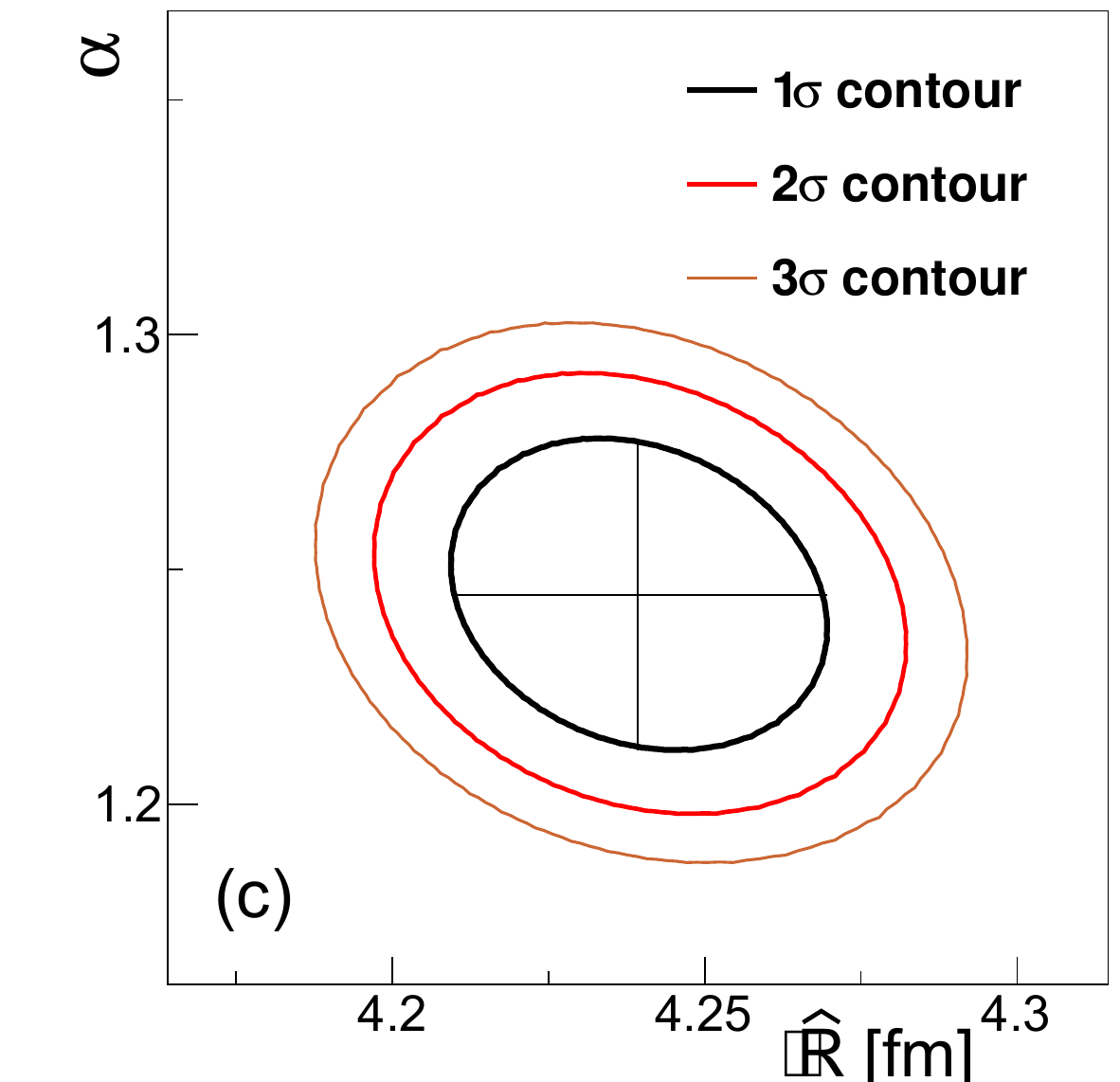}
\caption{Contour lines of the $\chi^2$ map in the (a) $\lambda,\hat{R}$ 
and (b) $\lambda,\alpha$ and (c) $\hat{R},\alpha$ planes for fits to 
$\pi^-\pi^-$ correlation functions of pairs with $\mT$ between 0.331 
and 0.349 GeV$/c^2$. The horizontal and vertical lines represent the 
MINOS fit uncertainties.}
\label{f:contours_rhatfit}
\end{figure*}

\begin{figure}
\includegraphics[width=1.0\linewidth]{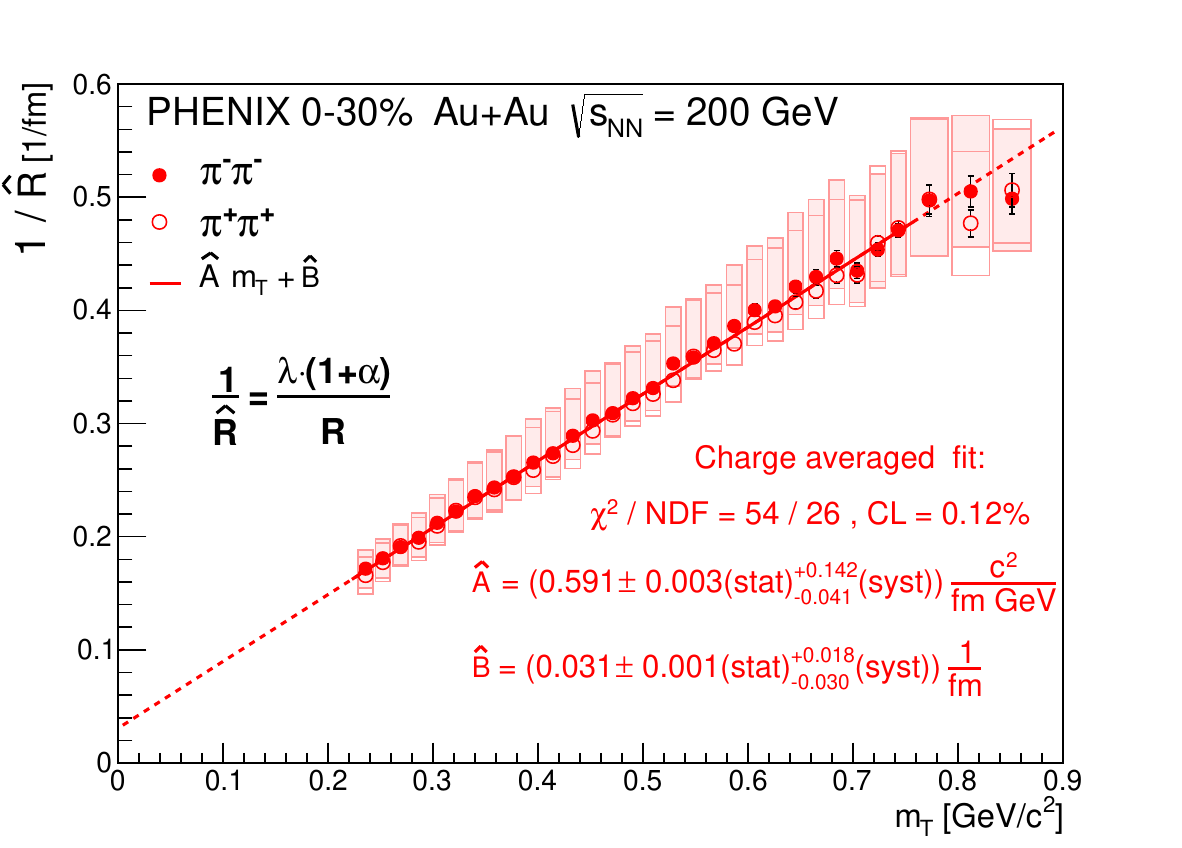}
\caption{New scale parameter $\hat{R}$ versus average $\mT$ of the 
pair, with a linear fit. Statistical and systematic uncertainties shown 
as bars and boxes, respectively.}
\label{f:Rhatmt}
\end{figure}

If this parameter is used as a fit parameter instead of the L\'evy 
scale parameter $R$ (which is calculated as 
$R=\hat{R}\lambda(1+\alpha)$), the obtained $\lambda$, $R$ and $\alpha$ 
parameters are the same as before, but the correlation coefficients for 
$(\lambda,\hat{R})$ and $(\hat{R},\alpha)$ are reduced substantially, 
to the region of 20\%--30\%, which indicates small correlation as 
compared to the $\approx$95\% values of the correlation coefficients 
between $(\lambda,R)$ and $(R,\alpha)$ (and all of them are negative in 
this case). The error contours obtained on the two-dimensional $\chi^2$ 
maps in the $(\lambda,\hat R)$, $(\lambda,\alpha)$ and $(\hat 
R,\alpha)$ planes for one example fit are shown in 
Fig.~\ref{f:contours_rhatfit}. Also note that due to the reduction of 
the correlation, the uncertainty of $\hat{R}$ is also significantly 
reduced compared to that of $R$, as indicated in Fig.~\ref{f:Rhatmt} 
and Table~\ref{t:rhat}.

%--------------------------------------------- Fig_9

It is interesting to observe that $1/\hat{R}$ scales linearly with 
$\mT$ , as shown in Fig.~\ref{f:Rhatmt}. The parameters of the linear 
$1/\hat{R}(\mT)=\hat{A}\mT+\hat{B}$ fit to the charge averaged 
$1/\hat{R}$ data are

\begin{align}\label{e:Rhatlinear}
\hat{A} &= (0.591 \pm 0.003\textnormal{ (stat)}^{+0.142}_{-0.041}\textnormal{ (syst)) } \frac{c^2}{\m{GeV fm}},\\
\hat{B} &= (0.031 \pm 0.001\textnormal{ (stat)}^{+0.018}_{-0.030}\textnormal{ (syst)) } \frac{1}{\m{fm}},
\end{align}
Statistical and systematic uncertainties were determined similarly to 
the fits to $1/R^2$ versus $\mT$ and $\lambda/\lambda_\m{\rm max}$ versus 
$\mT$.

The physical cause and possible interpretation of this remarkable 
affine linear dependence of $1/\hat R$ (not its square, as in the case 
of the scale parameter $R$) on $\mT$ is entirely unknown to us.

%--------------------------------------------- Fig_10

One still may try to explain the newly observed $\mT$ scaling of $\hat 
R$ by a simple $\mT$ scaling law for $\lambda$, based on the 
observation that both $1/R^2$ and $1/\hat R$ scale linearly with $\mT$, 
while $\alpha$ is approximately constant. It is important to note 
however that both of these scalings are affine linear, thus the ratio 
of the two is not constant. In particular, the linear parameters of 
\Eq{e:Rhatlinear} can be converted to a simple dependence of

\begin{align}
\hat R(\mT) = \frac{\hat R_\xi}{\mT/m_\pi+\hat \xi},
\end{align}
where then one gets $\hat R_\xi=(12.21\pm 0.06)$ fm and $\hat 
\xi=0.38\pm0.01$. This, together with the definition of $\hat R$ and 
\Eq{e:Rlinear_physical}, yields

\begin{align}
\lambda(\mT) = \frac{1}{1+\alpha} \frac{R_\xi}{\hat R_\xi} \frac{\mT/m_\pi+\hat \xi}{\sqrt{\mT/m_\pi+\xi}}
\end{align}
This (together with the assumption of $\alpha$ being constant in $\mT$) 
would imply that at large transverse masses $\lambda\approx\sqrt{\mT}$, 
however such a scaling is not meaningful, because $\lambda$, representing 
the fraction of pions contributing to Bose-Einstein correlations, 
typically cannot increase ad infinitum. In fact our data indicate a 
saturation of $\lambda(\mT)$ at large values of $\mT$.

%======================================================== Table_IV
\begin{table}
\caption{\label{t:rhat}
Value of $\hat{R}$ as a function of bin $\mT$, for $\pi^+\pi^+$ and 
$\pi^-\pi^-$ pairs, in fits where it replaced $R$ as a fit parameter. The 
other parameters of these fits ($\alpha$,$\lambda$) are the same as given in 
Table~\ref{t:pars}, and if one calculates $R$ from $\hat{R}$, one also 
obtains the same $R$ value. Also note that in this case, all statistical 
uncertainties turned out to be symmetric, so we denoted both of them by a 
single uncertainty, followed by systematic uncertainties.}
\begin{ruledtabular} \begin{tabular}{cccc}
$\mT$ (GeV$/c^2$) & $\hat{R}(\pi^-)$ (fm)  & $\hat{R}(\pi^+)$ (fm) & \\
\hline
0.236 & $5.94 \pm 0.06_{-0.60}^{+0.57} $ & $6.02 \pm 0.06_{-0.60}^{+0.58} $ \\ 
0.252 & $5.54 \pm 0.04_{-0.53}^{+0.53} $ & $5.74 \pm 0.05_{-0.55}^{+0.55} $ \\ 
0.269 & $5.12 \pm 0.04_{-0.46}^{+0.53} $ & $5.30 \pm 0.04_{-0.47}^{+0.54} $ \\ 
0.286 & $4.95 \pm 0.03_{-0.42}^{+0.55} $ & $5.04 \pm 0.03_{-0.43}^{+0.56} $ \\ 
0.304 & $4.71 \pm 0.03_{-0.38}^{+0.56} $ & $4.84 \pm 0.03_{-0.39}^{+0.57} $ \\ 
0.322 & $4.50 \pm 0.03_{-0.37}^{+0.56} $ & $4.55 \pm 0.03_{-0.37}^{+0.57} $ \\ 
0.340 & $4.24 \pm 0.03_{-0.35}^{+0.55} $ & $4.26 \pm 0.03_{-0.35}^{+0.55} $ \\ 
0.358 & $4.11 \pm 0.03_{-0.34}^{+0.56} $ & $4.13 \pm 0.03_{-0.34}^{+0.56} $ \\ 
0.377 & $3.90 \pm 0.03_{-0.32}^{+0.55} $ & $3.92 \pm 0.03_{-0.32}^{+0.56} $ \\ 
0.395 & $3.76 \pm 0.03_{-0.30}^{+0.55} $ & $3.86 \pm 0.03_{-0.31}^{+0.56} $ \\ 
0.414 & $3.67 \pm 0.03_{-0.28}^{+0.53} $ & $3.68 \pm 0.02_{-0.28}^{+0.54} $ \\ 
0.433 & $3.46 \pm 0.03_{-0.25}^{+0.50} $ & $3.56 \pm 0.03_{-0.26}^{+0.51} $ \\ 
0.452 & $3.31 \pm 0.03_{-0.23}^{+0.48} $ & $3.41 \pm 0.02_{-0.23}^{+0.49} $ \\ 
0.471 & $3.23 \pm 0.03_{-0.21}^{+0.46} $ & $3.25 \pm 0.02_{-0.21}^{+0.47} $ \\ 
0.490 & $3.10 \pm 0.03_{-0.19}^{+0.44} $ & $3.15 \pm 0.03_{-0.20}^{+0.45} $ \\ 
0.509 & $3.01 \pm 0.03_{-0.18}^{+0.43} $ & $3.07 \pm 0.03_{-0.18}^{+0.44} $ \\ 
0.529 & $2.83 \pm 0.03_{-0.16}^{+0.40} $ & $2.96 \pm 0.03_{-0.17}^{+0.42} $ \\ 
0.548 & $2.79 \pm 0.03_{-0.15}^{+0.39} $ & $2.78 \pm 0.03_{-0.15}^{+0.39} $ \\ 
0.567 & $2.69 \pm 0.03_{-0.13}^{+0.37} $ & $2.73 \pm 0.03_{-0.14}^{+0.38} $ \\ 
0.587 & $2.59 \pm 0.03_{-0.13}^{+0.36} $ & $2.70 \pm 0.03_{-0.14}^{+0.38} $ \\ 
0.606 & $2.50 \pm 0.03_{-0.13}^{+0.35} $ & $2.56 \pm 0.03_{-0.14}^{+0.36} $ \\ 
0.626 & $2.47 \pm 0.03_{-0.14}^{+0.37} $ & $2.53 \pm 0.03_{-0.14}^{+0.38} $ \\ 
0.645 & $2.38 \pm 0.03_{-0.14}^{+0.37} $ & $2.46 \pm 0.03_{-0.14}^{+0.38} $ \\ 
0.665 & $2.34 \pm 0.04_{-0.14}^{+0.37} $ & $2.40 \pm 0.04_{-0.14}^{+0.38} $ \\ 
0.684 & $2.25 \pm 0.04_{-0.13}^{+0.35} $ & $2.32 \pm 0.04_{-0.14}^{+0.36} $ \\ 
0.704 & $2.30 \pm 0.04_{-0.15}^{+0.35} $ & $2.33 \pm 0.04_{-0.15}^{+0.36} $ \\ 
0.724 & $2.20 \pm 0.03_{-0.16}^{+0.33} $ & $2.17 \pm 0.03_{-0.16}^{+0.32} $ \\ 
0.743 & $2.12 \pm 0.03_{-0.18}^{+0.31} $ & $2.11 \pm 0.03_{-0.18}^{+0.30} $ \\ 
0.773 & $2.01 \pm 0.06_{-0.20}^{+0.29} $ & $2.00 \pm 0.05_{-0.20}^{+0.29} $ \\ 
0.812 & $1.98 \pm 0.05_{-0.19}^{+0.26} $ & $2.09 \pm 0.05_{-0.20}^{+0.28} $ \\ 
0.852 & $2.01 \pm 0.05_{-0.19}^{+0.25} $ & $1.97 \pm 0.06_{-0.18}^{+0.24} $ \\ 
\end{tabular}  \end{ruledtabular} 
\end{table}

As discussed in Section~\ref{ss:lambdamtintro} and seen in 
Section~\ref{ss:mtdepresults}, the strength of the correlation 
functions is not equal to unity, and not even constant as a function of 
$\mT$, the reason for which may be the fact that a large fraction of 
low $\mT$ pions are produced from decays of long-lived resonances 
($\eta$, $\eta'$, $\omega$, $K^0_S$ mesons, etc). The detailed shape of 
$\lambda(\mT)$ may be compared to predictions based on various 
resonance cocktails, including models that incorporate modified 
in-medium resonance masses or calculations based on partially coherent 
pion production.

Earlier measurements or simulations were frequently done within the 
Gaussian approximation, usually yielding smaller $\lambda$ values 
compared to a L\'evy analysis. This can be explained by the 
anticorrelation between $\lambda$ and $\alpha$. If the correlation 
function has a nonzero slope at $Q=0$, then a Gaussian fit with zero 
slope at $Q=0$ artificially forces $\lambda$ to a lower value -- such 
fits do not capture a key feature of the data.

\begin{figure}
\includegraphics[width=1.0\linewidth]{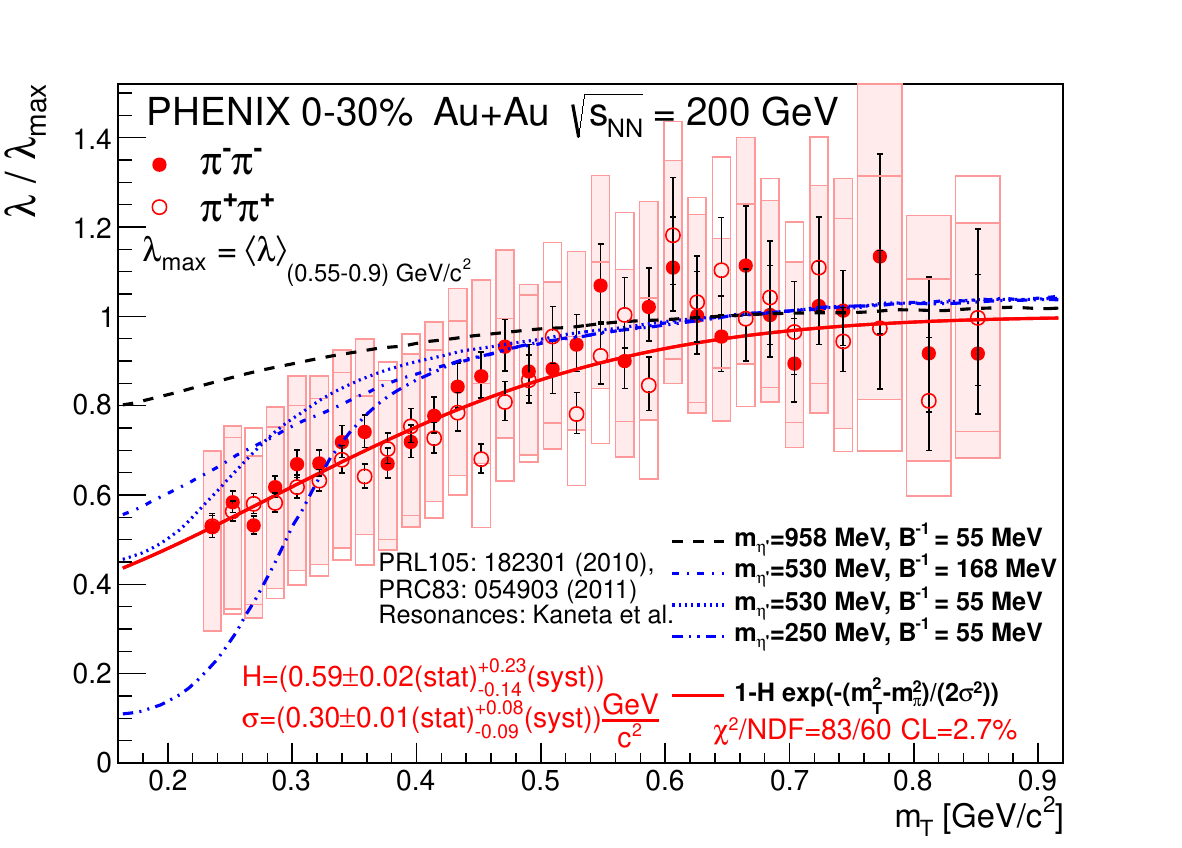}
\caption{Normalized correlation strength parameter 
$\lambda/\lambda_{\rm max}$ versus average $\mT$ of the pair. The data 
are compared with parameter scans from 
Refs.~\cite{Vertesi:2009wf,Csorgo:2009pa} using different values of 
in-medium $\eta'$ mass $m^{*}_{\eta'}$ and slope parameter 
$B^{-1}_{\eta'}$. A best fit with Eq.~\eqref{e:hole} and the resulting 
$H$ and $\sigma$ parameters are also shown.}
\label{f:lambdacontextmt}
\end{figure}

As seen in Fig.~\ref{f:lambdamt} $\lambda$ appears to increase with 
$\mT$ until it saturates around $\mT=0.6$ GeV$/c^2$. To further study 
the dependence of $\lambda$ on $\mT$ it is advantageous to use the 
ratio $\lambda/\lambda_{\rm max}$ where $\lambda_{\rm max}$ is the 
saturated value of $\lambda$, which we determine in the region 
$\mT>0.55$ GeV$/c^2$. This is advantageous for two reasons: (i) the 
systematic uncertainties largely cancel in the ratio, and (ii) the 
ratio is less sensitive to the assumed shape of Bose-Einstein 
correlation functions~\cite{Csanad:2005nr}. 
Figure~\ref{f:lambdacontextmt} shows the resulting $\lambda/\lambda_{\rm 
max}$ dependence on $\mT$.

To quantify this dependence the distribution is fit with the function

\begin{align}\label{e:hole}
\lambda(\mT)/\lambda_{\rm max} = 1-H\exp(-(\mT^2-m_\pi^2)/(2\sigma^2))
\end{align}
The parameters have a simple meaning.  Parameter $H$ measures the depth 
(intercept at $\mT=m_\pi$ i.e. $\kT=0$), while parameter $\sigma$ 
measures the width of the low-$\mT$ region of decrease. The following 
values of the parameters $(H,\sigma)$ were determined:

\begin{align}
H      &=  0.59 \pm 0.02\textnormal{ (stat)}^{+0.23}_{-0.14}\textnormal{ (syst)},\\
\sigma &= (0.30 \pm 0.01\textnormal{ (stat)}^{+0.08}_{-0.09}\textnormal{ (syst)) GeV}/c^2.
\end{align}
Only the statistical uncertainties of the $\lambda/\lambda_\m{\rm max}$ 
points were taken into account in the fit. Here the statistical 
uncertainty of $\lambda_\m{\rm max}$ is treated as a normalization 
uncertainty. This uncertainty and the systematic uncertainty caused by 
the choice of $\mT$ range when calculating $\lambda_\m{\rm max}$ (both 
$\approx$1\%) are negligible compared to other uncertainties. The 
systematic uncertainties of the fit parameters were determined by 
fitting $\lambda/\lambda_\m{\rm max}$ versus $\mT$ obtained from 
measurements and fits with varied settings (listed e.g. in 
Table~\ref{t:systtable}). It is important to note that the $(H,\sigma)$ 
values are significantly different from zero, so the existence of the 
decrease in the $\lambda(\mT)$ data is statistically significant.

Partial coherence effects may suppress the strength of the two-pion 
Bose-Einstein correlation functions. However, in the model of 
Ref.~\cite{Sinyukov:1994en} $\lambda$ is not expected to depend on 
$\mT$. An $\mT$ dependence given by Eq.~\eqref{e:hole} was derived in a 
pion-laser model~\cite{Pratt:1993uy,Csorgo:1997us}. However this model 
gives an upper limit of $H \le 0.06$ given our measured values of $R$ 
and $\sigma$. Measurements of higher order Bose-Einstein correlation 
functions could shed more light on the contributions of partial 
coherence.

It has been suggested~\cite{Vance:1998wd} that $U_A(1)$ symmetry 
restoration and its related in-medium mass reduction of the $\eta'$ 
meson in hot, dense hadronic matter would cause a reduction in the 
value of $\lambda$ at low $\mT$. In Fig.~\ref{f:lambdacontextmt}, our 
data are compared with parameter scans from 
Refs.~\cite{Vertesi:2009wf,Csorgo:2009pa} with the Kaneta-Xu model 
ratios of long-lived resonances~\cite{Kaneta:2004zr}, using different 
values for the in-medium $\eta'$ mass $m^{*}_{\eta'}$ and the $\eta'$ 
condensate temperature (slope parameter) $B^{-1}_{\eta'}$. Our data are 
seen to be suppressed compared to the prediction with no in-medium 
$\eta'$ mass modification, $m^{*}_{\eta'}=m_{\eta'}=958$ MeV. Within 
systematics, our data are not inconsistent with selected parameter scan 
results of Refs.~\cite{Vertesi:2009wf,Csorgo:2009pa} using a modified 
in-medium $\eta'$ mass. These data thus provide strong new constraints 
for more detailed theoretical studies on $U_A(1)$ symmetry restoration 
in hot and dense hadronic matter.

%--------------------------------------------- Fig_11

\section{Summary and conclusion\label{s:summary}}

In this paper we presented the measurement and analysis of two-pion 
Bose-Einstein correlations and their L\'evy parameters, measured in 
0\%--30\% centrality \auau collisions at \sqsntwo colliding energies in 
the PHENIX experiment at the RHIC accelerator. After selecting the 2.2 
billion 0\%--30\% centrality events from the 2010 data taking period, and 
after applying carefully chosen single track and two-track selection 
cuts, we performed a study of the proper variable and the shape of the 
two-pion Bose-Einstein correlation function and investigated their 
transverse mass dependence in 31 $\mT$ bins from 228 to 871 MeV$/c^2$.

We found that these data cannot be well represented by the usual 
Gaussian Bose-Einstein correlation functions. However, when Gaussian 
source distributions were generalized to L\'evy-stable source 
distributions, and the final state Coulomb interaction between 
like-sign pions emitted from L\'evy-stable source distributions was 
properly taken into account, the data could be described at a 
statistically acceptable level. We determined the $\mT$ dependence of 
the parameters of L\'evy-stable source distributions.

The L\'evy exponent $\alpha$ was found to be inconsistent not only with 
the Gaussian case of $\alpha=2$ and the exponential case of $\alpha=1$, 
but also with $\alpha\le0.5$, the conjectured value at the QCD critical 
point. We have found, that $\alpha$ is weakly dependent on the 
transverse momentum of the pair in 0\%--30\% centrality \auau collisions, 
in qualitative agreement with simulations based on anomalous diffusion 
in an expanding medium. However, a fit with a constant value of 
$\alpha$ to the $\alpha(m_T)$ data resulted in a statistically 
unacceptable confidence level.

Even though these $\alpha<2$ values may indicate a nonhydrodynamical 
component in the pion production processes in \sqsntwo \auau 
collisions, the bulk of pion production still seems to be of 
hydrodynamical origin.   A hydrodynamical type of $1/R^2 = A + B \mT$ 
scaling behavior is found to represent the measured data remarkably 
well, especially in the low $\mT$ region. However, we are not 
aware of theoretical predictions of $R(\mT)$ for L\'evy-stable source 
distributions with $\alpha<2$.

We found a statistically significant decrease of the intercept 
parameter $\lambda$ at low values of the transverse mass. Our new 
measurements are not consistent with predictions without in-medium 
$\eta'$ mass modification. Clearly additional measurements are needed 
in the soft ($\pT < 500$ MeV) region, including other decay channels of 
the $\eta'$ meson in order to clarify the role of $\eta'$ mass 
modification.

Surprisingly, we also found an unpredicted, empirical new scaling 
variable $\hat{R} = R/(\lambda(1+\alpha))$ that follows an $1/ \hat{R} 
\propto \mT$ affine linear scaling, which is stable against small 
variations of the exact value of the L\'evy exponent $\alpha$. The 
origin of this new empirical scaling law is unknown to us.

The methods described in this manuscript demonstrate that it is 
possible to measure the L\'evy exponent of the correlation function in 
high energy heavy ion reactions. Given that the value of the 
correlation exponent is expected to reach a specific value in second 
order phase transitions that is characteristic to the universality 
class of the given critical point, let us close this paper by proposing 
similar measurements at various collision energies, centralities, 
colliding system sizes and identified particle pair types, as well as 
analyses with two- or three-dimensional momentum difference variables, 
to improve our detailed understanding of the nature of the particle 
production in high energy heavy ion reactions, and to search for the 
vicinity of the critical end point of QCD, where the line of first 
order quark-hadron transitions in the $(\mu,T)$ plane ends, 
corresponding to a second order phase transition. Finally we emphasize 
the need for more detailed measurements, including measuring the 
centrality and collision energy, system size and particle type 
dependence of the L\'evy fit parameters $\lambda$, $\alpha$ and $R$.

\let\c=\oldc

\section*{ACKNOWLEDGMENTS}

We thank the staff of the Collider-Accelerator and Physics
Departments at Brookhaven National Laboratory and the staff of
the other PHENIX participating institutions for their vital
contributions.   We also thank S. Hegyi for significant
theoretical contributions and discussions.
We acknowledge support from the Office of Nuclear Physics in the
Office of Science of the Department of Energy,
the National Science Foundation,
Abilene Christian University Research Council,
Research Foundation of SUNY, and
Dean of the College of Arts and Sciences, Vanderbilt University
(U.S.A),
Ministry of Education, Culture, Sports, Science, and Technology
and the Japan Society for the Promotion of Science (Japan),
Conselho Nacional de Desenvolvimento Cient\'{\i}fico e
Tecnol{\'o}gico and Funda\c c{\~a}o de Amparo {\`a} Pesquisa do
Estado de S{\~a}o Paulo (Brazil),
Natural Science Foundation of China (People's Republic of China),
Croatian Science Foundation and
Ministry of Science and Education (Croatia),
Ministry of Education, Youth and Sports (Czech Republic),
Centre National de la Recherche Scientifique, Commissariat
{\`a} l'{\'E}nergie Atomique, and Institut National de Physique
Nucl{\'e}aire et de Physique des Particules (France),
Bundesministerium f\"ur Bildung und Forschung, Deutscher
Akademischer Austausch Dienst, and Alexander von Humboldt Stiftung (Germany),
J. Bolyai Research Scholarship, EFOP, the New National Excellence 
Program ({\'U}NKP), NKFIH, and OTKA (Hungary),
Department of Atomic Energy and Department of Science and Technology (India),
Israel Science Foundation (Israel),
Basic Science Research Program through NRF of the Ministry of Education (Korea),
Physics Department, Lahore University of Management Sciences (Pakistan),
Ministry of Education and Science, Russian Academy of Sciences,
Federal Agency of Atomic Energy (Russia),
VR and Wallenberg Foundation (Sweden),
the U.S. Civilian Research and Development Foundation for the
Independent States of the Former Soviet Union,
the Hungarian American Enterprise Scholarship Fund,
the US-Hungarian Fulbright Foundation, 
and the US-Israel Binational Science Foundation.

%\bibliography{ppg194x2}   

\begin{thebibliography}{89}%
\makeatletter
\providecommand \@ifxundefined [1]{%
 \@ifx{#1\undefined}
}%
\providecommand \@ifnum [1]{%
 \ifnum #1\expandafter \@firstoftwo
 \else \expandafter \@secondoftwo
 \fi
}%
\providecommand \@ifx [1]{%
 \ifx #1\expandafter \@firstoftwo
 \else \expandafter \@secondoftwo
 \fi
}%
\providecommand \natexlab [1]{#1}%
\providecommand \enquote  [1]{``#1''}%
\providecommand \bibnamefont  [1]{#1}%
\providecommand \bibfnamefont [1]{#1}%
\providecommand \citenamefont [1]{#1}%
\providecommand \href@noop [0]{\@secondoftwo}%
\providecommand \href [0]{\begingroup \@sanitize@url \@href}%
\providecommand \@href[1]{\@@startlink{#1}\@@href}%
\providecommand \@@href[1]{\endgroup#1\@@endlink}%
\providecommand \@sanitize@url [0]{\catcode `\\12\catcode `\$12\catcode
  `\&12\catcode `\#12\catcode `\^12\catcode `\_12\catcode `\%12\relax}%
\providecommand \@@startlink[1]{}%
\providecommand \@@endlink[0]{}%
\providecommand \url  [0]{\begingroup\@sanitize@url \@url }%
\providecommand \@url [1]{\endgroup\@href {#1}{\urlprefix }}%
\providecommand \urlprefix  [0]{URL }%
\providecommand \Eprint [0]{\href }%
\providecommand \doibase [0]{http://dx.doi.org/}%
\providecommand \selectlanguage [0]{\@gobble}%
\providecommand \bibinfo  [0]{\@secondoftwo}%
\providecommand \bibfield  [0]{\@secondoftwo}%
\providecommand \translation [1]{[#1]}%
\providecommand \BibitemOpen [0]{}%
\providecommand \bibitemStop [0]{}%
\providecommand \bibitemNoStop [0]{.\EOS\space}%
\providecommand \EOS [0]{\spacefactor3000\relax}%
\providecommand \BibitemShut  [1]{\csname bibitem#1\endcsname}%
\let\auto@bib@innerbib\@empty
%</preamble>
\bibitem [{\citenamefont {Lednicky}(2001)}]{Lednicky:2001qv}%
  \BibitemOpen
  \bibfield  {author} {\bibinfo {author} {\bibfnamefont {R.}~\bibnamefont
  {Lednicky}},\ }\bibfield  {title} {\enquote {\bibinfo {title} {{Femtoscopy
  with unlike particles}},}\ }in\ \href@noop {} {\emph {\bibinfo {booktitle}
  {{International Workshop on the Physics of the Quark Gluon Plasma Palaiseau,
  France, September 4-7, 2001}}}}\ (\bibinfo {year} {2001})\BibitemShut
  {NoStop}%
\bibitem [{\citenamefont {Hanbury~Brown}\ and\ \citenamefont
  {Twiss}(1956)}]{HanburyBrown:1956bqd}%
  \BibitemOpen
  \bibfield  {author} {\bibinfo {author} {\bibfnamefont {R.}~\bibnamefont
  {Hanbury~Brown}}\ and\ \bibinfo {author} {\bibfnamefont {R.~Q.}\ \bibnamefont
  {Twiss}},\ }\bibfield  {title} {\enquote {\bibinfo {title} {{A Test of a new
  type of stellar interferometer on Sirius}},}\ }\href {\doibase
  10.1038/1781046a0} {\bibfield  {journal} {\bibinfo  {journal} {Nature}\
  }\textbf {\bibinfo {volume} {178}},\ \bibinfo {pages} {1046} (\bibinfo {year}
  {1956})}\BibitemShut {NoStop}%
%%CITATION = NATUA,178,1046;%%
\bibitem [{\citenamefont {Glauber}(1963)}]{Glauber:1962tt}%
  \BibitemOpen
  \bibfield  {author} {\bibinfo {author} {\bibfnamefont {R.~J.}\ \bibnamefont
  {Glauber}},\ }\bibfield  {title} {\enquote {\bibinfo {title} {{Photon
  correlations}},}\ }\href {\doibase 10.1103/PhysRevLett.10.84} {\bibfield
  {journal} {\bibinfo  {journal} {Phys. Rev. Lett.}\ }\textbf {\bibinfo
  {volume} {10}},\ \bibinfo {pages} {84} (\bibinfo {year} {1963})}\BibitemShut
  {NoStop}%
%%CITATION = PRLTA,10,84;%%
\bibitem [{\citenamefont {Glauber}(2006{\natexlab{a}})}]{Glauber:2006zz}%
  \BibitemOpen
  \bibfield  {author} {\bibinfo {author} {\bibfnamefont {R.~J.}\ \bibnamefont
  {Glauber}},\ }\bibfield  {title} {\enquote {\bibinfo {title} {{Nobel Lecture:
  One hundred years of light quanta}},}\ }\href {\doibase
  10.1103/RevModPhys.78.1267} {\bibfield  {journal} {\bibinfo  {journal} {Rev.
  Mod. Phys.}\ }\textbf {\bibinfo {volume} {78}},\ \bibinfo {pages} {1267}
  (\bibinfo {year} {2006}{\natexlab{a}})}\BibitemShut {NoStop}%
%%CITATION = RMPHA,78,1267;%%
\bibitem [{\citenamefont {Glauber}(2006{\natexlab{b}})}]{Glauber:2006gd}%
  \BibitemOpen
  \bibfield  {author} {\bibinfo {author} {\bibfnamefont {R.~J.}\ \bibnamefont
  {Glauber}},\ }\bibfield  {title} {\enquote {\bibinfo {title} {{Quantum Optics
  and Heavy Ion Physics}},}\ }\bibfield  {booktitle} {\emph {\bibinfo
  {booktitle} {{Proceedings, 18th International Conference on
  Ultra-Relativistic Nucleus-Nucleus Collisions (Quark Matter 2005): Budapest,
  Hungary, August 4-9, 2005}}},\ }\href {\doibase
  10.1016/j.nuclphysa.2006.06.009} {\bibfield  {journal} {\bibinfo  {journal}
  {Nucl. Phys. A}\ }\textbf {\bibinfo {volume} {774}},\ \bibinfo {pages} {3}
  (\bibinfo {year} {2006}{\natexlab{b}})}\BibitemShut {NoStop}%
\bibitem [{\citenamefont {Goldhaber}\ \emph {et~al.}(1959)\citenamefont
  {Goldhaber}, \citenamefont {Fowler}, \citenamefont {Goldhaber},\ and\
  \citenamefont {Hoang}}]{Goldhaber:1959mj}%
  \BibitemOpen
  \bibfield  {author} {\bibinfo {author} {\bibfnamefont {G.}~\bibnamefont
  {Goldhaber}}, \bibinfo {author} {\bibfnamefont {W.~B.}\ \bibnamefont
  {Fowler}}, \bibinfo {author} {\bibfnamefont {S.}~\bibnamefont {Goldhaber}}, \
  and\ \bibinfo {author} {\bibfnamefont {T.~F.}\ \bibnamefont {Hoang}},\
  }\bibfield  {title} {\enquote {\bibinfo {title} {{Pion-pion correlations in
  antiproton annihilation events}},}\ }\href {\doibase
  10.1103/PhysRevLett.3.181} {\bibfield  {journal} {\bibinfo  {journal} {Phys.
  Rev. Lett.}\ }\textbf {\bibinfo {volume} {3}},\ \bibinfo {pages} {181}
  (\bibinfo {year} {1959})}\BibitemShut {NoStop}%
%%CITATION = PRLTA,3,181;%%
\bibitem [{\citenamefont {Goldhaber}\ \emph {et~al.}(1960)\citenamefont
  {Goldhaber}, \citenamefont {Goldhaber}, \citenamefont {Lee},\ and\
  \citenamefont {Pais}}]{Goldhaber:1960sf}%
  \BibitemOpen
  \bibfield  {author} {\bibinfo {author} {\bibfnamefont {G.}~\bibnamefont
  {Goldhaber}}, \bibinfo {author} {\bibfnamefont {S.}~\bibnamefont
  {Goldhaber}}, \bibinfo {author} {\bibfnamefont {W.-Y.}\ \bibnamefont {Lee}},
  \ and\ \bibinfo {author} {\bibfnamefont {A.}~\bibnamefont {Pais}},\
  }\bibfield  {title} {\enquote {\bibinfo {title} {{Influence of Bose-Einstein
  statistics on the anti-proton proton annihilation process}},}\ }\href
  {\doibase 10.1103/PhysRev.120.300} {\bibfield  {journal} {\bibinfo  {journal}
  {Phys. Rev.}\ }\textbf {\bibinfo {volume} {120}},\ \bibinfo {pages} {300}
  (\bibinfo {year} {1960})}\BibitemShut {NoStop}%
%%CITATION = PHRVA,120,300;%%
\bibitem [{\citenamefont {Adcox}\ \emph {et~al.}(2005)\citenamefont {Adcox}
  \emph {et~al.}}]{Adcox:2004mh}%
  \BibitemOpen
  \bibfield  {author} {\bibinfo {author} {\bibfnamefont {K.}~\bibnamefont
  {Adcox}} \emph {et~al.} (\bibinfo {collaboration} {PHENIX Collaboration}),\
  }\bibfield  {title} {\enquote {\bibinfo {title} {{Formation of dense partonic
  matter in relativistic nucleus-nucleus collisions at RHIC: Experimental
  evaluation by the PHENIX collaboration}},}\ }\href {\doibase
  10.1016/j.nuclphysa.2005.03.086} {\bibfield  {journal} {\bibinfo  {journal}
  {Nucl. Phys. A}\ }\textbf {\bibinfo {volume} {757}},\ \bibinfo {pages} {184}
  (\bibinfo {year} {2005})}\BibitemShut {NoStop}%
\bibitem [{\citenamefont {Adams}\ \emph
  {et~al.}(2005{\natexlab{a}})\citenamefont {Adams} \emph
  {et~al.}}]{Adams:2005dq}%
  \BibitemOpen
  \bibfield  {author} {\bibinfo {author} {\bibfnamefont {J.}~\bibnamefont
  {Adams}} \emph {et~al.} (\bibinfo {collaboration} {STAR Collaboration}),\
  }\bibfield  {title} {\enquote {\bibinfo {title} {{Experimental and
  theoretical challenges in the search for the quark gluon plasma: The STAR
  Collaboration's critical assessment of the evidence from RHIC collisions}},}\
  }\href {\doibase 10.1016/j.nuclphysa.2005.03.085} {\bibfield  {journal}
  {\bibinfo  {journal} {Nucl. Phys. A}\ }\textbf {\bibinfo {volume} {757}},\
  \bibinfo {pages} {102} (\bibinfo {year} {2005}{\natexlab{a}})}\BibitemShut
  {NoStop}%
\bibitem [{\citenamefont {Arsene}\ \emph {et~al.}(2005)\citenamefont {Arsene}
  \emph {et~al.}}]{Arsene:2004fa}%
  \BibitemOpen
  \bibfield  {author} {\bibinfo {author} {\bibfnamefont {I.}~\bibnamefont
  {Arsene}} \emph {et~al.} (\bibinfo {collaboration} {BRAHMS Collaboration}),\
  }\bibfield  {title} {\enquote {\bibinfo {title} {{Quark gluon plasma and
  color glass condensate at RHIC? The Perspective from the BRAHMS
  experiment}},}\ }\href {\doibase 10.1016/j.nuclphysa.2005.02.130} {\bibfield
  {journal} {\bibinfo  {journal} {Nucl. Phys. A}\ }\textbf {\bibinfo {volume}
  {757}},\ \bibinfo {pages} {1} (\bibinfo {year} {2005})}\BibitemShut {NoStop}%
\bibitem [{\citenamefont {Back}\ \emph {et~al.}(2005)\citenamefont {Back} \emph
  {et~al.}}]{Back:2004je}%
  \BibitemOpen
  \bibfield  {author} {\bibinfo {author} {\bibfnamefont {B.~B.}\ \bibnamefont
  {Back}} \emph {et~al.},\ }\bibfield  {title} {\enquote {\bibinfo {title}
  {{The PHOBOS perspective on discoveries at RHIC}},}\ }\href {\doibase
  10.1016/j.nuclphysa.2005.03.084} {\bibfield  {journal} {\bibinfo  {journal}
  {Nucl. Phys. A}\ }\textbf {\bibinfo {volume} {757}},\ \bibinfo {pages} {28}
  (\bibinfo {year} {2005})}\BibitemShut {NoStop}%
\bibitem [{\citenamefont {Adler}\ \emph {et~al.}(2004)\citenamefont {Adler}
  \emph {et~al.}}]{Adler:2004rq}%
  \BibitemOpen
  \bibfield  {author} {\bibinfo {author} {\bibfnamefont {S.~S.}\ \bibnamefont
  {Adler}} \emph {et~al.} (\bibinfo {collaboration} {PHENIX Collaboration}),\
  }\bibfield  {title} {\enquote {\bibinfo {title} {{Bose-Einstein correlations
  of charged pion pairs in Au + Au collisions at $\sqrt{s_{NN}}=$ 200 GeV}},}\
  }\href@noop {} {\bibfield  {journal} {\bibinfo  {journal} {Phys. Rev. Lett.}\
  }\textbf {\bibinfo {volume} {93}},\ \bibinfo {pages} {152302} (\bibinfo
  {year} {2004})}\BibitemShut {NoStop}%
\bibitem [{\citenamefont {Afanasiev}\ \emph {et~al.}(2009)\citenamefont
  {Afanasiev} \emph {et~al.}}]{Afanasiev:2009ii}%
  \BibitemOpen
  \bibfield  {author} {\bibinfo {author} {\bibfnamefont {S.}~\bibnamefont
  {Afanasiev}} \emph {et~al.} (\bibinfo {collaboration} {PHENIX
  Collaboration}),\ }\bibfield  {title} {\enquote {\bibinfo {title} {{Kaon
  interferometric probes of space-time evolution in Au+Au collisions at
  $\sqrt{s_{NN}} =$ 200 GeV}},}\ }\href {\doibase
  10.1103/PhysRevLett.103.142301} {\bibfield  {journal} {\bibinfo  {journal}
  {Phys. Rev. Lett.}\ }\textbf {\bibinfo {volume} {103}},\ \bibinfo {pages}
  {142301} (\bibinfo {year} {2009})}\BibitemShut {NoStop}%
\bibitem [{\citenamefont {Makhlin}\ and\ \citenamefont
  {Sinyukov}(1988)}]{Makhlin:1987gm}%
  \BibitemOpen
  \bibfield  {author} {\bibinfo {author} {\bibfnamefont {A.~N.}\ \bibnamefont
  {Makhlin}}\ and\ \bibinfo {author} {\bibfnamefont {{\relax Yu}.~M.}\
  \bibnamefont {Sinyukov}},\ }\bibfield  {title} {\enquote {\bibinfo {title}
  {{Hydrodynamics of Hadron Matter Under Pion Interferometric Microscope}},}\
  }\href {\doibase 10.1007/BF01560393} {\bibfield  {journal} {\bibinfo
  {journal} {Z. Phys. C}\ }\textbf {\bibinfo {volume} {39}},\ \bibinfo {pages}
  {69} (\bibinfo {year} {1988})}\BibitemShut {NoStop}%
%%CITATION = ZEPYA,C39,69;%%
\bibitem [{\citenamefont {Cs{\"o}rg\H{o}}\ and\ \citenamefont
  {L{\"o}rstad}(1996)}]{Csorgo:1995bi}%
  \BibitemOpen
  \bibfield  {author} {\bibinfo {author} {\bibfnamefont {T.}~\bibnamefont
  {Cs{\"o}rg\H{o}}}\ and\ \bibinfo {author} {\bibfnamefont {B.}~\bibnamefont
  {L{\"o}rstad}},\ }\bibfield  {title} {\enquote {\bibinfo {title}
  {{Bose-Einstein correlations for three-dimensionally expanding, cylindrically
  symmetric, finite systems}},}\ }\href {\doibase 10.1103/PhysRevC.54.1390}
  {\bibfield  {journal} {\bibinfo  {journal} {Phys. Rev. C}\ }\textbf {\bibinfo
  {volume} {54}},\ \bibinfo {pages} {1390} (\bibinfo {year}
  {1996})}\BibitemShut {NoStop}%
\bibitem [{\citenamefont {Chapman}\ \emph
  {et~al.}(1995{\natexlab{a}})\citenamefont {Chapman}, \citenamefont {Scotto},\
  and\ \citenamefont {Heinz}}]{Chapman:1994yv}%
  \BibitemOpen
  \bibfield  {author} {\bibinfo {author} {\bibfnamefont {S.}~\bibnamefont
  {Chapman}}, \bibinfo {author} {\bibfnamefont {P.}~\bibnamefont {Scotto}}, \
  and\ \bibinfo {author} {\bibfnamefont {U.~W.}\ \bibnamefont {Heinz}},\
  }\bibfield  {title} {\enquote {\bibinfo {title} {{A new cross term in the two
  particle HBT correlation function}},}\ }\href {\doibase
  10.1103/PhysRevLett.74.4400} {\bibfield  {journal} {\bibinfo  {journal}
  {Phys. Rev. Lett.}\ }\textbf {\bibinfo {volume} {74}},\ \bibinfo {pages}
  {4400} (\bibinfo {year} {1995}{\natexlab{a}})}\BibitemShut {NoStop}%
\bibitem [{\citenamefont {Chapman}\ \emph
  {et~al.}(1995{\natexlab{b}})\citenamefont {Chapman}, \citenamefont {Scotto},\
  and\ \citenamefont {Heinz}}]{Chapman:1994ax}%
  \BibitemOpen
  \bibfield  {author} {\bibinfo {author} {\bibfnamefont {S.}~\bibnamefont
  {Chapman}}, \bibinfo {author} {\bibfnamefont {P.}~\bibnamefont {Scotto}}, \
  and\ \bibinfo {author} {\bibfnamefont {U.~W.}\ \bibnamefont {Heinz}},\
  }\bibfield  {title} {\enquote {\bibinfo {title} {{Model independent features
  of the two particle correlation function}},}\ }\href@noop {} {\bibfield
  {journal} {\bibinfo  {journal} {Heavy Ion Phys.}\ }\textbf {\bibinfo {volume}
  {1}},\ \bibinfo {pages} {1} (\bibinfo {year}
  {1995}{\natexlab{b}})}\BibitemShut {NoStop}%
\bibitem [{\citenamefont {Csan\'ad}\ \emph {et~al.}(2004)\citenamefont
  {Csan\'ad}, \citenamefont {Cs{\"o}rg{\H{o}}}, \citenamefont {L{\"o}rstad},\
  and\ \citenamefont {Ster}}]{Csanad:2004mm}%
  \BibitemOpen
  \bibfield  {author} {\bibinfo {author} {\bibfnamefont {M.}~\bibnamefont
  {Csan\'ad}}, \bibinfo {author} {\bibfnamefont {T.}~\bibnamefont
  {Cs{\"o}rg{\H{o}}}}, \bibinfo {author} {\bibfnamefont {B.}~\bibnamefont
  {L{\"o}rstad}}, \ and\ \bibinfo {author} {\bibfnamefont {A.}~\bibnamefont
  {Ster}},\ }\bibfield  {title} {\enquote {\bibinfo {title} {{Indication of
  quark deconfinement and evidence for a Hubble flow in 130-GeV and 200-GeV
  Au+Au collisions}},}\ }\bibfield  {booktitle} {\emph {\bibinfo {booktitle}
  {{Ultra-relativistic nucleus-nucleus collisions. Proceedings, 17th
  International Conference, Quark Matter 2004, Oakland, USA, January 11-17,
  2004}}},\ }\href {\doibase 10.1088/0954-3899/30/8/062} {\bibfield  {journal}
  {\bibinfo  {journal} {J. Phys. G}\ }\textbf {\bibinfo {volume} {30}},\
  \bibinfo {pages} {S1079} (\bibinfo {year} {2004})}\BibitemShut {NoStop}%
\bibitem [{\citenamefont {Bekele}\ \emph {et~al.}(2007)\citenamefont {Bekele}
  \emph {et~al.}}]{Bekele:2007ee}%
  \BibitemOpen
  \bibfield  {author} {\bibinfo {author} {\bibfnamefont {S.}~\bibnamefont
  {Bekele}} \emph {et~al.},\ }\bibfield  {title} {\enquote {\bibinfo {title}
  {{Status and Promise of Particle Interferometry in Heavy-Ion Collisions}},}\
  }\href {\doibase 10.1590/S0103-97332007000600002} {\bibfield  {journal}
  {\bibinfo  {journal} {Braz. J. Phys.}\ }\textbf {\bibinfo {volume} {37}},\
  \bibinfo {pages} {31} (\bibinfo {year} {2007})}\BibitemShut {NoStop}%
\bibitem [{\citenamefont {Lisa}\ and\ \citenamefont
  {Pratt}(2008)}]{Lisa:2008gf}%
  \BibitemOpen
  \bibfield  {author} {\bibinfo {author} {\bibfnamefont {M.~A.}\ \bibnamefont
  {Lisa}}\ and\ \bibinfo {author} {\bibfnamefont {S.}~\bibnamefont {Pratt}},\
  }\href@noop {} {\enquote {\bibinfo {title} {{Femtoscopically Probing the
  Freeze-out Configuration in Heavy Ion Collisions}},}\ } (\bibinfo {year}
  {2008}),\ \bibinfo {note} {arXiv:0811.1352}\BibitemShut {NoStop}%
\bibitem [{\citenamefont {Pratt}(2009)}]{Pratt:2008qv}%
  \BibitemOpen
  \bibfield  {author} {\bibinfo {author} {\bibfnamefont {S.}~\bibnamefont
  {Pratt}},\ }\bibfield  {title} {\enquote {\bibinfo {title} {{Resolving the
  HBT Puzzle in Relativistic Heavy Ion Collision}},}\ }\href {\doibase
  10.1103/PhysRevLett.102.232301} {\bibfield  {journal} {\bibinfo  {journal}
  {Phys. Rev. Lett.}\ }\textbf {\bibinfo {volume} {102}},\ \bibinfo {pages}
  {232301} (\bibinfo {year} {2009})}\BibitemShut {NoStop}%
\bibitem [{\citenamefont {Heinz}(2010)}]{Heinz:2009xj}%
  \BibitemOpen
  \bibfield  {author} {\bibinfo {author} {\bibfnamefont {U.~W.}\ \bibnamefont
  {Heinz}},\ }\enquote {\bibinfo {title} {{Early collective expansion:
  Relativistic hydrodynamics and the transport properties of QCD matter}},}\
  in\ \href {\doibase 10.1007/978-3-642-01539-7-9} {\emph {\bibinfo {booktitle}
  {{Landolt-B{\"o}rnstein- Group I Elementary Particles, Nuclei and Atoms 23
  (Relativistic Heavy Ion Physics)}}}},\ \bibinfo {editor} {edited by\ \bibinfo
  {editor} {\bibfnamefont {R.}~\bibnamefont {Stock}}}\ (\bibinfo  {publisher}
  {Springer-Verlag Berlin Heidelberg},\ \bibinfo {year} {2010})\ Chap.\
  \bibinfo {chapter} {Primordial Bulk Plasma Dynamics in Nuclear Collisions at
  RHIC}\BibitemShut {NoStop}%
\bibitem [{\citenamefont {Bo\.zek}(2012)}]{Bozek:2011ua}%
  \BibitemOpen
  \bibfield  {author} {\bibinfo {author} {\bibfnamefont {P.}~\bibnamefont
  {Bo\.zek}},\ }\bibfield  {title} {\enquote {\bibinfo {title} {{Flow and
  interferometry in 3+1 dimensional viscous hydrodynamics}},}\ }\href {\doibase
  10.1103/PhysRevC.85.034901} {\bibfield  {journal} {\bibinfo  {journal} {Phys.
  Rev. C}\ }\textbf {\bibinfo {volume} {85}},\ \bibinfo {pages} {034901}
  (\bibinfo {year} {2012})}\BibitemShut {NoStop}%
\bibitem [{\citenamefont {Boal}\ \emph {et~al.}(1990)\citenamefont {Boal},
  \citenamefont {Gelbke},\ and\ \citenamefont {Jennings}}]{Boal:1990yh}%
  \BibitemOpen
  \bibfield  {author} {\bibinfo {author} {\bibfnamefont {D.~H.}\ \bibnamefont
  {Boal}}, \bibinfo {author} {\bibfnamefont {C.~K.}\ \bibnamefont {Gelbke}}, \
  and\ \bibinfo {author} {\bibfnamefont {B.~K.}\ \bibnamefont {Jennings}},\
  }\bibfield  {title} {\enquote {\bibinfo {title} {{Intensity interferometry in
  subatomic physics}},}\ }\href {\doibase 10.1103/RevModPhys.62.553} {\bibfield
   {journal} {\bibinfo  {journal} {Rev. Mod. Phys.}\ }\textbf {\bibinfo
  {volume} {62}},\ \bibinfo {pages} {553} (\bibinfo {year} {1990})}\BibitemShut
  {NoStop}%
%%CITATION = RMPHA,62,553;%%
\bibitem [{\citenamefont {Weiner}(2000)}]{Weiner:1999th}%
  \BibitemOpen
  \bibfield  {author} {\bibinfo {author} {\bibfnamefont {R.~M.}\ \bibnamefont
  {Weiner}},\ }\bibfield  {title} {\enquote {\bibinfo {title} {{Boson
  interferometry in high-energy physics}},}\ }\href {\doibase
  10.1016/S0370-1573(99)00114-3} {\bibfield  {journal} {\bibinfo  {journal}
  {Phys. Rept.}\ }\textbf {\bibinfo {volume} {327}},\ \bibinfo {pages} {249}
  (\bibinfo {year} {2000})}\BibitemShut {NoStop}%
\bibitem [{\citenamefont {Wiedemann}\ and\ \citenamefont
  {Heinz}(1999)}]{Wiedemann:1999qn}%
  \BibitemOpen
  \bibfield  {author} {\bibinfo {author} {\bibfnamefont {U.~A.}\ \bibnamefont
  {Wiedemann}}\ and\ \bibinfo {author} {\bibfnamefont {U.~W.}\ \bibnamefont
  {Heinz}},\ }\bibfield  {title} {\enquote {\bibinfo {title} {{Particle
  interferometry for relativistic heavy ion collisions}},}\ }\href {\doibase
  10.1016/S0370-1573(99)00032-0} {\bibfield  {journal} {\bibinfo  {journal}
  {Phys. Rept.}\ }\textbf {\bibinfo {volume} {319}},\ \bibinfo {pages} {145}
  (\bibinfo {year} {1999})}\BibitemShut {NoStop}%
\bibitem [{\citenamefont {Cs{\"o}rg\H{o}}(2002)}]{Csorgo:1999sj}%
  \BibitemOpen
  \bibfield  {author} {\bibinfo {author} {\bibfnamefont {T.}~\bibnamefont
  {Cs{\"o}rg\H{o}}},\ }\bibfield  {title} {\enquote {\bibinfo {title}
  {{Particle interferometry from 40-MeV to 40-TeV}},}\ }\bibfield  {booktitle}
  {\emph {\bibinfo {booktitle} {{NATO Advanced Study Institute on Particle
  Production Spanning MeV and TeV Energies (Nijmegen 99) Nijmegen, Netherlands,
  August 8-20, 1999}}},\ }\href {\doibase 10.1556/APH.15.2002.1-2.1} {\bibfield
   {journal} {\bibinfo  {journal} {Heavy Ion Phys.}\ }\textbf {\bibinfo
  {volume} {15}},\ \bibinfo {pages} {1} (\bibinfo {year} {2002})}\BibitemShut
  {NoStop}%
\bibitem [{\citenamefont {Lisa}\ \emph {et~al.}(2005)\citenamefont {Lisa},
  \citenamefont {Pratt}, \citenamefont {Soltz},\ and\ \citenamefont
  {Wiedemann}}]{Lisa:2005dd}%
  \BibitemOpen
  \bibfield  {author} {\bibinfo {author} {\bibfnamefont {M.~A.}\ \bibnamefont
  {Lisa}}, \bibinfo {author} {\bibfnamefont {S.}~\bibnamefont {Pratt}},
  \bibinfo {author} {\bibfnamefont {R.}~\bibnamefont {Soltz}}, \ and\ \bibinfo
  {author} {\bibfnamefont {U.}~\bibnamefont {Wiedemann}},\ }\bibfield  {title}
  {\enquote {\bibinfo {title} {{Femtoscopy in relativistic heavy ion
  collisions}},}\ }\href {\doibase 10.1146/annurev.nucl.55.090704.151533}
  {\bibfield  {journal} {\bibinfo  {journal} {Ann. Rev. Nucl. Part. Sci.}\
  }\textbf {\bibinfo {volume} {55}},\ \bibinfo {pages} {357} (\bibinfo {year}
  {2005})}\BibitemShut {NoStop}%
\bibitem [{\citenamefont {Tannenbaum}(2006)}]{Tannenbaum:2006ch}%
  \BibitemOpen
  \bibfield  {author} {\bibinfo {author} {\bibfnamefont {M.~J.}\ \bibnamefont
  {Tannenbaum}},\ }\bibfield  {title} {\enquote {\bibinfo {title} {{Recent
  results in relativistic heavy ion collisions: From 'a new state of matter' to
  'the perfect fluid'}},}\ }\href {\doibase 10.1088/0034-4885/69/7/R01}
  {\bibfield  {journal} {\bibinfo  {journal} {Rept. Prog. Phys.}\ }\textbf
  {\bibinfo {volume} {69}},\ \bibinfo {pages} {2005} (\bibinfo {year}
  {2006})}\BibitemShut {NoStop}%
\bibitem [{\citenamefont {{A. Kisiel, for the ALICE
  Collaboration}}(2011)}]{Kisiel:2011jt}%
  \BibitemOpen
  \bibfield  {author} {\bibinfo {author} {\bibnamefont {{A. Kisiel, for the
  ALICE Collaboration}}},\ }\bibfield  {title} {\enquote {\bibinfo {title}
  {{Overview of the femtoscopy studies in Pb Pb and p p collisions at the LHC
  by the ALICE experiment}},}\ }\bibfield  {booktitle} {\emph {\bibinfo
  {booktitle} {{Proceedings, 7th Workshop on Particle Correlations and
  Femtoscopy (WPCF 2011): Tokyo, Japan, September 20-24, 2011}}},\ }\href@noop
  {} {\bibfield  {journal} {\bibinfo  {journal} {PoS}\ }\textbf {\bibinfo
  {volume} {WPCF2011}},\ \bibinfo {pages} {003} (\bibinfo {year}
  {2011})}\BibitemShut {NoStop}%
%%CITATION = POSCI,WPCF2011,003;%%
\bibitem [{\citenamefont {Heinz}\ and\ \citenamefont
  {Snellings}(2013)}]{Heinz:2013th}%
  \BibitemOpen
  \bibfield  {author} {\bibinfo {author} {\bibfnamefont {U.}~\bibnamefont
  {Heinz}}\ and\ \bibinfo {author} {\bibfnamefont {R.}~\bibnamefont
  {Snellings}},\ }\bibfield  {title} {\enquote {\bibinfo {title} {{Collective
  flow and viscosity in relativistic heavy-ion collisions}},}\ }\href {\doibase
  10.1146/annurev-nucl-102212-170540} {\bibfield  {journal} {\bibinfo
  {journal} {Ann. Rev. Nucl. Part. Sci.}\ }\textbf {\bibinfo {volume} {63}},\
  \bibinfo {pages} {123} (\bibinfo {year} {2013})}\BibitemShut {NoStop}%
\bibitem [{\citenamefont {Adamczyk}\ \emph {et~al.}(2015)\citenamefont
  {Adamczyk} \emph {et~al.}}]{Adamczyk:2014mxp}%
  \BibitemOpen
  \bibfield  {author} {\bibinfo {author} {\bibfnamefont {L.}~\bibnamefont
  {Adamczyk}} \emph {et~al.} (\bibinfo {collaboration} {STAR Collaboration}),\
  }\bibfield  {title} {\enquote {\bibinfo {title} {{Beam-energy-dependent
  two-pion interferometry and the freeze-out eccentricity of pions measured in
  heavy ion collisions at the STAR detector}},}\ }\href {\doibase
  10.1103/PhysRevC.92.014904} {\bibfield  {journal} {\bibinfo  {journal} {Phys.
  Rev. C}\ }\textbf {\bibinfo {volume} {92}},\ \bibinfo {pages} {014904}
  (\bibinfo {year} {2015})}\BibitemShut {NoStop}%
\bibitem [{\citenamefont {Achard}\ \emph {et~al.}(2011)\citenamefont {Achard}
  \emph {et~al.}}]{Achard:2011zza}%
  \BibitemOpen
  \bibfield  {author} {\bibinfo {author} {\bibfnamefont {P.}~\bibnamefont
  {Achard}} \emph {et~al.} (\bibinfo {collaboration} {L3 Collaboration}),\
  }\bibfield  {title} {\enquote {\bibinfo {title} {{Test of the
  \boldmath{$\tau$}-Model of Bose-Einstein Correlations and Reconstruction of
  the Source Function in Hadronic Z-boson Decay at LEP}},}\ }\href {\doibase
  10.1140/epjc/s10052-011-1648-8} {\bibfield  {journal} {\bibinfo  {journal}
  {Eur. Phys. J. C}\ }\textbf {\bibinfo {volume} {71}},\ \bibinfo {pages}
  {1648} (\bibinfo {year} {2011})}\BibitemShut {NoStop}%
\bibitem [{\citenamefont {Khachatryan}\ \emph {et~al.}(2011)\citenamefont
  {Khachatryan} \emph {et~al.}}]{Khachatryan:2011hi}%
  \BibitemOpen
  \bibfield  {author} {\bibinfo {author} {\bibfnamefont {V.}~\bibnamefont
  {Khachatryan}} \emph {et~al.} (\bibinfo {collaboration} {CMS
  Collaboration}),\ }\bibfield  {title} {\enquote {\bibinfo {title}
  {{Measurement of Bose-Einstein Correlations in $pp$ Collisions at
  $\sqrt{s}=0.9$ and 7 TeV}},}\ }\href {\doibase 10.1007/JHEP05(2011)029}
  {\bibfield  {journal} {\bibinfo  {journal} {JHEP}\ }\textbf {\bibinfo
  {volume} {05}},\ \bibinfo {pages} {029} (\bibinfo {year} {2011})}\BibitemShut
  {NoStop}%
\bibitem [{\citenamefont {{F. Sikl\'er, for the CMS
  Collaboration}}(2014)}]{Sikler:2014aea}%
  \BibitemOpen
  \bibfield  {author} {\bibinfo {author} {\bibnamefont {{F. Sikl\'er, for the
  CMS Collaboration}}},\ }\href@noop {} {\enquote {\bibinfo {title}
  {{Femtoscopy with identified hadrons in pp, pPb, and peripheral PbPb
  collisions in CMS}},}\ } (\bibinfo {year} {2014}),\ \bibinfo {note}
  {arXiv:1411.6609}\BibitemShut {NoStop}%
\bibitem [{\citenamefont {Astalos}(2015)}]{Astalos:2015}%
  \BibitemOpen
  \bibfield  {author} {\bibinfo {author} {\bibfnamefont {R.}~\bibnamefont
  {Astalos}},\ }\emph {\bibinfo {title} {{Bose-Einstein correlations in 7 TeV
  proton-proton collisions in the ATLAS experiment}}},\ \href@noop {} {Ph.D.
  thesis},\ \bibinfo  {school} {Radboud University} (\bibinfo {year}
  {2015})\BibitemShut {NoStop}%
\bibitem [{\citenamefont {Bolz}\ \emph {et~al.}(1993)\citenamefont {Bolz},
  \citenamefont {Ornik}, \citenamefont {Pl{\"u}mer}, \citenamefont {Schlei},\
  and\ \citenamefont {Weiner}}]{Bolz:1992hc}%
  \BibitemOpen
  \bibfield  {author} {\bibinfo {author} {\bibfnamefont {J.}~\bibnamefont
  {Bolz}}, \bibinfo {author} {\bibfnamefont {U.}~\bibnamefont {Ornik}},
  \bibinfo {author} {\bibfnamefont {M.}~\bibnamefont {Pl{\"u}mer}}, \bibinfo
  {author} {\bibfnamefont {B.~R.}\ \bibnamefont {Schlei}}, \ and\ \bibinfo
  {author} {\bibfnamefont {R.~M.}\ \bibnamefont {Weiner}},\ }\bibfield  {title}
  {\enquote {\bibinfo {title} {{Resonance decays and partial coherence in
  Bose-Einstein correlations}},}\ }\href {\doibase 10.1103/PhysRevD.47.3860}
  {\bibfield  {journal} {\bibinfo  {journal} {Phys. Rev. D}\ }\textbf {\bibinfo
  {volume} {47}},\ \bibinfo {pages} {3860} (\bibinfo {year}
  {1993})}\BibitemShut {NoStop}%
%%CITATION = PHRVA,D47,3860;%%
\bibitem [{\citenamefont {Cs{\"o}rg\H{o}}\ \emph {et~al.}(1996)\citenamefont
  {Cs{\"o}rg\H{o}}, \citenamefont {L{\"o}rstad},\ and\ \citenamefont
  {Zim\'anyi}}]{Csorgo:1994in}%
  \BibitemOpen
  \bibfield  {author} {\bibinfo {author} {\bibfnamefont {T.}~\bibnamefont
  {Cs{\"o}rg\H{o}}}, \bibinfo {author} {\bibfnamefont {B.}~\bibnamefont
  {L{\"o}rstad}}, \ and\ \bibinfo {author} {\bibfnamefont {J.}~\bibnamefont
  {Zim\'anyi}},\ }\bibfield  {title} {\enquote {\bibinfo {title}
  {{Bose-Einstein correlations for systems with large halo}},}\ }\href
  {\doibase 10.1007/BF02907008, 10.1007/s002880050195} {\bibfield  {journal}
  {\bibinfo  {journal} {Z. Phys. C}\ }\textbf {\bibinfo {volume} {71}},\
  \bibinfo {pages} {491} (\bibinfo {year} {1996})}\BibitemShut {NoStop}%
\bibitem [{\citenamefont {Cs{\"o}rg\H{o}}\ \emph {et~al.}(2005)\citenamefont
  {Cs{\"o}rg\H{o}}, \citenamefont {Hegyi}, \citenamefont {Nov\'ak},\ and\
  \citenamefont {Zajc}}]{Csorgo:2004sr}%
  \BibitemOpen
  \bibfield  {author} {\bibinfo {author} {\bibfnamefont {T.}~\bibnamefont
  {Cs{\"o}rg\H{o}}}, \bibinfo {author} {\bibfnamefont {S.}~\bibnamefont
  {Hegyi}}, \bibinfo {author} {\bibfnamefont {T.}~\bibnamefont {Nov\'ak}}, \
  and\ \bibinfo {author} {\bibfnamefont {W.~A.}\ \bibnamefont {Zajc}},\
  }\bibfield  {title} {\enquote {\bibinfo {title} {{Bose-Einstein or HBT
  correlations and the anomalous dimension of QCD}},}\ }\bibfield  {booktitle}
  {\emph {\bibinfo {booktitle} {{Proceedings, 34th International Symposium on
  Multiparticle dynamics (ISMD 2004): Rohnert Park, USA, July 26-August 1,
  2004}}},\ }\href@noop {} {\bibfield  {journal} {\bibinfo  {journal} {Acta
  Phys. Polon. B}\ }\textbf {\bibinfo {volume} {36}},\ \bibinfo {pages} {329}
  (\bibinfo {year} {2005})}\BibitemShut {NoStop}%
\bibitem [{\citenamefont {Cs{\"o}rg\H{o}}\ \emph {et~al.}(2004)\citenamefont
  {Cs{\"o}rg\H{o}}, \citenamefont {Hegyi},\ and\ \citenamefont
  {Zajc}}]{Csorgo:2003uv}%
  \BibitemOpen
  \bibfield  {author} {\bibinfo {author} {\bibfnamefont {T.}~\bibnamefont
  {Cs{\"o}rg\H{o}}}, \bibinfo {author} {\bibfnamefont {S.}~\bibnamefont
  {Hegyi}}, \ and\ \bibinfo {author} {\bibfnamefont {W.~A.}\ \bibnamefont
  {Zajc}},\ }\bibfield  {title} {\enquote {\bibinfo {title} {{Bose-Einstein
  correlations for L\'evy stable source distributions}},}\ }\href@noop {}
  {\bibfield  {journal} {\bibinfo  {journal} {Eur. Phys. J. C}\ }\textbf
  {\bibinfo {volume} {36}},\ \bibinfo {pages} {67} (\bibinfo {year}
  {2004})}\BibitemShut {NoStop}%
\bibitem [{\citenamefont {Metzler}\ \emph {et~al.}(1999)\citenamefont
  {Metzler}, \citenamefont {Barkai},\ and\ \citenamefont
  {Klafter}}]{Metzler:1999zz}%
  \BibitemOpen
  \bibfield  {author} {\bibinfo {author} {\bibfnamefont {R.}~\bibnamefont
  {Metzler}}, \bibinfo {author} {\bibfnamefont {E.}~\bibnamefont {Barkai}}, \
  and\ \bibinfo {author} {\bibfnamefont {J.}~\bibnamefont {Klafter}},\
  }\bibfield  {title} {\enquote {\bibinfo {title} {{Anomalous Diffusion and
  Relaxation Close to Thermal Equilibrium: A Fractional Fokker-Planck Equation
  Approach}},}\ }\href {\doibase 10.1103/PhysRevLett.82.3563} {\bibfield
  {journal} {\bibinfo  {journal} {Phys. Rev. Lett.}\ }\textbf {\bibinfo
  {volume} {82}},\ \bibinfo {pages} {3563} (\bibinfo {year}
  {1999})}\BibitemShut {NoStop}%
%%CITATION = PRLTA,82,3563;%%
\bibitem [{\citenamefont {Adcox}\ \emph
  {et~al.}(2003{\natexlab{a}})\citenamefont {Adcox} \emph
  {et~al.}}]{Adcox:2003zm}%
  \BibitemOpen
  \bibfield  {author} {\bibinfo {author} {\bibfnamefont {K.}~\bibnamefont
  {Adcox}} \emph {et~al.} (\bibinfo {collaboration} {PHENIX Collaboration}),\
  }\bibfield  {title} {\enquote {\bibinfo {title} {{PHENIX detector
  overview}},}\ }\href {\doibase 10.1016/S0168-9002(02)01950-2} {\bibfield
  {journal} {\bibinfo  {journal} {Nucl. Instrum. Methods Phys. Res., Sec. A}\
  }\textbf {\bibinfo {volume} {499}},\ \bibinfo {pages} {469} (\bibinfo {year}
  {2003}{\natexlab{a}})}\BibitemShut {NoStop}%
%%CITATION = NUIMA,A499,469;%%
\bibitem [{\citenamefont {Adare}\ \emph {et~al.}(2013)\citenamefont {Adare}
  \emph {et~al.}}]{Adare:2013esx}%
  \BibitemOpen
  \bibfield  {author} {\bibinfo {author} {\bibfnamefont {A.}~\bibnamefont
  {Adare}} \emph {et~al.} (\bibinfo {collaboration} {PHENIX Collaboration}),\
  }\bibfield  {title} {\enquote {\bibinfo {title} {{Spectra and ratios of
  identified particles in Au+Au and $d$+Au collisions at $\sqrt{s_{NN}}=200$
  GeV}},}\ }\href {\doibase 10.1103/PhysRevC.88.024906} {\bibfield  {journal}
  {\bibinfo  {journal} {Phys. Rev. C}\ }\textbf {\bibinfo {volume} {88}},\
  \bibinfo {pages} {024906} (\bibinfo {year} {2013})}\BibitemShut {NoStop}%
\bibitem [{\citenamefont {Adcox}\ \emph
  {et~al.}(2003{\natexlab{b}})\citenamefont {Adcox} \emph
  {et~al.}}]{Adcox:2003zp}%
  \BibitemOpen
  \bibfield  {author} {\bibinfo {author} {\bibfnamefont {K.}~\bibnamefont
  {Adcox}} \emph {et~al.} (\bibinfo {collaboration} {PHENIX Collaboration}),\
  }\bibfield  {title} {\enquote {\bibinfo {title} {{PHENIX central arm tracking
  detectors}},}\ }\href {\doibase 10.1016/S0168-9002(02)01952-6} {\bibfield
  {journal} {\bibinfo  {journal} {Nucl. Instrum. Methods Phys. Res., Sec. A}\
  }\textbf {\bibinfo {volume} {499}},\ \bibinfo {pages} {489} (\bibinfo {year}
  {2003}{\natexlab{b}})}\BibitemShut {NoStop}%
%%CITATION = NUIMA,A499,489;%%
\bibitem [{\citenamefont {Anderson}\ \emph {et~al.}(2011)\citenamefont
  {Anderson} \emph {et~al.}}]{Anderson:2011jw}%
  \BibitemOpen
  \bibfield  {author} {\bibinfo {author} {\bibfnamefont {W.}~\bibnamefont
  {Anderson}} \emph {et~al.},\ }\bibfield  {title} {\enquote {\bibinfo {title}
  {{Design, Construction, Operation and Performance of a Hadron Blind Detector
  for the PHENIX Experiment}},}\ }\href {\doibase 10.1016/j.nima.2011.04.015}
  {\bibfield  {journal} {\bibinfo  {journal} {Nucl. Instrum. Methods Phys.
  Res., Sec. A}\ }\textbf {\bibinfo {volume} {646}},\ \bibinfo {pages} {35}
  (\bibinfo {year} {2011})}\BibitemShut {NoStop}%
\bibitem [{\citenamefont {Adare}\ \emph {et~al.}(2012)\citenamefont {Adare}
  \emph {et~al.}}]{Adare:2012vq}%
  \BibitemOpen
  \bibfield  {author} {\bibinfo {author} {\bibfnamefont {A.}~\bibnamefont
  {Adare}} \emph {et~al.} (\bibinfo {collaboration} {PHENIX Collaboration}),\
  }\bibfield  {title} {\enquote {\bibinfo {title} {{Deviation from quark-number
  scaling of the anisotropy parameter $v_2$ of pions, kaons, and protons in
  Au+Au collisions at $\sqrt{s_{NN}} = 200$ GeV}},}\ }\href {\doibase
  10.1103/PhysRevC.85.064914} {\bibfield  {journal} {\bibinfo  {journal} {Phys.
  Rev. C}\ }\textbf {\bibinfo {volume} {85}},\ \bibinfo {pages} {064914}
  (\bibinfo {year} {2012})}\BibitemShut {NoStop}%
\bibitem [{\citenamefont {Aizawa}\ \emph {et~al.}(2003)\citenamefont {Aizawa}
  \emph {et~al.}}]{Aizawa:2003zq}%
  \BibitemOpen
  \bibfield  {author} {\bibinfo {author} {\bibfnamefont {M.}~\bibnamefont
  {Aizawa}} \emph {et~al.} (\bibinfo {collaboration} {PHENIX Collaboration}),\
  }\bibfield  {title} {\enquote {\bibinfo {title} {{PHENIX central arm particle
  ID detectors}},}\ }\href {\doibase 10.1016/S0168-9002(02)01953-8} {\bibfield
  {journal} {\bibinfo  {journal} {Nucl. Instrum. Methods Phys. Res., Sec. A}\
  }\textbf {\bibinfo {volume} {499}},\ \bibinfo {pages} {508} (\bibinfo {year}
  {2003})}\BibitemShut {NoStop}%
%%CITATION = NUIMA,A499,508;%%
\bibitem [{\citenamefont {Adare}\ \emph {et~al.}(2016)\citenamefont {Adare}
  \emph {et~al.}}]{Adare:2015ila}%
  \BibitemOpen
  \bibfield  {author} {\bibinfo {author} {\bibfnamefont {A.}~\bibnamefont
  {Adare}} \emph {et~al.} (\bibinfo {collaboration} {PHENIX Collaboration}),\
  }\bibfield  {title} {\enquote {\bibinfo {title} {{Dielectron production in
  Au$+$Au collisions at $\sqrt{s_{NN}}$=200 GeV}},}\ }\href {\doibase
  10.1103/PhysRevC.93.014904} {\bibfield  {journal} {\bibinfo  {journal} {Phys.
  Rev. C}\ }\textbf {\bibinfo {volume} {93}},\ \bibinfo {pages} {014904}
  (\bibinfo {year} {2016})}\BibitemShut {NoStop}%
\bibitem [{\citenamefont {Yano}\ and\ \citenamefont
  {Koonin}(1978)}]{Yano:1978gk}%
  \BibitemOpen
  \bibfield  {author} {\bibinfo {author} {\bibfnamefont {F.~B.}\ \bibnamefont
  {Yano}}\ and\ \bibinfo {author} {\bibfnamefont {S.~E.}\ \bibnamefont
  {Koonin}},\ }\bibfield  {title} {\enquote {\bibinfo {title} {{Determining
  Pion Source Parameters in Relativistic Heavy Ion Collisions}},}\ }\href
  {\doibase 10.1016/0370-2693(78)90638-X} {\bibfield  {journal} {\bibinfo
  {journal} {Phys. Lett. B}\ }\textbf {\bibinfo {volume} {78}},\ \bibinfo
  {pages} {556} (\bibinfo {year} {1978})}\BibitemShut {NoStop}%
%%CITATION = PHLTA,B78,556;%%
\bibitem [{\citenamefont {Pratt}\ \emph {et~al.}(1990)\citenamefont {Pratt},
  \citenamefont {Cs{\"o}rg{\H{o}}},\ and\ \citenamefont
  {Zim\'anyi}}]{Pratt:1990zq}%
  \BibitemOpen
  \bibfield  {author} {\bibinfo {author} {\bibfnamefont {S.}~\bibnamefont
  {Pratt}}, \bibinfo {author} {\bibfnamefont {T.}~\bibnamefont
  {Cs{\"o}rg{\H{o}}}}, \ and\ \bibinfo {author} {\bibfnamefont
  {J.}~\bibnamefont {Zim\'anyi}},\ }\bibfield  {title} {\enquote {\bibinfo
  {title} {{Detailed predictions for two pion correlations in ultrarelativistic
  heavy ion collisions}},}\ }\href {\doibase 10.1103/PhysRevC.42.2646}
  {\bibfield  {journal} {\bibinfo  {journal} {Phys. Rev. C}\ }\textbf {\bibinfo
  {volume} {42}},\ \bibinfo {pages} {2646} (\bibinfo {year}
  {1990})}\BibitemShut {NoStop}%
%%CITATION = PHRVA,C42,2646;%%
\bibitem [{\citenamefont {Pratt}(1986)}]{Pratt:1986ev}%
  \BibitemOpen
  \bibfield  {author} {\bibinfo {author} {\bibfnamefont {S.}~\bibnamefont
  {Pratt}},\ }\bibfield  {title} {\enquote {\bibinfo {title} {{Coherence and
  Coulomb Effects on Pion Interferometry}},}\ }\href {\doibase
  10.1103/PhysRevD.33.72} {\bibfield  {journal} {\bibinfo  {journal} {Phys.
  Rev. D}\ }\textbf {\bibinfo {volume} {33}},\ \bibinfo {pages} {72} (\bibinfo
  {year} {1986})}\BibitemShut {NoStop}%
%%CITATION = PHRVA,D33,72;%%
\bibitem [{\citenamefont {Bertsch}\ \emph {et~al.}(1988)\citenamefont
  {Bertsch}, \citenamefont {Gong},\ and\ \citenamefont
  {Tohyama}}]{Bertsch:1988db}%
  \BibitemOpen
  \bibfield  {author} {\bibinfo {author} {\bibfnamefont {G.}~\bibnamefont
  {Bertsch}}, \bibinfo {author} {\bibfnamefont {M.}~\bibnamefont {Gong}}, \
  and\ \bibinfo {author} {\bibfnamefont {M.}~\bibnamefont {Tohyama}},\
  }\bibfield  {title} {\enquote {\bibinfo {title} {{Pion Interferometry in
  Ultrarelativistic Heavy Ion Collisions}},}\ }\href {\doibase
  10.1103/PhysRevC.37.1896} {\bibfield  {journal} {\bibinfo  {journal} {Phys.
  Rev. C}\ }\textbf {\bibinfo {volume} {37}},\ \bibinfo {pages} {1896}
  (\bibinfo {year} {1988})}\BibitemShut {NoStop}%
%%CITATION = PHRVA,C37,1896;%%
\bibitem [{\citenamefont {Adams}\ \emph
  {et~al.}(2005{\natexlab{b}})\citenamefont {Adams} \emph
  {et~al.}}]{Adams:2004yc}%
  \BibitemOpen
  \bibfield  {author} {\bibinfo {author} {\bibfnamefont {J.}~\bibnamefont
  {Adams}} \emph {et~al.} (\bibinfo {collaboration} {STAR Collaboration}),\
  }\bibfield  {title} {\enquote {\bibinfo {title} {{Pion interferometry in
  Au+Au collisions at $\sqrt{s_{NN}} =$ 200 GeV}},}\ }\href {\doibase
  10.1103/PhysRevC.71.044906} {\bibfield  {journal} {\bibinfo  {journal} {Phys.
  Rev. C}\ }\textbf {\bibinfo {volume} {71}},\ \bibinfo {pages} {044906}
  (\bibinfo {year} {2005}{\natexlab{b}})}\BibitemShut {NoStop}%
\bibitem [{\citenamefont {Afanasiev}\ \emph {et~al.}(2008)\citenamefont
  {Afanasiev} \emph {et~al.}}]{Afanasiev:2007kk}%
  \BibitemOpen
  \bibfield  {author} {\bibinfo {author} {\bibfnamefont {S.}~\bibnamefont
  {Afanasiev}} \emph {et~al.} (\bibinfo {collaboration} {PHENIX
  Collaboration}),\ }\bibfield  {title} {\enquote {\bibinfo {title} {{Source
  breakup dynamics in Au+Au Collisions at $\sqrt{s_{NN}} =$ 200 GeV via
  three-dimensional two-pion source imaging}},}\ }\href {\doibase
  10.1103/PhysRevLett.100.232301} {\bibfield  {journal} {\bibinfo  {journal}
  {Phys. Rev. Lett.}\ }\textbf {\bibinfo {volume} {100}},\ \bibinfo {pages}
  {232301} (\bibinfo {year} {2008})}\BibitemShut {NoStop}%
\bibitem [{\citenamefont {Nov\'ak}\ \emph {et~al.}(2016)\citenamefont
  {Nov\'ak}, \citenamefont {Cs{\"o}rg\H{o}}, \citenamefont {Eggers},\ and\
  \citenamefont {de~Kock}}]{Novak:2016cyc}%
  \BibitemOpen
  \bibfield  {author} {\bibinfo {author} {\bibfnamefont {T.}~\bibnamefont
  {Nov\'ak}}, \bibinfo {author} {\bibfnamefont {T.}~\bibnamefont
  {Cs{\"o}rg\H{o}}}, \bibinfo {author} {\bibfnamefont {H.~C.}\ \bibnamefont
  {Eggers}}, \ and\ \bibinfo {author} {\bibfnamefont {M.}~\bibnamefont
  {de~Kock}},\ }\bibfield  {title} {\enquote {\bibinfo {title} {{Model
  independent analysis of nearly L\'evy correlations}},}\ }\bibfield
  {booktitle} {\emph {\bibinfo {booktitle} {{Proceedings, 11th Workshop on
  Particle Correlations and Femtoscopy and NICA Days 2015 (WPCF 2015): Warsaw,
  Poland,November 3-7, 2015}}},\ }\href {\doibase 10.5506/APhysPolBSupp.9.289}
  {\bibfield  {journal} {\bibinfo  {journal} {Acta Phys. Polon. Supp.}\
  }\textbf {\bibinfo {volume} {9}},\ \bibinfo {pages} {289} (\bibinfo {year}
  {2016})}\BibitemShut {NoStop}%
\bibitem [{\citenamefont {Kapusta}\ \emph {et~al.}(1996)\citenamefont
  {Kapusta}, \citenamefont {Kharzeev},\ and\ \citenamefont
  {McLerran}}]{Kapusta:1995ww}%
  \BibitemOpen
  \bibfield  {author} {\bibinfo {author} {\bibfnamefont {J.~I.}\ \bibnamefont
  {Kapusta}}, \bibinfo {author} {\bibfnamefont {D.}~\bibnamefont {Kharzeev}}, \
  and\ \bibinfo {author} {\bibfnamefont {L.~D.}\ \bibnamefont {McLerran}},\
  }\bibfield  {title} {\enquote {\bibinfo {title} {{The return of the prodigal
  Goldstone boson}},}\ }\href@noop {} {\bibfield  {journal} {\bibinfo
  {journal} {Phys. Rev. D}\ }\textbf {\bibinfo {volume} {53}},\ \bibinfo
  {pages} {5028} (\bibinfo {year} {1996})}\BibitemShut {NoStop}%
\bibitem [{\citenamefont {Vance}\ \emph {et~al.}(1998)\citenamefont {Vance},
  \citenamefont {Cs{\"o}rg\H{o}},\ and\ \citenamefont
  {Kharzeev}}]{Vance:1998wd}%
  \BibitemOpen
  \bibfield  {author} {\bibinfo {author} {\bibfnamefont {S.~E.}\ \bibnamefont
  {Vance}}, \bibinfo {author} {\bibfnamefont {T.}~\bibnamefont
  {Cs{\"o}rg\H{o}}}, \ and\ \bibinfo {author} {\bibfnamefont {D.}~\bibnamefont
  {Kharzeev}},\ }\bibfield  {title} {\enquote {\bibinfo {title} {{Partial
  U(A)(1) restoration from Bose-Einstein correlations}},}\ }\href@noop {}
  {\bibfield  {journal} {\bibinfo  {journal} {Phys. Rev. Lett.}\ }\textbf
  {\bibinfo {volume} {81}},\ \bibinfo {pages} {2205} (\bibinfo {year}
  {1998})}\BibitemShut {NoStop}%
\bibitem [{\citenamefont {Cs{\"o}rg\H{o}}\ \emph {et~al.}(2010)\citenamefont
  {Cs{\"o}rg\H{o}}, \citenamefont {V\'ertesi},\ and\ \citenamefont
  {Sziklai}}]{Csorgo:2009pa}%
  \BibitemOpen
  \bibfield  {author} {\bibinfo {author} {\bibfnamefont {T.}~\bibnamefont
  {Cs{\"o}rg\H{o}}}, \bibinfo {author} {\bibfnamefont {R.}~\bibnamefont
  {V\'ertesi}}, \ and\ \bibinfo {author} {\bibfnamefont {J.}~\bibnamefont
  {Sziklai}},\ }\bibfield  {title} {\enquote {\bibinfo {title} {{Indirect
  observation of an in-medium $\eta$ ' mass reduction in $\sqrt{s_{NN}}=200$
  GeV Au+Au collisions}},}\ }\href {\doibase 10.1103/PhysRevLett.105.182301}
  {\bibfield  {journal} {\bibinfo  {journal} {Phys. Rev. Lett.}\ }\textbf
  {\bibinfo {volume} {105}},\ \bibinfo {pages} {182301} (\bibinfo {year}
  {2010})}\BibitemShut {NoStop}%
\bibitem [{\citenamefont {V\'ertesi}\ \emph {et~al.}(2011)\citenamefont
  {V\'ertesi}, \citenamefont {Cs{\"o}rg{\H{o}}},\ and\ \citenamefont
  {Sziklai}}]{Vertesi:2009wf}%
  \BibitemOpen
  \bibfield  {author} {\bibinfo {author} {\bibfnamefont {R.}~\bibnamefont
  {V\'ertesi}}, \bibinfo {author} {\bibfnamefont {T.}~\bibnamefont
  {Cs{\"o}rg{\H{o}}}}, \ and\ \bibinfo {author} {\bibfnamefont
  {J.}~\bibnamefont {Sziklai}},\ }\bibfield  {title} {\enquote {\bibinfo
  {title} {{Significant in-medium $\eta$ ' mass reduction in
  $\sqrt{s_{NN}}=200$ GeV Au+Au collisions at the BNL Relativistic Heavy Ion
  Collider}},}\ }\href {\doibase 10.1103/PhysRevC.83.054903} {\bibfield
  {journal} {\bibinfo  {journal} {Phys. Rev. C}\ }\textbf {\bibinfo {volume}
  {83}},\ \bibinfo {pages} {054903} (\bibinfo {year} {2011})}\BibitemShut
  {NoStop}%
\bibitem [{\citenamefont {Adam}\ \emph {et~al.}(2016)\citenamefont {Adam} \emph
  {et~al.}}]{Adam:2015pbc}%
  \BibitemOpen
  \bibfield  {author} {\bibinfo {author} {\bibfnamefont {J.}~\bibnamefont
  {Adam}} \emph {et~al.} (\bibinfo {collaboration} {ALICE Collaboration}),\
  }\bibfield  {title} {\enquote {\bibinfo {title} {{Multipion Bose-Einstein
  correlations in p p, p Pb, and Pb Pb collisions at energies available at the
  CERN Large Hadron Collider}},}\ }\href {\doibase 10.1103/PhysRevC.93.054908}
  {\bibfield  {journal} {\bibinfo  {journal} {Phys. Rev. C}\ }\textbf {\bibinfo
  {volume} {93}},\ \bibinfo {pages} {054908} (\bibinfo {year}
  {2016})}\BibitemShut {NoStop}%
\bibitem [{\citenamefont {Adler}\ \emph {et~al.}(2007)\citenamefont {Adler}
  \emph {et~al.}}]{Adler:2006as}%
  \BibitemOpen
  \bibfield  {author} {\bibinfo {author} {\bibfnamefont {S.~S.}\ \bibnamefont
  {Adler}} \emph {et~al.} (\bibinfo {collaboration} {PHENIX Collaboration}),\
  }\bibfield  {title} {\enquote {\bibinfo {title} {{Evidence for a long-range
  component in the pion emission source in Au + Au collisions at
  $\sqrt{s_{NN}}=200$ GeV}},}\ }\href {\doibase 10.1103/PhysRevLett.98.132301}
  {\bibfield  {journal} {\bibinfo  {journal} {Phys. Rev. Lett.}\ }\textbf
  {\bibinfo {volume} {98}},\ \bibinfo {pages} {132301} (\bibinfo {year}
  {2007})}\BibitemShut {NoStop}%
\bibitem [{\citenamefont {Cs{\"o}rg\H{o}}(2008)}]{Csorgo:2009gb}%
  \BibitemOpen
  \bibfield  {author} {\bibinfo {author} {\bibfnamefont {T.}~\bibnamefont
  {Cs{\"o}rg\H{o}}},\ }\bibfield  {title} {\enquote {\bibinfo {title}
  {{Correlation Probes of a QCD Critical Point}},}\ }\bibfield  {booktitle}
  {\emph {\bibinfo {booktitle} {{High-p(T) physics at LHC. Proceedings, 3rd
  International Workshop, HIGH-pTLHC, Tokaj, Hungary, March 16-19, 2008}}},\
  }\href@noop {} {\bibfield  {journal} {\bibinfo  {journal} {PoS}\ }\textbf
  {\bibinfo {volume} {HIGH-PTLHC08}},\ \bibinfo {pages} {027} (\bibinfo {year}
  {2008})}\BibitemShut {NoStop}%
\bibitem [{\citenamefont {Aoki}\ \emph {et~al.}(2006)\citenamefont {Aoki},
  \citenamefont {Endr\H{o}di}, \citenamefont {Fodor}, \citenamefont {Katz},\
  and\ \citenamefont {Szab\'o}}]{Aoki:2006we}%
  \BibitemOpen
  \bibfield  {author} {\bibinfo {author} {\bibfnamefont {Y.}~\bibnamefont
  {Aoki}}, \bibinfo {author} {\bibfnamefont {G.}~\bibnamefont {Endr\H{o}di}},
  \bibinfo {author} {\bibfnamefont {Z.}~\bibnamefont {Fodor}}, \bibinfo
  {author} {\bibfnamefont {S.~D.}\ \bibnamefont {Katz}}, \ and\ \bibinfo
  {author} {\bibfnamefont {K.~K.}\ \bibnamefont {Szab\'o}},\ }\bibfield
  {title} {\enquote {\bibinfo {title} {The order of the quantum chromodynamics
  transition predicted by the standard model of particle physics},}\
  }\href@noop {} {\bibfield  {journal} {\bibinfo  {journal} {Nature}\ }\textbf
  {\bibinfo {volume} {443}},\ \bibinfo {pages} {675} (\bibinfo {year}
  {2006})}\BibitemShut {NoStop}%
\bibitem [{\citenamefont {Bhattacharya}\ \emph {et~al.}(2014)\citenamefont
  {Bhattacharya} \emph {et~al.}}]{Bhattacharya:2014ara}%
  \BibitemOpen
  \bibfield  {author} {\bibinfo {author} {\bibfnamefont {T.}~\bibnamefont
  {Bhattacharya}} \emph {et~al.},\ }\bibfield  {title} {\enquote {\bibinfo
  {title} {{QCD Phase Transition with Chiral Quarks and Physical Quark
  Masses}},}\ }\href {\doibase 10.1103/PhysRevLett.113.082001} {\bibfield
  {journal} {\bibinfo  {journal} {Phys. Rev. Lett.}\ }\textbf {\bibinfo
  {volume} {113}},\ \bibinfo {pages} {082001} (\bibinfo {year}
  {2014})}\BibitemShut {NoStop}%
\bibitem [{\citenamefont {Soltz}\ \emph {et~al.}(2015)\citenamefont {Soltz},
  \citenamefont {DeTar}, \citenamefont {Karsch}, \citenamefont {Mukherjee},\
  and\ \citenamefont {Vranas}}]{Soltz:2015ula}%
  \BibitemOpen
  \bibfield  {author} {\bibinfo {author} {\bibfnamefont {R.~A.}\ \bibnamefont
  {Soltz}}, \bibinfo {author} {\bibfnamefont {C.}~\bibnamefont {DeTar}},
  \bibinfo {author} {\bibfnamefont {F.}~\bibnamefont {Karsch}}, \bibinfo
  {author} {\bibfnamefont {S.}~\bibnamefont {Mukherjee}}, \ and\ \bibinfo
  {author} {\bibfnamefont {P.}~\bibnamefont {Vranas}},\ }\bibfield  {title}
  {\enquote {\bibinfo {title} {{Lattice QCD Thermodynamics with Physical Quark
  Masses}},}\ }\href {\doibase 10.1146/annurev-nucl-102014-022157} {\bibfield
  {journal} {\bibinfo  {journal} {Ann. Rev. Nucl. Part. Sci.}\ }\textbf
  {\bibinfo {volume} {65}},\ \bibinfo {pages} {379} (\bibinfo {year}
  {2015})}\BibitemShut {NoStop}%
\bibitem [{\citenamefont {El-Showk}\ \emph {et~al.}(2014)\citenamefont
  {El-Showk}, \citenamefont {Paulos}, \citenamefont {Poland}, \citenamefont
  {Rychkov}, \citenamefont {Simmons-Duffin},\ and\ \citenamefont
  {Vichi}}]{El-Showk:2014dwa}%
  \BibitemOpen
  \bibfield  {author} {\bibinfo {author} {\bibfnamefont {S.}~\bibnamefont
  {El-Showk}}, \bibinfo {author} {\bibfnamefont {M.~F.}\ \bibnamefont
  {Paulos}}, \bibinfo {author} {\bibfnamefont {D.}~\bibnamefont {Poland}},
  \bibinfo {author} {\bibfnamefont {S.}~\bibnamefont {Rychkov}}, \bibinfo
  {author} {\bibfnamefont {D.}~\bibnamefont {Simmons-Duffin}}, \ and\ \bibinfo
  {author} {\bibfnamefont {A.}~\bibnamefont {Vichi}},\ }\bibfield  {title}
  {\enquote {\bibinfo {title} {{Solving the 3d Ising Model with the Conformal
  Bootstrap II. c-Minimization and Precise Critical Exponents}},}\ }\href
  {\doibase 10.1007/s10955-014-1042-7} {\bibfield  {journal} {\bibinfo
  {journal} {J. Stat. Phys.}\ }\textbf {\bibinfo {volume} {157}},\ \bibinfo
  {pages} {869} (\bibinfo {year} {2014})}\BibitemShut {NoStop}%
\bibitem [{\citenamefont {Rieger}(1995)}]{Rieger:1995aa}%
  \BibitemOpen
  \bibfield  {author} {\bibinfo {author} {\bibfnamefont {H.}~\bibnamefont
  {Rieger}},\ }\bibfield  {title} {\enquote {\bibinfo {title} {{Critical
  behavior of the three-dimensional random-field Ising model: Two-exponent
  scaling and discontinuous transition}},}\ }\href@noop {} {\bibfield
  {journal} {\bibinfo  {journal} {Phys. Rev. B}\ }\textbf {\bibinfo {volume}
  {52}},\ \bibinfo {pages} {6659} (\bibinfo {year} {1995})}\BibitemShut
  {NoStop}%
\bibitem [{\citenamefont {Halasz}\ \emph {et~al.}(1998)\citenamefont {Halasz},
  \citenamefont {Jackson}, \citenamefont {Shrock}, \citenamefont {Stephanov},\
  and\ \citenamefont {Verbaarschot}}]{Halasz:1998qr}%
  \BibitemOpen
  \bibfield  {author} {\bibinfo {author} {\bibfnamefont {M.~A.}\ \bibnamefont
  {Halasz}}, \bibinfo {author} {\bibfnamefont {A.~D.}\ \bibnamefont {Jackson}},
  \bibinfo {author} {\bibfnamefont {R.~E.}\ \bibnamefont {Shrock}}, \bibinfo
  {author} {\bibfnamefont {M.~A.}\ \bibnamefont {Stephanov}}, \ and\ \bibinfo
  {author} {\bibfnamefont {J.~J.~M.}\ \bibnamefont {Verbaarschot}},\ }\bibfield
   {title} {\enquote {\bibinfo {title} {{On the phase diagram of QCD}},}\
  }\href {\doibase 10.1103/PhysRevD.58.096007} {\bibfield  {journal} {\bibinfo
  {journal} {Phys. Rev. D}\ }\textbf {\bibinfo {volume} {58}},\ \bibinfo
  {pages} {096007} (\bibinfo {year} {1998})}\BibitemShut {NoStop}%
\bibitem [{\citenamefont {Stephanov}\ \emph {et~al.}(1998)\citenamefont
  {Stephanov}, \citenamefont {Rajagopal},\ and\ \citenamefont
  {Shuryak}}]{Stephanov:1998dy}%
  \BibitemOpen
  \bibfield  {author} {\bibinfo {author} {\bibfnamefont {M.~A.}\ \bibnamefont
  {Stephanov}}, \bibinfo {author} {\bibfnamefont {K.}~\bibnamefont
  {Rajagopal}}, \ and\ \bibinfo {author} {\bibfnamefont {Edward~V.}\
  \bibnamefont {Shuryak}},\ }\bibfield  {title} {\enquote {\bibinfo {title}
  {{Signatures of the tricritical point in QCD}},}\ }\href {\doibase
  10.1103/PhysRevLett.81.4816} {\bibfield  {journal} {\bibinfo  {journal}
  {Phys. Rev. Lett.}\ }\textbf {\bibinfo {volume} {81}},\ \bibinfo {pages}
  {4816} (\bibinfo {year} {1998})}\BibitemShut {NoStop}%
\bibitem [{\citenamefont {Sinyukov}\ \emph {et~al.}(1998)\citenamefont
  {Sinyukov}, \citenamefont {Lednicky}, \citenamefont {Akkelin}, \citenamefont
  {Pluta},\ and\ \citenamefont {Erazmus}}]{Sinyukov:1998fc}%
  \BibitemOpen
  \bibfield  {author} {\bibinfo {author} {\bibfnamefont {{\relax
  Yu}.}~\bibnamefont {Sinyukov}}, \bibinfo {author} {\bibfnamefont
  {R.}~\bibnamefont {Lednicky}}, \bibinfo {author} {\bibfnamefont {S.~V.}\
  \bibnamefont {Akkelin}}, \bibinfo {author} {\bibfnamefont {J.}~\bibnamefont
  {Pluta}}, \ and\ \bibinfo {author} {\bibfnamefont {B.}~\bibnamefont
  {Erazmus}},\ }\bibfield  {title} {\enquote {\bibinfo {title} {{Coulomb
  corrections for interferometry analysis of expanding hadron systems}},}\
  }\href {\doibase 10.1016/S0370-2693(98)00653-4} {\bibfield  {journal}
  {\bibinfo  {journal} {Phys. Lett. B}\ }\textbf {\bibinfo {volume} {432}},\
  \bibinfo {pages} {248} (\bibinfo {year} {1998})}\BibitemShut {NoStop}%
%%CITATION = PHLTA,B432,248;%%
\bibitem [{\citenamefont {Bowler}(1991)}]{Bowler:1991vx}%
  \BibitemOpen
  \bibfield  {author} {\bibinfo {author} {\bibfnamefont {M.~G.}\ \bibnamefont
  {Bowler}},\ }\bibfield  {title} {\enquote {\bibinfo {title} {{Coulomb
  corrections to Bose-Einstein correlations have been greatly exaggerated}},}\
  }\href {\doibase 10.1016/0370-2693(91)91541-3} {\bibfield  {journal}
  {\bibinfo  {journal} {Phys. Lett. B}\ }\textbf {\bibinfo {volume} {270}},\
  \bibinfo {pages} {69} (\bibinfo {year} {1991})}\BibitemShut {NoStop}%
%%CITATION = PHLTA,B270,69;%%
\bibitem [{\citenamefont {Boggild}\ \emph {et~al.}(1995)\citenamefont {Boggild}
  \emph {et~al.}}]{Boggild:1994vk}%
  \BibitemOpen
  \bibfield  {author} {\bibinfo {author} {\bibfnamefont {H.}~\bibnamefont
  {Boggild}} \emph {et~al.} (\bibinfo {collaboration} {NA44 Collaboration}),\
  }\bibfield  {title} {\enquote {\bibinfo {title} {{Directional dependence of
  the pion source in high-energy heavy ion collisions}},}\ }\href {\doibase
  10.1016/0370-2693(95)00305-5} {\bibfield  {journal} {\bibinfo  {journal}
  {Phys. Lett. B}\ }\textbf {\bibinfo {volume} {349}},\ \bibinfo {pages} {386}
  (\bibinfo {year} {1995})}\BibitemShut {NoStop}%
%%CITATION = PHLTA,B349,386;%%
\bibitem [{\citenamefont {Zajc}(1982)}]{Zajc:1982vf}%
  \BibitemOpen
  \bibfield  {author} {\bibinfo {author} {\bibfnamefont {W.~A.}\ \bibnamefont
  {Zajc}},\ }\emph {\bibinfo {title} {{Two pion correlations in heavy ion
  collisions}}},\ \href@noop {} {Ph.D. thesis},\ \bibinfo  {school} {LBL,
  Berkeley} (\bibinfo {year} {1982})\BibitemShut {NoStop}%
%%CITATION = LBL-14864;%%
\bibitem [{\citenamefont {James}\ and\ \citenamefont
  {Roos}(1975)}]{James:1975dr}%
  \BibitemOpen
  \bibfield  {author} {\bibinfo {author} {\bibfnamefont {F.}~\bibnamefont
  {James}}\ and\ \bibinfo {author} {\bibfnamefont {M.}~\bibnamefont {Roos}},\
  }\bibfield  {title} {\enquote {\bibinfo {title} {{Minuit: A System for
  Function Minimization and Analysis of the Parameter Errors and
  Correlations}},}\ }\href {\doibase 10.1016/0010-4655(75)90039-9} {\bibfield
  {journal} {\bibinfo  {journal} {Comput. Phys. Commun.}\ }\textbf {\bibinfo
  {volume} {10}},\ \bibinfo {pages} {343} (\bibinfo {year} {1975})}\BibitemShut
  {NoStop}%
%%CITATION = CPHCB,10,343;%%
\bibitem [{\citenamefont {Cs{\"o}rg\H{o}}(2009)}]{Csorgo:2009wc}%
  \BibitemOpen
  \bibfield  {author} {\bibinfo {author} {\bibfnamefont {T.}~\bibnamefont
  {Cs{\"o}rg\H{o}}},\ }\bibfield  {title} {\enquote {\bibinfo {title}
  {{Critical Opalescence: An Optical Signature for a QCD Critical Point}},}\
  }\bibfield  {booktitle} {\emph {\bibinfo {booktitle} {{Proceedings, 5th
  International Workshop on Critical point and onset of deconfinement (CPOD
  2009): Upton, USA, June 8-12, 2009}}},\ }\href@noop {} {\bibfield  {journal}
  {\bibinfo  {journal} {PoS}\ }\textbf {\bibinfo {volume} {CPOD2009}},\
  \bibinfo {pages} {035} (\bibinfo {year} {2009})}\BibitemShut {NoStop}%
\bibitem [{\citenamefont {Akkelin}\ and\ \citenamefont
  {Sinyukov}(1996)}]{Akkelin:1996sg}%
  \BibitemOpen
  \bibfield  {author} {\bibinfo {author} {\bibfnamefont {S.~V.}\ \bibnamefont
  {Akkelin}}\ and\ \bibinfo {author} {\bibfnamefont {{\relax Yu}.~M.}\
  \bibnamefont {Sinyukov}},\ }\bibfield  {title} {\enquote {\bibinfo {title}
  {{The HBT-interferometry of expanding inhomogeneous sources}},}\ }\href
  {\doibase 10.1007/s002880050271} {\bibfield  {journal} {\bibinfo  {journal}
  {Z. Phys. C}\ }\textbf {\bibinfo {volume} {72}},\ \bibinfo {pages} {501}
  (\bibinfo {year} {1996})}\BibitemShut {NoStop}%
%%CITATION = ZEPYA,C72,501;%%
\bibitem [{\citenamefont {Akkelin}\ and\ \citenamefont
  {Sinyukov}(1995)}]{Akkelin:1995gh}%
  \BibitemOpen
  \bibfield  {author} {\bibinfo {author} {\bibfnamefont {S.~V.}\ \bibnamefont
  {Akkelin}}\ and\ \bibinfo {author} {\bibfnamefont {{\relax Yu}.~M.}\
  \bibnamefont {Sinyukov}},\ }\bibfield  {title} {\enquote {\bibinfo {title}
  {{The HBT interferometry of expanding sources}},}\ }\href {\doibase
  10.1016/0370-2693(95)00765-D} {\bibfield  {journal} {\bibinfo  {journal}
  {Phys. Lett. B}\ }\textbf {\bibinfo {volume} {356}},\ \bibinfo {pages} {525}
  (\bibinfo {year} {1995})}\BibitemShut {NoStop}%
%%CITATION = PHLTA,B356,525;%%
\bibitem [{\citenamefont {Cs{\"o}rg{\H{o}}}\ \emph {et~al.}(1994)\citenamefont
  {Cs{\"o}rg{\H{o}}}, \citenamefont {L{\"o}rstad},\ and\ \citenamefont
  {Zim\'anyi}}]{Csorgo:1994fg}%
  \BibitemOpen
  \bibfield  {author} {\bibinfo {author} {\bibfnamefont {T.}~\bibnamefont
  {Cs{\"o}rg{\H{o}}}}, \bibinfo {author} {\bibfnamefont {B.}~\bibnamefont
  {L{\"o}rstad}}, \ and\ \bibinfo {author} {\bibfnamefont {J.}~\bibnamefont
  {Zim\'anyi}},\ }\bibfield  {title} {\enquote {\bibinfo {title} {{Quantum
  statistical correlations for slowly expanding systems}},}\ }\href {\doibase
  10.1016/0370-2693(94)91356-0} {\bibfield  {journal} {\bibinfo  {journal}
  {Phys. Lett. B}\ }\textbf {\bibinfo {volume} {338}},\ \bibinfo {pages} {134}
  (\bibinfo {year} {1994})}\BibitemShut {NoStop}%
\bibitem [{\citenamefont {Csizmadia}\ \emph {et~al.}(1998)\citenamefont
  {Csizmadia}, \citenamefont {Cs{\"o}rg{\H{o}}},\ and\ \citenamefont
  {Luk\'acs}}]{Csizmadia:1998ef}%
  \BibitemOpen
  \bibfield  {author} {\bibinfo {author} {\bibfnamefont {P.}~\bibnamefont
  {Csizmadia}}, \bibinfo {author} {\bibfnamefont {T.}~\bibnamefont
  {Cs{\"o}rg{\H{o}}}}, \ and\ \bibinfo {author} {\bibfnamefont
  {B.}~\bibnamefont {Luk\'acs}},\ }\bibfield  {title} {\enquote {\bibinfo
  {title} {{New analytic solutions of the nonrelativistic hydrodynamical
  equations}},}\ }\href {\doibase 10.1016/S0370-2693(98)01297-0} {\bibfield
  {journal} {\bibinfo  {journal} {Phys. Lett. B}\ }\textbf {\bibinfo {volume}
  {443}},\ \bibinfo {pages} {21} (\bibinfo {year} {1998})}\BibitemShut
  {NoStop}%
\bibitem [{\citenamefont {Csan\'ad}\ and\ \citenamefont
  {Vargyas}(2010)}]{Csanad:2009wc}%
  \BibitemOpen
  \bibfield  {author} {\bibinfo {author} {\bibfnamefont {M.}~\bibnamefont
  {Csan\'ad}}\ and\ \bibinfo {author} {\bibfnamefont {M.}~\bibnamefont
  {Vargyas}},\ }\bibfield  {title} {\enquote {\bibinfo {title} {{Observables
  from a solution of 1+3 dimensional relativistic hydrodynamics}},}\ }\href
  {\doibase 10.1140/epja/i2010-10973-3} {\bibfield  {journal} {\bibinfo
  {journal} {Eur. Phys. J. A}\ }\textbf {\bibinfo {volume} {44}},\ \bibinfo
  {pages} {473} (\bibinfo {year} {2010})}\BibitemShut {NoStop}%
\bibitem [{\citenamefont {Cs{\"o}rg{\H{o}}}(2001)}]{Csorgo:2000vs}%
  \BibitemOpen
  \bibfield  {author} {\bibinfo {author} {\bibfnamefont {T.}~\bibnamefont
  {Cs{\"o}rg{\H{o}}}},\ }\bibfield  {title} {\enquote {\bibinfo {title}
  {{Particle interferometry, binary sources and oscillations in two particle
  correlations}},}\ }\bibfield  {booktitle} {\emph {\bibinfo {booktitle}
  {{Multiparticle production: New frontiers in soft physics and correlations on
  the threshold of the third millennium. Proceedings, 9th International
  Workshop, Torino, Italy, June 12-17, 2000}}},\ }\href {\doibase
  10.1016/S0920-5632(00)01037-9} {\bibfield  {journal} {\bibinfo  {journal}
  {Nucl. Phys. Proc. Suppl.}\ }\textbf {\bibinfo {volume} {92}},\ \bibinfo
  {pages} {223} (\bibinfo {year} {2001})}\BibitemShut {NoStop}%
\bibitem [{\citenamefont {Akiba}\ \emph {et~al.}(1993)\citenamefont {Akiba}
  \emph {et~al.}}]{Akiba:1992cj}%
  \BibitemOpen
  \bibfield  {author} {\bibinfo {author} {\bibfnamefont {Y.}~\bibnamefont
  {Akiba}} \emph {et~al.} (\bibinfo {collaboration} {E-802 Collaboration}),\
  }\bibfield  {title} {\enquote {\bibinfo {title} {{Bose-Einstein correlation
  of kaons in Si + Au collisions at 14.6-A/GeV/c}},}\ }\href {\doibase
  10.1103/PhysRevLett.70.1057} {\bibfield  {journal} {\bibinfo  {journal}
  {Phys. Rev. Lett.}\ }\textbf {\bibinfo {volume} {70}},\ \bibinfo {pages}
  {1057} (\bibinfo {year} {1993})}\BibitemShut {NoStop}%
%%CITATION = PRLTA,70,1057;%%
\bibitem [{\citenamefont {Csan\'ad}\ \emph {et~al.}(2007)\citenamefont
  {Csan\'ad}, \citenamefont {Cs{\"o}rg\H{o}},\ and\ \citenamefont
  {Nagy}}]{Csanad:2007fr}%
  \BibitemOpen
  \bibfield  {author} {\bibinfo {author} {\bibfnamefont {M.}~\bibnamefont
  {Csan\'ad}}, \bibinfo {author} {\bibfnamefont {T.}~\bibnamefont
  {Cs{\"o}rg\H{o}}}, \ and\ \bibinfo {author} {\bibfnamefont {M.}~\bibnamefont
  {Nagy}},\ }\bibfield  {title} {\enquote {\bibinfo {title} {{Anomalous
  diffusion of pions at RHIC}},}\ }\bibfield  {booktitle} {\emph {\bibinfo
  {booktitle} {{Particle correlations and femtoscopy. Proceedings, 2nd
  Workshop, WPCF 2006, Sao Paulo, Brazil, September 9-11, 2006}}},\ }\href
  {\doibase 10.1590/S0103-97332007000600018} {\bibfield  {journal} {\bibinfo
  {journal} {Braz. J. Phys.}\ }\textbf {\bibinfo {volume} {37}},\ \bibinfo
  {pages} {1002} (\bibinfo {year} {2007})}\BibitemShut {NoStop}%
\bibitem [{\citenamefont {Zajc}(1993)}]{Zajc:1992sz}%
  \BibitemOpen
  \bibfield  {author} {\bibinfo {author} {\bibfnamefont {W.~A.}\ \bibnamefont
  {Zajc}},\ }\bibfield  {title} {\enquote {\bibinfo {title} {{A pedestrian's
  guide to interferometry}},}\ }\bibfield  {booktitle} {\emph {\bibinfo
  {booktitle} {{NATO Advanced Study Institute on Particle Production in Highly
  Excited Matter Castelvecchio Pascoli, Italy, July 12-24, 1992}}},\
  }\href@noop {} {\bibfield  {journal} {\bibinfo  {journal} {NATO Sci. Ser. B}\
  }\textbf {\bibinfo {volume} {303}},\ \bibinfo {pages} {435} (\bibinfo {year}
  {1993})}\BibitemShut {NoStop}%
%%CITATION = INSPIRE-349197;%%
\bibitem [{\citenamefont {{M. Csan\'ad for the PHENIX
  Collaboration}}(2006)}]{Csanad:2005nr}%
  \BibitemOpen
  \bibfield  {author} {\bibinfo {author} {\bibnamefont {{M. Csan\'ad for the
  PHENIX Collaboration}}},\ }\bibfield  {title} {\enquote {\bibinfo {title}
  {Measurement and analysis of two- and three-particle correlations},}\
  }\href@noop {} {\bibfield  {journal} {\bibinfo  {journal} {Nucl. Phys. A}\
  }\textbf {\bibinfo {volume} {774}},\ \bibinfo {pages} {611} (\bibinfo {year}
  {2006})}\BibitemShut {NoStop}%
\bibitem [{\citenamefont {Sinyukov}\ and\ \citenamefont
  {Tolstykh}(1994)}]{Sinyukov:1994en}%
  \BibitemOpen
  \bibfield  {author} {\bibinfo {author} {\bibfnamefont {{\relax Yu}.~M.}\
  \bibnamefont {Sinyukov}}\ and\ \bibinfo {author} {\bibfnamefont {Y.~{\relax
  Yu}.}\ \bibnamefont {Tolstykh}},\ }\bibfield  {title} {\enquote {\bibinfo
  {title} {{Coherence influence on the Bose-Einstein correlations}},}\ }\href
  {\doibase 10.1007/BF01552626} {\bibfield  {journal} {\bibinfo  {journal} {Z.
  Phys. C}\ }\textbf {\bibinfo {volume} {61}},\ \bibinfo {pages} {593}
  (\bibinfo {year} {1994})}\BibitemShut {NoStop}%
%%CITATION = ZEPYA,C61,593;%%
\bibitem [{\citenamefont {Pratt}(1993)}]{Pratt:1993uy}%
  \BibitemOpen
  \bibfield  {author} {\bibinfo {author} {\bibfnamefont {S.}~\bibnamefont
  {Pratt}},\ }\bibfield  {title} {\enquote {\bibinfo {title} {{Pion lasers from
  high-energy collisions}},}\ }\href {\doibase 10.1016/0370-2693(93)90682-8}
  {\bibfield  {journal} {\bibinfo  {journal} {Phys. Lett. B}\ }\textbf
  {\bibinfo {volume} {301}},\ \bibinfo {pages} {159} (\bibinfo {year}
  {1993})}\BibitemShut {NoStop}%
%%CITATION = PHLTA,B301,159;%%
\bibitem [{\citenamefont {Cs{\"o}rg{\H{o}}}\ and\ \citenamefont
  {Zim\'anyi}(1998)}]{Csorgo:1997us}%
  \BibitemOpen
  \bibfield  {author} {\bibinfo {author} {\bibfnamefont {T.}~\bibnamefont
  {Cs{\"o}rg{\H{o}}}}\ and\ \bibinfo {author} {\bibfnamefont {J.}~\bibnamefont
  {Zim\'anyi}},\ }\bibfield  {title} {\enquote {\bibinfo {title} {{Analytic
  solution of the pion- laser model}},}\ }\href {\doibase
  10.1103/PhysRevLett.80.916} {\bibfield  {journal} {\bibinfo  {journal} {Phys.
  Rev. Lett.}\ }\textbf {\bibinfo {volume} {80}},\ \bibinfo {pages} {916}
  (\bibinfo {year} {1998})}\BibitemShut {NoStop}%
\bibitem [{\citenamefont {Kaneta}\ and\ \citenamefont
  {Xu}(2004)}]{Kaneta:2004zr}%
  \BibitemOpen
  \bibfield  {author} {\bibinfo {author} {\bibfnamefont {M.}~\bibnamefont
  {Kaneta}}\ and\ \bibinfo {author} {\bibfnamefont {N.}~\bibnamefont {Xu}},\
  }\bibfield  {title} {\enquote {\bibinfo {title} {{Centrality dependence of
  chemical freeze-out in Au+Au collisions at RHIC}},}\ }in\ \href@noop {}
  {\emph {\bibinfo {booktitle} {{Write-up of a poster presented at the 17th
  International Conference on Ultra-relativistic Nucleus-Nucleus Collisions,
  Oakland, USA}}}}\ (\bibinfo {year} {2004})\BibitemShut {NoStop}%
\end{thebibliography}

 
\end{document}